\documentclass[a4paper,11pt]{article}
\pdfoutput=1
\usepackage{amssymb,amsmath,amsfonts,makeidx,placeins,pbox,multirow}
\usepackage{graphicx,rotate,subcaption,color,slashed,cite,caption,verbatim}
\usepackage{longtable,tabu}
\usepackage{array}
\usepackage{pstricks}
\usepackage{color}
\usepackage{amsmath}
\usepackage{mathtools}
\usepackage{amssymb}
\usepackage[normalem]{ulem}
\newcolumntype{P}[1]{>{\centering\arraybackslash}p{#1}}
\usepackage[colorlinks=true,
            linkcolor=magenta,
            urlcolor=blue,
            citecolor=blue]{hyperref}

\def\lapp{\mathrel{\rlap{\raise.5ex\hbox{$<$}}
                    {\lower.5ex\hbox{$\sim$}}}}
\def\gapp{\mathrel{\rlap{\raise.5ex\hbox{$>$}}
                    {\lower.5ex\hbox{$\sim$}}}}

\usepackage{color}
\long\def\/*#1*/{}
\usepackage{color}
 \usepackage[normalem]{ulem}
\definecolor{darkgreen}{cmyk}{1,0,1,0.4}
\definecolor{darkred}{cmyk}{0,1,1,0.4}

\evensidemargin=\oddsidemargin
\def\bar {\overline}

\def\bea {\begin{eqnarray}}
\def\eea {\end{eqnarray}}

\def\beq{\begin{equation}}
\def\eeq{\end{equation}}
\def\barr{\begin{array}}
\def\earr{\end{array}}
\def\dis{\displaystyle}

\def\gev{\,\ensuremath{\mathrm{Ge\kern -0.1em V}}}
\def\tev{\,\ensuremath{\mathrm{Te\kern -0.1em V}}}

\DeclareUnicodeCharacter{2212}{-}
\UseRawInputEncoding
\begin{document}
\begin{center}
{\Large {\bf Looking for a vectorlike B quark at LHC using jet substructure }} \\
\vspace*{0.8cm} {\sf  Debajyoti Choudhury\footnote{debchou.physics@gmail.com}, Kuldeep Deka\footnote{kuldeepdeka.physics@gmail.com}, Nilanjana Kumar\footnote{nilanjana.kumar@gmail.com}} \\
\vspace{10pt} {\small } {\em Department of Physics and Astrophysics, University of Delhi, Delhi 110007, India.} \\
\normalsize
\end{center}
\bigskip
\begin{abstract}
  Vectorlike quarks have been shown to resolve certain long-standing
  discrepancies pertaining to the bottom sector. We investigate, here,
  the prospects of identifying the existence of a topless
  vectorlike doublet $(B,~Y)$, as is preferred by
  the electroweak precision measurements.
  Concentrating on single production, {\em viz.} $B \bar b$ with $B
  \to b + Z/H$ subsequently, we find that the fully hadronic
  decay-channel is susceptible to discovery provided jet substructure
  observables are used. At the 13 TeV LHC with an integrated
  luminosity of 300 fb$^{-1}$, a modest value of the chromomagnetic
  transition moments allows for the exclusion of $M \lapp 1.8 \, (2.2)
  \tev$ in the $Z$ and $H$ channels respectively.

\end{abstract}
\tableofcontents
\section{Introduction}
\label{sec:intro}
Over the last three decades, the Standard Model (SM) has been
vindicated to an unprecedented degree of accuracy by several
experiments, taken singly or in conjunction. These include, but are
not limited to, the electroweak precision tests~\cite{ALEPH:2010aa},
the measurement of the top mass~\cite{10.1093/ptep/ptaa104}
culminating with the discovery of the Higgs
particle~\cite{Aad:2012tfa,Chatrchyan:2012ufa}. Yet, certain anomalies
persist, starting from the $2.9 \sigma$ discrepancy in the
forward-backward asymmetry ($A^b_{FB}$) of the $b$-quark at
LEP/SLC~\cite{ALEPH:2010aa}, a newly updated $4.2\sigma$ deviation in the anomalous
magnetic moment of the muon~\cite{PhysRevLett.126.141801} to the more
recent ones in both charged and neutral-current decays of
$B$-mesons~\cite{Amhis:2016xyh}. These are accompanied by the SM's failure
in explaining neutrino masses and mixings on the one hand and, on the other, to
address questions related to the lightness of the Higgs or
the preponderance of matter over antimatter.

Many diverse scenarios going beyond the SM have been proposed to
address such issues. Near-ubiquitous in all of them are extra fermion
fields. With most such fields not being gauge-singlets,
non-observation at collider experiments can only mean that they are
very heavy. A chiral assignment of quantum numbers (as is the case
within the SM) for heavy fields would imply uncomfortably large Yukawa
couplings and, hence, an acute tension with not only the
electroweak precision tests, but also with Higgs production and decay.
Non-chiral fermions, since they allow for
gauge-invariant bare mass terms, easily evade such restrictions.

Such vector-like assignments abound in many diverse
scenarios\cite{Aguilar-Saavedra:2013qpa,Ellis:2014dza}.  Even the
simplest supersymmetric construct has them in the form of the
Higgsinos. Additional vector-like matter may arise in the quest of
enlarging the symmetry, whether
gauge~\cite{Kang:2007ib,Dermisek:2012ke,Bhattacherjee:2017cxh,EmmanuelCosta:2005nh,Barger:2006fm,Dorsner:2014wva},
or space-time~\cite{Choi:2010gc}, with the Dirac-like gauginos in the
latter suppressing pair production channels and cascade decays that
are possible for the usual Majorana-gaugino
theories. Similarly, extending the Higgs-sector to obtain an 
$R$-symmetric theory has very interesting
ramifications~\cite{Choi:2010an}.  Vector-like matter can also help
raise the mass of the SM-like Higgs~\cite{Martin:2009bg,Martin:2010dc}
in supersymmetric theories or to alleviate the little hierarchy
problem~\cite{Graham:2009gy}. And, finally, many of the problems that
beset gauge-mediated breaking of supersymmetry may be alleviated
too~\cite{Moroi:2011aa,Endo:2011xq,Martin:2012dg,Fischler:2013tva}.

Non-chiral fermions abound in nonsupersymmetric
scenarios as well, historically interesting examples being provided by
models wherein electroweak symmetry
breaking proceeds dynamically through the
condensation of top quarks (or its
partners)~\cite{Dobrescu:1997nm,Chivukula:1998wd,He:1999vp}.  And
while they are, almost by definition, ubiquitous in any
extra-dimensional model wherein the SM fields venture into the bulk
(in the form of Kaluza-Klein excitations), they are also present in
composite
Higgs~\cite{Contino:2006qr,Anastasiou:2009rv,Vignaroli:2012sf,
DeSimone:2012fs,Delaunay:2013iia,Gillioz:2013pba,Banerjee:2017wmg}, as
well as little Higgs
scenarios~\cite{Han:2003wu,Carena:2006jx,Choudhury:2006mp,Choudhury:2006sq,
Matsumoto:2008fq,Choudhury:2012xc,Berger:2012ec}, or those advocating
a Higgs-portal to obtain the correct relic abundance for dark
matter~\cite{Patt:2006fw,Gopalakrishna:2009yz,Baek:2011aa}.

As argued above, to address different issues, a plethora of
vector-like fermions have been advocated, either singly\footnote{Note
that anomaly cancellation is not an issue.}  or in combinations.
For example, these have been invoked to explain the longstanding
discrepancy in the muon anomalous moment~\cite{Chun:2020uzw,Kowalska:2017iqv}
or to address flavour discrepancing pertaining to $b$-quarks~\cite{Ishiwata:2015cga,Barman:2018jhz,Crivellin:2020oup,Cheung:2020vqm}.
In
the present analysis, we would be concentrating on a single such
vector-like quark (VLQ) doublet. Since substantial mixings with the SM
quarks are liable to set up too large a FCNC, we assume that it is
only with the third-generation that it may participate in Yukawa
interactions\footnote{In the absence of any such mixing, the lighter
of the two VLQs would be stable, and, hence run afoul of constraints from
the nonexistence of exotic bound states.}. Such scenarios had been
shown, in Ref.\cite{Choudhury:2001hs}, to explain the long-standing
problem with $A^b_{FB}$ without coming into conflict with any other
constraint pertaining to electroweak precision tests.  Interestingly,
while, of the two quantum number assignments proposed in
Ref.\cite{Choudhury:2001hs}, one preferred relatively light VLQs (and,
hence, would be disfavoured by the absence of signals at the LHC), the
other is not only more attractive on theoretical grounds, but also
preferred VLQs in the TeV-range.

In the present paper, we are concerned with a VLQ that, amongst other
things, is useful in ameliorating the longstanding tension between the
global fits to electroweak precision tests~\cite{ALEPH:2010aa} and
$A^b_{FB}$. The latter, taken in conjunction with the $Z\to \bar b
b$ branching ratio and the forward-backward asymmetry for the
charm-quark requires a substantial new physics effect in the
$b$-sector~\cite{Chanowitz:2001bv}. This is best brought forward by
considering an effective $Z\bar b b$ vertex parameterised by
\[
g_{L,R} \to g_{L,R} + \delta g_{L,R}
\]
where the tree-level SM couplings are $g_L=1/2+s_W^2/3 \approx -0.42$ and
$g_R= s_W^2/3 \approx -0.077$, with $s_W$ being the Weinberg angle.
The deviations $\delta g_{L,R}$ could
arise either from quantum corrections (as in the SM) or new physics
effects. In terms of the effective couplings, the expressions for 
$ R_b \equiv \Gamma(Z \to \bar b b)/\Gamma(Z \to {\rm hadrons})$ and $A^b_{FB}$ 
are approximated to the leading order by
\begin{equation}
R_b = R_b^{SM}(1-1.820 \delta g_L +0.336 \delta g_R)\ , \qquad \quad 
A^b_{FB} = A^{b,SM}_{FB}(1-0.164\delta g_L-0.8877\delta g_R) \ .
\label{zpole}
\end{equation}
With the measured value of $A^b_{FB}$ being substantially less than
the SM prediction, but that for $R_b$ being marginally higher, a
positive $\delta g_R$ is called for. As was argued in
Ref.\cite{Choudhury:2001hs}, the required magnitude of deviation is
too large to be attributable to loop corrections. On the other hand,
a tree-level $\delta g_R$ (along with a tiny, but welcome, $\delta g_L$)
is naturally generated if the $b$ quark mixes with a VLQ that
is part of a $SU(2)_L$-doublet. The standard choice of the
hypercharge ($Y = 1/6$) would require a very large $\delta g_R$, such that
the sign of the effective $g_R$ is reversed~\cite{Choudhury:2001hs}.
The ensuing large mixing has severe ramifications in low-energy physics.
This, along with LHC constraints as well as several theoretical
considerations~\cite{Choudhury:2001hs}
implies that $Y = -5/6$ (leading to a top-less doublet, $\psi_{L,R}^T=(B ~Y)$) is favoured. 

The production and subsequent decays of such VLQs at the LHC has been
widely studied in the literature and constraints are imposed from
nonobservation~\cite{Buchkremer:2013bha, Nutter:2012an,
Gong:2019zws,Cai:2012ji,
ATLAS:2018qxs,Aaboud:2018pii,Sirunyan:2018ncp,Sirunyan:2017fho}.
While QCD-driven pair-production is bereft of any model-dependence
and, naively, expected to be the dominant mode at the LHC, for large
masses, the kinematical suppression as well as the suppression due to
falling gluon densities quickly limit the sensitivity. Single
production, being model-dependent, has the added advantage of
providing a window to the ultraviolet completion. A particularly
interesting set of couplings that this may probe are the transition
magnetic moments, whether of electroweak nature or
chromomagnetic. Simple strategies to probe this was developed in
Refs.\cite{Bhattacharya:2007gs,Bhattacharya:2009xg} and, subsequently,
exploited by the CMS
collaboration~\cite{Khachatryan:2014aka,Sirunyan:2017fho} to look at a
singly-produced VLQ decaying into its SM counterpart and a
photon. Somewhat analogous is the case of a VLQ decaying into a
bottom quark and a $Z$~\cite{Gong:2019zws} or a Higgs
boson~\cite{ATLAS:2018qxs}.

While both the pair-production and single-production modes have already been used by the ATLAS and CMS
collaboration~\cite{ATLAS:2018qxs,Aaboud:2018pii,Sirunyan:2018qau,Sirunyan:2018fjh,Sirunyan:2018omb}
to impose a lower limit of $m_B \gapp
1-1.3$~TeV~\cite{Buckley:2020wzk} for such a VLQ coupling primarily to
the $b$-quark, it still behoves us to look for complementary
channels. To this end, we consider the production of such a VLQ, in
association with $b$ quark and subsequent decays of VLQ into either a
$b+Z$ or a $b+H$ pair, as these modes are associated with a very
substantial branching fraction for a large part of the parameter
space.  With a large $m_B$ implying that both the daughters would be
highly boosted, the $Z$ (or $H$) is more likely to manifest itself as
a single fat jet rather than two resolved jets.  Notwithstanding the
subtleties involved in the use of jet substructures, the much larger
hadronic branching ratio for the $Z$ and $H$ renders these channels
very competitive with the leptonic ones. While fatjet signatures have
been investigated in the context of VLQs~\cite{Chatrchyan:2013uxa,
Khachatryan:2015axa, Aad:2015voa, Aad:2016qpo}, ours is the first
effort to use it for this all hadronic final state.

The rest of the paper has been structured as follows. We begin, in
Sec.\ref{sec:theory}, by setting up the formalism and detailing the
relevant aspects of the parameter space that we would explore. This is
followed, in Sec.\ref{sec:substructure}, by a brief recounting of jet
substructure as this would be the bedrock of our analysis. The details
of the analysis are presented in Sec.\ref{sec:collider} followed by
the concluding Sec.\ref{sec:outlook}. 

\section{Theoretical Backbone}
\label{sec:theory}
As we have explained in the preceding section, we would be
concentrating on a non-chiral colored $SU(2)_L$ fermion field with a
hypercharge of $-5/6$, {\rm viz.}, $\Psi_{L,R} \equiv (B',
Y')_{L,R}^T$.  With there being no charge-2/3 quark, several
constraints are automatically relaxed. Our motivation being to
develop a particular search algorithm, we desist from incorporating
the other fields included in Ref.\cite{Choudhury:2001hs}, and
begin by
setting up the formalism and reviewing some immediate consequences. 
\subsection{The Model}
The kinetic (including gauge interactions) term for the $\Psi_{L,R}$
fields are exactly analogous to those for the SM quarks. Being
vectorial in nature, these quarks can acquire a bare mass
$M$. Furthermore, in view of the strong constraints from flavour
physics, the only Yukawa coupling allowed to $\Psi_L$
is that with the SM field
$b_R$ alone(with gauge invariance preventing any such term of
$\Psi_R$). Denoting all the gauge (mass) eigenstates by primed
(unprimed) fields, the relevant additional term in the Lagrangian can
be written as
\begin{equation}
\mathcal{-L} \ni y_1 \bar{Q}_{3L} b'_R H + y_2 \bar{\Psi'}_L b'_R H  + M  \bar{\Psi'}_L \Psi'_R+ h.c,
\end{equation}
where $H$ is the SM Higgs doublet. 
After spontaneous symmetry breaking, the relevant mass terms are expressible as
\begin{equation}
  \mathcal{L}_{\rm mass} \ni
  \left[ \bar{b'}_L \quad \bar{B'}_L\,  \right] \, {\cal M} \,
  \begin{bmatrix}
b'_R \\
B'_R
\end{bmatrix} \ , \qquad \quad 
{\cal M} =  \begin{bmatrix}
y_1 v & 0 \\
y_2 v & M \\
\end{bmatrix} 
\end{equation}
where $v$ is the SM vev. For the sake of simplicity, we have neglected
here the much smaller terms connecting $b'$ to $s',d'$. Considerations
of electroweak precision tests (more specifically, the
$\rho$-parameter) stipulate that $y_2 v \ll M$.  The mass matrix can
be diagonalized by a bi-unitary transformation defined by
\begin{equation}
\begin{bmatrix}
b_{L,R} \\
B_{L,R}
\end{bmatrix}=\begin{bmatrix}
c_{L,R} & -s_{L,R}  \\
s_{L,R} & c_{L,R} \\
\end{bmatrix}\begin{bmatrix}
b'_{L,R} \\
B'_{L,R}
\end{bmatrix}
\end{equation}
where $c_{L, R} \equiv \cos \theta_{L,R}$ and $s_{L,R}\equiv\sin \theta_{L,R}$.
The physical masses and the mixing angles are given by
\begin{eqnarray}
  m_b\approx y_1 v \left[1+\frac{y_2^2 v^2}{M^2}\right]^{-1/2} \ ,
  \qquad \quad
  \tan\theta_R\approx \frac {-y_2 v}{M} \ ,
  \qquad \quad
\tan\theta_L\approx \frac{-y_1 y_2 v^2}{(M^2+y_2^2 v^2)} \ .
\label{eq:value}
\end{eqnarray}
It is worth remembering that, for small mixing angles,
\begin{equation}
s_L \approx \frac{m_b}{M} s_R \ .
\label{sl}
\end{equation}
It is easy to ascertain (for example, from the very structure of ${\cal M}^2$)
that 
\begin{eqnarray}
m_Y^2 = M^2= m_B^2 c_R^2+m_b^2 s_R^2 \ .
\end{eqnarray}
While $m_B > m_Y$, the smallness of the ratio $(y_2 v / M)$ ensures
that the splitting is too small to permit $B \to Y + W^-$, and
even $B \to Y + \bar f + f'$ is very small indeed. The electroweak
interactions of the mass-eigenstate $B$ are obtained
trivially. Restricting ourselves to those that could, potentially,
lead to the decay of the $B$, we have
\begin{equation}
\barr{rcl}
\dis \mathcal{L_W}^{(B)}& = & \dis \frac{-g s_L}{\sqrt{2}} \;
      \bar{t} \gamma^{\mu} P_L \, B \; W_{\mu}^{+} + h.c\\[2.5ex]
\dis \mathcal{L_Z}^{(B)}& = & \dis \frac{g}{2 c_w} \,
\bar{b} \gamma^{\mu}\,
\left[2 s_L c_L P_L + s_R c_R P_R \right] \, B \; Z_{\mu}+ h.c
\\[2.5ex]
\dis \mathcal{L_H}^{(B)}& = & \dis \frac{-g s_R c_R}{2 m_W}\,
         \bar{b} \, \left[M_B \, P_L + m_b P_R \right]\, B \; H + h.c;
\earr
\label{eq:weak_int}
\end{equation}         
where $P_{L,R} = (1 \mp \gamma_5)/2$ and $g$ is the $SU(2)_L$ gauge coupling.
The mixing of quark fields of different isospin (and hypercharge) that
generated the interactions of eq.(\ref{eq:weak_int}) would also result
in an alteration of the effective gauge couplings of
the eigenstates dominated by the SM quarks. In particular, both
of the $\bar b b Z$ couplings $g_{L,R}(b)$ receive nonzero
corrections, leading, in turn, to corrections in $R_b$ and $A^b_{FB}$
(see eq.(\ref{zpole})). The most important parameter, in this context,
turns of to be $s_R$~\cite{Choudhury:2001hs}, with $s_L$ playing only
a subservient role is ensuring a correct $R_b$.  While observables
such as $R_c$ and $\Gamma_{\rm had}(Z)$ are affected too, these play
even more subservient roles. Thus, $R_b$ and $A^b_{FB}$ can be used to delineate the
appropriate part of the paramater space.  With the present scenario
relating $s_L$ and $s_R$ (with $s_L \ll s_R$, {\em vide}
eq.(\ref{sl})), only one of them is a free parameter. Furthermore,
with the aforementioned observables being essentially independent of
$m_B$ (since one-loop corrections due to the new quarks are too small
be of any consequence), the observables under discussion constrain
only $s_R$.

We see from eq.(\ref{eq:value}) that a perturbative limit on the
Yukawa coupling $y_2$ would translate to a bound on $s_R$, scaling
approximately as $m_B^{-1}$.  However, stronger constraints arise from observables such as 
$A^b_{FB}$ and $R_b$, as discussed in the preceding section. Although the other precision measurements,
  such as $A_b \, \equiv [(g_L^b)^2 - (g_R^b)^2] / [(g_L^b)^2 +
    (g_R^b)^2]$ and $R_c$ too contribute to the constraints, these
  have relatively minor roles\footnote{It should be remembered
      that with the beams being unpolarized at LEP, $A_b$ and
      $A^b_{FB}$ differ essentially by a factor of $A_e$. At the SLD,
      though, they differed on account of the polarization, but in a
      straightforward manner.} to play as far as this sector is
  concerned. For completeness, we quote here the
  ``SM'' values of the parameters, defined as the values inferred from
  the global best-fit~\cite{10.1093/ptep/ptaa104} to all electroweak
  precision observables (with quantum corrections incorporated).
  Restricting ourselves, for brevity's sake, to the most relevant
  ones, these are $R_b^{SM} = 0.21576 \, (0.21629 \pm, 0.00066)$,
$A_{FB}^{SM} = 0.1034\, (0.0992 \pm 0.0016)$, $A_b^{SM} = 0.9348
\,(0.923 \pm 0.020)$, $R_c^{SM} = 0.17227 \,(0.1721 \pm 0.003)$,
where the numbers in the parentheses refer to the direct
  experimental measurements~\cite{ALEPH:2005ab}. While all the
  electroweak precision measurements were considered in
  ref.~\cite{Choudhury:2001hs} in deriving the preferred values of the
  parameters, ref.~\cite{Aguilar-Saavedra:2013qpa} concentrated on
$R_b$ and $A_{FB}^{SM}$ in obtaining the ``best-fit'' value of $s_R$ ,
along with error bands. It should be realised,
  though, that the other precision electroweak
measurements\cite{ALEPH:2005ab}, too have an important role to
  play.  In particular, the oblique parameter $T$
(equivalently $\rho$), receives an additional contribution proportional to $s_R^2
m^2_B$ and this translates to a stringent
bound on $s_R \propto m_B^{-1}$ especially for larger
$m_B$ values. The exact results are shown in
\cite{Aguilar-Saavedra:2013qpa} where experimental value of $\Delta T$
and $\Delta S$ are taken to be $0.07 \pm 0.08$ and $0.04 \pm 0.07$
from \cite{10.1093/ptep/ptaa104}.

Typically, for $m_B \lapp 1 \tev$, it is the $Z \to b \bar b$
observables that impose stronger constraints, while for $m_B \gapp
1 \tev$, it is the constraints from the oblique parameters which are
stronger, and, for $m_B \gapp 1.7 \tev$, these tend to rule out the
values that best fit the $Z \to b \bar b$
observables~\cite{Aguilar-Saavedra:2013qpa}.  Being interested
primarily in $m_B \gapp 1.2 \tev$, we use the corresponding $1\sigma$
upper limits.

  The presence of such exotic quarks would also serve to alter the
  effective gluon-gluon-Higgs and photon-photon-Higgs vertices,
  nominally enhancing the former and suppressing the latter (on
  account of destructive interference with the $W$-loop). However,
  with $B$ being vector-like, its contribution to either loop can only
  be proportional to the mass term connect it to the SM quark.
  In other words, in the limit of large $m_B$, the new physics
contribution is proportional to $s_R$; even for
  its maximum possible value of $0.18$(as constrained by the $Z$-pole observables), a
  maximal enhancement of $3.2\%$ is admissible in the $ggH$ vertex and
  a maximal suppression of $0.2\%$ in the $\gamma\gamma H$ vertex. The
  partial decay width for $H \to b \bar b$ suffers a change on two
  accounts, primarily from the alteration of the $Hb \bar b$ vertex(by a factor of $c_R^2$)
  due to the mixing and, to a much smaller extent, the aforementioned
  change in $H \to gg$. The cumulative effect is a suppression by less
  than $6.4\%$. 
Consequently, the Higgs branching ratios into other final
states are enhanced, at most, by an extra 3.8\%.
On the experimental side, the production through $gg$ fusion has an
uncertainity of about $9\%$ and $H\rightarrow b \bar{b}$ has an
uncertainity of $10 \%$\cite{Aad:2019mbh}. If we consider the production of $H$ through
Vector-Boson fusion, then the uncertainty in the production cross
section is close to $20 \%$ \cite{Aad:2019mbh}.
In summary, the LHC measurements in the Higgs sector leads to virtually
  no worthwhile constraint on this sector. The situation would, of course
  improve considerably at the HL-LHC.

    A very important issue is that of FCNCs involving either of
    the down-- or strange--quarks. The existence of such terms would
    lead to enhancement of suppressed decays (that, normally, occur
    only at the loop-level) as also diverse meson-antimeson
    oscillations. As we have already mentioned in Sec.\ref{sec:intro},
    such FCNCs can be suppressed if one posits that the $B$ mixes only
    with the $b$. Of course, the fact that, in general, mass terms
    such as $m_{bs} \bar b s $ or $m_{d b}^* \bar d b$ are nonzero (in
    the flavour basis) one would still expect some $FCNC$s involving the
    light flavours. The strength thereof would, however,
    be suppressed. Given the structure of the
    Cabibbo-Kobayashi-Maskawa matrix, these suppressions, barring
    large cancellations in the flavour sector, would be ${\cal
      O}(\lambda^4)$ or even more severe, where $\lambda$ is the
    Cabibbo angle. Such a minimal texture in the quark mixing matrix
    (as indicated above) is, thus, more than enough to protect one
    from the FCNC bounds. Indeed, the texture could be perturbed a
    little to induce tiny tree-level FCNCs connecting the $b$ to the
    $d/s$ in an effort to address the recently observed anomalies in
    $B$-decays. However, since the latter also seemingly involve
    lepton-flavour violations, we deliberately desist from treading
    that path.

\subsection{Transition Chromomagnetic Moment}
A consequence of the mixing between fields with different quantum
numbers has been the generation of $Z$-mediated flavour-changing
neutral currents.  This also explains why the mixing in
  the left-handed sector is allowed to be much larger than in the
  right-handed one. Understandably, the extant $SU(3)_c \otimes
U(1)_{em}$ gauge symmetry precludes any such tree-level currents
coupling to the photon or the gluons. This, however, does not prevent
the generation, as a result of quantum corrections, of gauge invariant
terms of the form 
\begin{equation}
  \mathcal{L}_g=\frac{g_s}{2\Lambda}G^a_{\mu\nu}\bar{b}\sigma^{\mu\nu} T^a
  \left[\kappa_L^b  B_L+\kappa_R^b B_R \right] +h.c \ ,
\label{eq:chromo}  
\end{equation}
where $g_s$ is the strong coupling constant and $G^a_{\mu\nu}$ is the
field strength tensor for the gluon.
Analogous to electromagnetic transition moments, the dimensionless constants $\kappa_{L,R}$ are
determined by the couplings and mass spectrum of the full theory,
including all those above the mass scale $\Lambda$ that have been
integrated out. For the sake of simplicity, we
consider these to be be real and equal, thereby also eliminating
additional sources of CP violation.
For our purposes, only the magnitude of $\kappa \equiv \kappa_{L,R}$ is
relevant (for more details see  
ref.~\cite{PhysRevD.86.094006} and references therein), and in the absence of any theoretical knowledge, we shall only assume a phenomenologically guided limit
of $|\kappa| < 0.8$. 
\subsection{Decay Width and Branching Ratio}
\label{subsec_decay_width}
With its isospin partner being nearly degenerate with it and all the
SM particles being much lighter, $B$ decays are dominated by two-body
final states, namely $B \rightarrow bg$, $B \rightarrow Zb$ and
$B \rightarrow Hb$. The first of these proceeds through the
chromomagnetic moment $\kappa$. The other two are driven by the
mixings\footnote{Since $s_L \ll s_R$, the decay $B \rightarrow Wt$ is
highly suppressed.} $s_{L,R}$, and for a very large
$m_B$, have nearly equal partial widths, a consequence
of the Goldstone equivalence theorem. For relatively modest values of
$m_B$, though, the two may differ by as much as $20\%$. 

\begin{figure}[!h]
\begin{center}
\includegraphics[width=6.8cm,height=6.0cm]{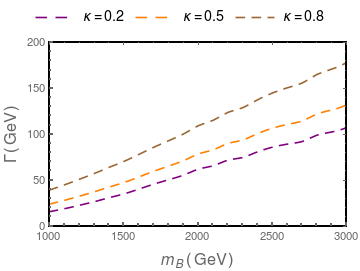}
\includegraphics[width=6.8cm,height=5.7cm]{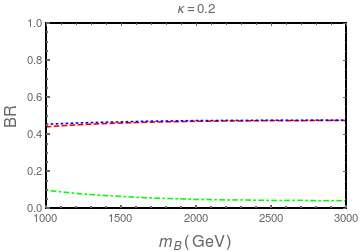}\\
\includegraphics[width=6.8cm,height=5.7cm]{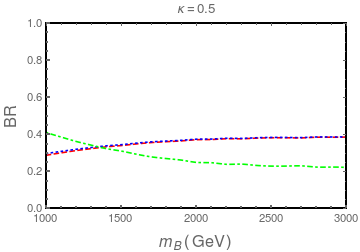}
\includegraphics[width=6.8cm,height=5.7cm]{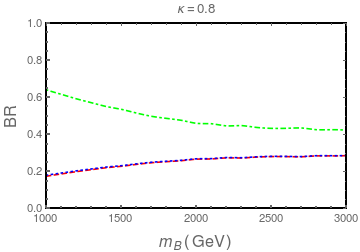}
\includegraphics[width=8cm,height=1.2cm]{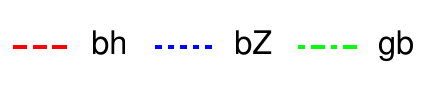}
\end{center}
\caption{\em The upper left plot shows total decay
  width as a function of $m_B$ using the limit on $s_R$ mainly coming
  from oblique parameters. The other three plots show the branching
  ratio of $B$ to $bH$, $bZ$ and $bg$ as a function of $m_B$, for three different values of
  $\kappa$.}
\label{fig:branching}
\end{figure}

In Fig.\ref{fig:branching}, we depict the total decay width and the
major branching fractions for a value of $s_R$ that best fits
$A^b_{\rm FB}$ while being consistent with the oblique parameter
constraints. As the top left panel shows, even for a very large value
of $\kappa \equiv \kappa_{L,R}$, the ratio $\Gamma/m_B \lapp 0.06$,
thereby validating the narrow width approximation. We would be using
smaller $\kappa$ values, though. 
In this paper we study the case of a particularly heavy $B$ decaying  
to $Z/H$ and a $b$-jet, with the $Z/H$ manifesting itself as a
fatjet. Before going to the detailed analysis, we briefly discuss,
in the next section the jet
substructure techniques that are useful to detect the properties of
such a fatjet.

\section{Jet-substructure Techniques}
\label{sec:substructure}
Jets are nothing but a cluster of hadrons produced as a result of
showering and hadronization following high energy particle collisions
and identified using an appropriate jet reconstruction algorithm. With
the $B$ quark being very heavy, and its daughters (say, $b$ and $Z/H$)
relatively light, the latter would, typically, be highly
boosted. Consequently, the angular separation between the subsequent
decay products (of the $Z/H$ in this case) would tend to be small. While
this would still not present a problem for the detection of $Z \to
\mu^+ \mu^-, e^+ e^-$, for the dominant decay modes, {\em viz.}, $Z
\to q \bar q$ , the closeness would, typically, result in the ensuing
jets merging into a single, albeit fat, jet. Fat jets(along with their
subjet structures) have been studied in the context of many search
strategies, such as those for strongly interacting $W$
bosons~\cite{Skiba:2007fw, Khachatryan:2014vla}, supersymmetric
particles~\cite{Butterworth:2009qa,ATLAS:2009typ}, heavy resonances
decaying to strongly boosted top quarks \cite{Schaetzel:2014kha}, as well as the Higgs
production in association with a
$W/Z$~\cite{Butterworth:2008iy,ATLAS:2009elr}.

To reconstruct the W/Z/H jets, different algorithms have been used in
the literature, whether it be anti-$k_T$~\cite{Sirunyan:2016cao},
Cambridge-Aachen ~\cite{CMS:2009lxa} or a combination of
both~\cite{Khachatryan:2014vla}. In the present study, the hadrons are
clustered using the\footnote{The specificity of the hierarchical
clustering renders this algorithm more useful for the kind of fat-jets
that we are interested in.}  anti-$k_T$
algorithm~\cite{Cacciari:2008gp}.  We investigate the variation of the
signal efficiency and the reconstruction of jet substructure variables
with two choices of the distance parameter $R$ in the next section.
This distance parameter, also known as the jet radius, is an angular
cut-off such that a splitting of a parent particle to two daughter
particles $P \rightarrow ij$ will never be combined by the jet
algorithm if $\Delta R_{ij} > R$ where, $\Delta
R \equiv \sqrt{(\Delta \eta)^2+ (\Delta \phi)^2}$ is the (angular)
separation in the azimuth-rapidity plane.  Not only should the fat jet
arising from the hadronic decay of the $Z$($H$) have a reconstructed
jet mass close to $m_Z$($m_H$), but it is also expected to resolve
into a two-prong substructure. However, some of these properties can
be degraded on account of extraneous objects (owing to QCD radiation)
that can vitiate jet-reconstruction. Consequently, one needs to use
appropriate jet grooming techniques that help in eliminating soft and
large-angle QCD radiation. Of many such, we choose to use the {\it jet
pruning} algorithm~\cite{Ellis:2009su,Ellis:2009me}, a popular choice
for reconstructing $W/Z/H$ fatjets~\cite{Sirunyan:2016cao}. While
alternative algorithms such as {\it softdrop} ~\cite{Larkoski:2014wba}
have also been used~\cite{Sirunyan:2017acf}, it has been shown that
the discrimination afforded is very similar~\cite{CMS:2017wyc}.


\subsection{Jet Pruning}
In a typical event at the LHC, a very large number of hadronic
entities impinge on the detectors. Furthermore, there exist
colour
reconnections between objects emanating from disparate fundamental
processes, hard or soft.  The jet reconstruction algorithm, being
statistical in nature, cannot always effect an exact
differentiation. Pruning is one of the jet-grooming methods to remove
those constituents from the jets that carry no significant or useful
information. For example, the (invariant) mass of a jet ought to
reflect the mass of the primary object, small for a pure QCD jet, and
close to the particle mass for a heavy particle. This is facilitated
by removing the soft yet large angle constituents of the jet because
of the statistically smaller likelihood of their being correlated with
the energetic constituents of the jet. Mathematically, at each merging
step ($j + k \rightarrow l$), we define two constraints given by
\begin{itemize}
\item softness: $p_k/p_l < z_{\rm cut}$ for $(p_k < p_j)$, and
\item separation: $\Delta R_{jk} > R_{\rm cut} $.
\end{itemize}
 If both these conditions are met, then we prune (remove) the
constituent $k$ and proceed for the next merging.  A larger (smaller)
value for $z_{\rm cut}
\, (R_{\rm cut})$ will result in more aggressive pruning. Clearly, the
level of pruning is largely determined by the less aggressive of the two
parameters.  As for $z_{\rm cut}$ and $R_{\rm cut}$, the parameters of
the pruning algorithm, we choose the default values\footnote{While using
  a $R_{\rm cut}$ that is dependent on the natural angular size of the
  jet, viz $m_J / p_j$ is tempting, this introduces a degree of
  sophistication not commensurate with the nature of our analysis.},
namely, $z_{\rm cut} =0.1$ and $R_{\rm cut} = 0.5$ as suggested in
Ref.\cite{Ellis:2009su}. The pruned jet mass, $m_J$, is computed from
the sum of the four-momenta of the components that remain
after
pruning, and the resultant jet is considered to be a
$Z$($H$)-fatjet candidate if $m_J$ is commensurate with the $Z$($H$)-mass
  within the detector resolution limits. And while the irreducible SM
  backgrounds (with a high-momentum $Z$($H$) would also show similar
  characteristics, the overwhelmingly stronger pure QCD background
  would, typically, have jets with much lower masses.

\subsection{N-subjettiness}
Heavy VLQ decays tend to produce top quarks and/or a $W$, $Z$, or a
Higgs boson with high momenta, causing their respective decay products
to merge into a single fat
jet~\cite{Asquith:2018igt,Sirunyan:2017usq,Sirunyan:2016ipo}. If the
latter is to be resolved into subjets, a naturally important question
relates to the precise number of such
putative subjets. A good measure of this is
N-subjettiness~\cite{Thaler:2010tr} defined as
\beq
\tau_N=\dfrac{1}{d_0} \sum_{k} p_{\rm T,k} \, {\rm min}\Big(\Delta R_{1k},\Delta R_{2k},....\Delta R_{Nk}\Big)
\label{N_subjettiness}
\eeq
where $N$ is the number of candidate subjets of the jet to be
reconstructed. $k$ runs over constituent particles in a given jet with
$p_{\rm T,k}$ being their transverse momenta and $\Delta R_{j,k}$ the
angular separation between a candidate subjet $j$ and a constituent
particle $k$. Furthermore,
\beq
d_0=\sum_{k} p_{\rm T,k} R_0
\eeq
where $R_0$ is the characteristic jet-radius. 
Physically, $\tau_N$ provides a dimensionless measure of whether a jet
can be regarded to be composed of $N$-subjets.  In particular, ratios
$\tau_N/\tau_{N-1}$ are powerful discriminants between jets predicted
to have $N$ internal energy clusters and those with fewer
clusters.  As applied to our case, jets coming from the hadronic
decays of the $Z$($H$) tend to have lower values for the ratio $\tau_{21}
\equiv \tau_2/\tau_1$ as compared to QCD or top-jets and, hence,
constitutes a good discriminator. 

\section{Collider Signatures}
\label{sec:collider}
As discussed above, we are interested in the single production of
the vector-like quark $B$ at the LHC. Several parton-level
processes may contribute. In decreasing order of the production
cross sections, these are 
\begin{itemize}
\item $bg\rightarrow B $, 
\item $bg \rightarrow B g$
\item $q \bar q\rightarrow \bar b B$ and $gg \rightarrow \bar b B$,
\item $bg\rightarrow (Z/H) B$,
\end{itemize}
with the conjugate process being understood in each case. We would be
concentrating here on the third, {\em viz.} $\bar b B$ production (dominated by the
gluon-fusion contribution). The reason is easy to understand. Since we would be
concentrating on fully hadronic cascade decays of the $B$, each of these channels suffer from large QCD backgrounds. The $\bar b B$ mode, with $B \to b + Z/H$
would admit two hard $b$-tagged jets, thereby offering additional discriminatory power.

  To this end, we begin by a detailed discussion, in the next
  subsection, of the process $pp\rightarrow b B$, $B\rightarrow Z b$
  involving the $Z$-fatjet.  The other channel, namely $pp\rightarrow
  b B$, $B\rightarrow H b$ shows analogous behaviour and, hence, for
  the sake of brevity, we do not provide details but only highlight the differences.

\subsection{Signal: $pp\rightarrow b B$, $B\rightarrow Z b \, (H b)$}
\label{subsec:signal}

We calculate the leading-order on-shell production
cross-section\footnote{As discussed earlier, the ratio of the decay
  width and the mass of $B$ in this model is, at best, 6\%. Consequently, even
  for the full  calculation of
  $p p\rightarrow b \bar b Z$, the contribution from the interference of the SM
  amplitude with the $B$-mediated one (including possible
  contributions from an off-shell $B$) is very small, and is further suppressed
  once all kinematic cuts are imposed.} in the five flavor
scheme (5FS), using the NNPDF23LO1~\cite{Ball:2012cx} parton
distributions holding the factorisation scale to $\mu_F^2 = 2
m_B^2$. The final state, being comprised of jets only, largely depends
on additional jet radiations. Within this scheme, inclusion of the
next-to-leading order effects suppresses the cross-sections by a
factor $\sim 0.9$~\cite{Cacciapaglia:2018qep} over the mass-range of
interest to us. It should also be pointed out at this stage that owing
to the relatively modest reach of this mode, uncertainties due to the
choice of the parton distributions are relatively small.

\begin{figure}[!h]
\begin{center}
\includegraphics[width=8.1cm,height=7.0cm]{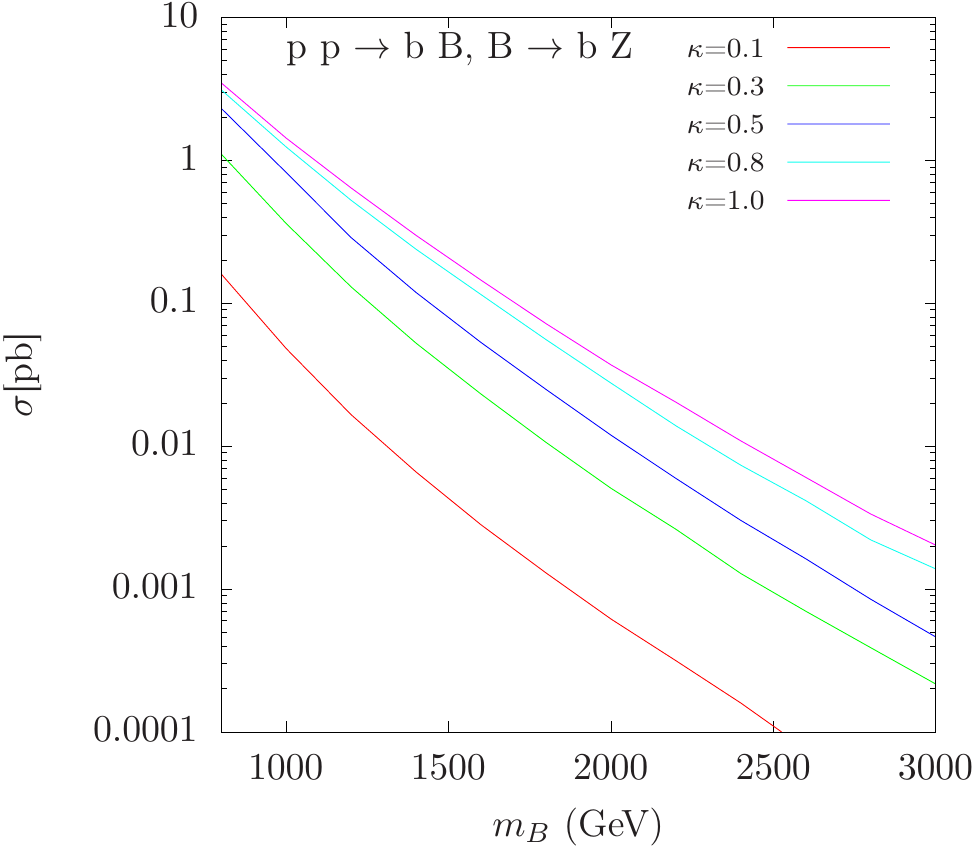}
\end{center}
\caption{\em The LO cross section in 5FS (using
    NNPDF23LO1) for $p p \rightarrow B b \rightarrow Z b b$ at 13 TeV
    LHC for different values of $\kappa$. 
    The cross sections for $p p \rightarrow B b \rightarrow H b b$ are
    virtually the same, 
owing to a nearly identical branching ratio, as in Fig:\ref{fig:branching}. We choose the value of $s_R$
    from Fig:\ref{fig:branching}(left) as a function of vectorlike $B$ mass.}
\label{cs_signal}
\end{figure}
In Fig.~\ref{cs_signal}, we plot, as a function of $m_B$, the product
$\sigma(pp \to b B) \times BR(B \to bZ)$. The production cross section
is largely dominated by gluon fusion, with both the leading and the
subleading ({\em i.e.}, $q \bar q \to b B$) contributions scaling as
$\kappa^2$.  On the other hand, increasing $\kappa$ would also enhance
the partial width $\Gamma(B \to b g)$, thereby suppressing the
branching fraction for the decay mode of interest, {\em viz.} $B \to b
Z$. This explains the relatively slow growth of the product with
$\kappa$ in Fig.~\ref{cs_signal}.  As a reference point, we choose to work
with a benchmark value of $\kappa = 0.5$. Analogously, our benchmark
points (BP) in the mass-axis read $m_B=$1.2, 1.8 and 2.2 TeV.  At the
very end, though, we will present the reach in the $m_B$--$\kappa$
plane.

The signal, thus, comprises of a pair of $b$-jets accompanied by a
$Z$, whose high momentum would, typically, imply that its decay
products coalesce into a further fat jet. Of the two putative
$b$-tagged jets, one would, typically, be attributed with a
transvere momemtum similar in magnitude and in a direction nearly
opposite to that of the fat-$Z$.  If an excess in this channel is
seen, $m_B$ could, presumably, be reconstructed from such pairings.

The second mode, {\em viz.} $p p \rightarrow B b \rightarrow H b b$, 
apart from having a nearly identical strength, also has
a very similar event topology. The only difference is that, owing to a
slightly higher mass of the $H$ (as compared to the $Z$), its momentum
would be marginally smaller and the fraction of events being identified
as a fat jet a little lower.

\subsection{Backgrounds}
\label{subsec:bkgd}
While the SM backgrounds to the two aforementioned signal channels are
similar, they, understandably, contribute differently in the two cases.
We begin by discussing the case that suffers larger backgrounds, namely
the $Z b \bar b $ channel. The major SM backgrounds arise 
from the processes detailed beow.
\begin{itemize}
\item \underline{$W/Z$ production accompanied by multijets :} As with
  the signal, the fatjet would arise, mostly, from the hadronic decays
  of the vector boson, with a rather subdominant contribution from
  events wherein some of the jets accompanying the $W/Z$ are
  reconstructed as a fatjet. While the corresponding SM cross sections
  are very large, {\em viz.}, for $Z$+jets, it is $6.3\times 10^4$ pb
  (NNLO)~\cite{Catani:2009sm,Catani:2007vq} and $1.95\times 10^5$ pb
  (NLO) for $W$+jets~\cite{Balossini:2009sa}, note that requiring
  these additional jets to be $b$-tagged would result in a severe
  suppression.
\item \underline{$t\bar{t}+jets$ :} If the radius parameter $R$ is
  small enough, then the two-prong hadronic decay of one of $W$'s
  emanating from the top decay could be reconstructed as a
  fatjet. Despite the smaller production cross section ($\approx 990$
  pb at N$^3$LO~\cite{Muselli:2015kba}), this would be expected to be
  a major background owing to the automatic presence of two
  $b$-jets.

\item \underline{QCD $n$-jets $(n \geq 4)$ with none being $b$- or $c$-like:}
  This, of course, is very large indeed ($\sim 10^8$ pb). And even
  with the requirement that two (or more) of these jets should
  reconstruct to close to $m_Z$, the wide jetmass distribution means
  that a non-negligible fraction would satisfy the cuts. However, with
  the cross-section being an even stronger function of $p_T^{\rm min}$
  than the preceding background contribution, a strong cut is likely
  to be useful. However, once we impose the condition that at least two jets are $b$-like,
  the small mistagging probability ($\lapp 10^{-3}$ for each of
  $u,d,s,g$\cite {Chatrchyan:2012jua}) ensures that this background is not of any importance.

\item \underline{Semi-inclusive $b\bar{b}$+ $n$ ($n >1$) jets :}
Mainly QCD driven, the cross section is very large ($\sim 10^5$ pb)
and depends crucially on the number of jets in the final state and on
the minimum $p_T$ ($p_T^{\rm min}$) allowed to the jets. Calculated
using MadGraph, the cross sections have been validated against
Refs.\cite{ATLAS:2011ac,Alwall:2014hca}. To this, we also add the contribution from
$c\bar{c}$+ $n$ ($n >1$) jets, where the charm is misstagged as a $b$-jet.

\item \underline{Semi-inclusive $VV$ production ($V = W/Z$) :} 
  Diboson ($WW$, $WZ$ and $ZZ$) production processes  with a boosted gauge boson
  decaying hadronically can 
  also mimic the fatjet signal. The cross sections for the aforementioned
  processes, at the NLO level, respectively are 
  119 pb, 47 pb and 16 pb~\cite {Campbell:2011bn}. 

  For our analysis, though, we consider the semi-inclusive version with upto two extra jets, with the latter
  being required to satisfy $p_T(j) > 30$ GeV and $|\eta (j)| < 2.5$
  and, in the case of two jets, $\Delta R (j,j) \geq 0.4$ as well.
  Considered at the LO level, using the 5 flavour scheme and MLM
  matching the corresponding cross sections are found to be $\sim$ 120
  pb, $\sim$ 55 pb and $\sim$ 16 pb respectively.  As we would see
  later, these processes lead only to a minor component of the total
  background, mainly from $WZ$ and $ZZ$ events.

\item \underline{Semi-inclusive $HV$ production:} Quite analogous
to the previous set, these processes may contribute to the background
  for either signal channels, depending on whether the hadronic decays
  of Higgs or the W/Z boson are identified as the fatjet. The cross
  section for inclusive $HZ$ process is $\sim$ 1 pb and $HW$ is $\sim$
  1.8 pb at LO (when the cuts mentioned in the context of $VV$
  production above are imposed). While the higher
  order corrections can be found in ref.~\cite{Aad:2020eiv} and the
  references therein, this contribution is subdominant even in the
  context of the $H b \bar b$ mode.

\item \underline{Single top production:} Despite being semiweak processes, the
  total cross section ($\sigma_{tW} = 83.1$~pb, $\sigma_{tb} = 248$~pb and
  $\sigma_{tj} = 12.35$~pb at NNLO~\cite{Kidonakis:2015nna}) is quite comparable to the QCD-driven $t \bar t$ cross section, in a large part on account of the
  phase space. However, once the large $p_T^{\rm min}$ criterion is imposed,
  the contribution is suppressed to a great degree.
 
\end{itemize}

As the kinematic distributions for the inclusive $n$-jet cross
sections are somewhat similar to the inclusive $b\bar b$ events (both
being largely QCD-driven), we, henceforth, merge the two and term it
``Inclusive $b \bar b$''. Of course, in doing this, tagging efficiency
and/or mistagging probability (as the case may be) are duly taken care
of.  Furthermore, we would subsume the relatively smaller
contributions accruing from inclusive-$VV$, $VH$ and single top events into
an ``Others'' category.

For the $H$-fatjet channel, some quantitative changes in the
background profile turn out to be crucial. For
example, with $m_H$ varying significantly from $m_Z$, the fraction of
the electroweak gauge bosons (whether emanating from hard $V$+jets or
hard $VV +X$ production or from the decay of tops), reconstructing as
a $H$-fatjet reduces considerably. On the other hand, new processes
such as inclusive $H$-production (whether QCD-initiated or as a result
of vector-boson fusion) and Higgsstrahlung processes ($WH$, $ZH$ or
even $t \bar t H$) now need to be considered. Fortunately, the
corresponding cross sections are much lower, especially for the event
topology that we need to consider. Naively, this would already
suggest that the Higgs-channel would be the more sensitive one.

\subsection{Details of Simulation}
\label{subsec:simulation}
Implementing the model in
FeynRules~\cite{Alloul:2013bka,Christensen:2008py}, we generate signal
and background events at the tree order using
MadGraph~\cite{Alwall:2011uj} interfaced with
PYTHIA8~\cite{Sjostrand:2006za} for parton showering and
fragmentation. A given hard process contributing to either signal or
background is generated with up to two additional partons to account
for QCD radiations.  To avoid possible inconsistencies (such as
overcounting) arising from interfacing such matrix elements with
parton showering algorithms, we employ the MLM matching
scheme~\cite{Mangano:2006rw,Hoche:2006ph} with matching parameters as
suggested in refs.\cite{Alwall:2007fs,Alwall:2008qv}. Signal events
are simulated for mass points in the range $m_B \in
[0.8,3]$~TeV.  The
events are passed through DELPHES~3~\cite{deFavereau:2013fsa}, in
order to implement detector effects and applying reconstruction
algorithms. Jets are reconstructed using the anti-$k_T$
algorithm~\cite{Cacciari:2008gp} in FastJet~\cite{Cacciari:2011ma},
with $p_T > 30$ GeV and $R=0.5 (0.8)$. We impose $\Delta R (j,j) >
0.4$ and $\eta< 2.5$ on the jets.  For other parameters, we use the
default CMS card. To identify $b$-jets, we use a $b$-tagging module
inside DELPHES, with the ($p_T, \eta$)--dependent tagging efficiency
being in the 70--80\% range. The probability of mistagging a charm as
$b$-jet is 10$\%$ while for the other quarks and gluons, it is 0.1$\%$
or less.  A lepton $\ell$ is considered to be visible and isolated
only if it simultaneously satisfies $p_T
> 10$ GeV, $|\eta|< 2.5$ and $\Delta R (\ell,j) > 0.4$ for all the
jets in the event. 

We implement jet substructure techniques to distinguish the
fatjets from ordinary ones.  The former are reconstructed and
identified with the FastJet module demanding 
$p_T > 200$ GeV and $\eta < 2.4$.  As for the radius parameter $R$, we
investigate the performance for different values but, for brevity's sake, present
the result for only two choices. The first,
namely, $R = 0.8$, is a more conventional choice and enables tracking
down the decay products of $Z$ even when the latter is not highly
boosted. The second choice, {\em viz.} $R = 0.5$, is aimed towards
reconstructing a heavily boosted $Z$. This is expected to be
particularly useful for large $m_B$.  For other parameters of the
jet-substructure observables, as mentioned in Section 3, we follow
Ref.~\cite{Sirunyan:2018omb}.

Since jet energy measurements are very crucial for our analysis,
it is important to consider the energy resolutions, expecially in the
context of the hadronic calorimeter. This, however, depends on the
particular detector being considered, and, furthermore on whether one
is considering the barrel, endcap or the forward
region~\cite{Aaboud:2018scw,GoyLopez:2017kkg}. Multiple sources
(such as stochastic term, white noise {\em etc.}) contribute:

\[
\frac{\sigma_E}{E} = \frac{a_s}{\sqrt{E / 1 \gev}} \oplus \frac{a_w}{E / 1 \gev} \oplus a_C \ ,
\]
with the terms to be added in quadrature. The constants $a_{s, w, C}$
depend on the details of the detector (and the geometric location
within), and typically $a_s \in (0.5, 0.65)$, $a_C \in (0.03, 0.06)$
and $a_w \lapp 5$. Similar expressions hold for the electromagnetic
calorimeter as well, but with smaller $a_i$. However, since we are not
interested in either photons or $e^\pm$, the resolution of the
electromagnetic calorimeter is of little interest to us.

In addition, the
granularity of the detectors implies a finite angular
resolution. This, in principle, would be significant for the
measurement of both the missing transverse momentum, and, more
importantly, the invariant mass of a jet pair. However, in practice,
these uncertainties are of little concern in view of the relatively
large error in the jet energy measurements.

\subsection{Signal and Background Profiles for Different Choices of $R$}
\label{subsec:sig_bkgd_prop}
With the profiles (both signal and backgrounds) for the two channels,
{\em viz.} $Z$ or $H$, being quite similar, for the sake of brevity, we
discuss only the former. This choice is also occasioned by the fact that this
would turn out to be the one with the worse signal-to-background ratio.

\begin{figure}[!h]
\begin{center}
\includegraphics[width=5.5cm,height=5.0cm]{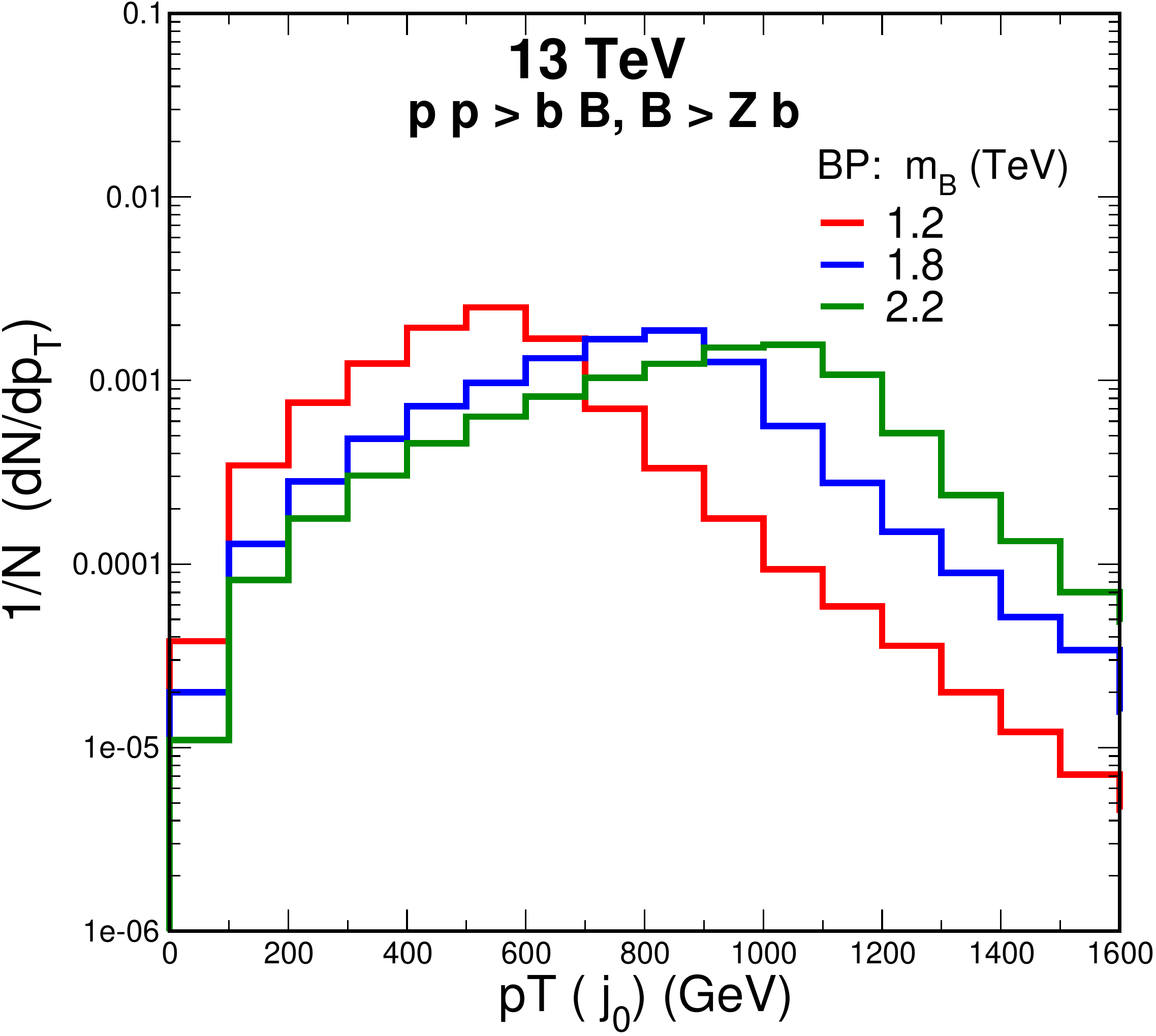}
\includegraphics[width=5.5cm,height=5.0cm]{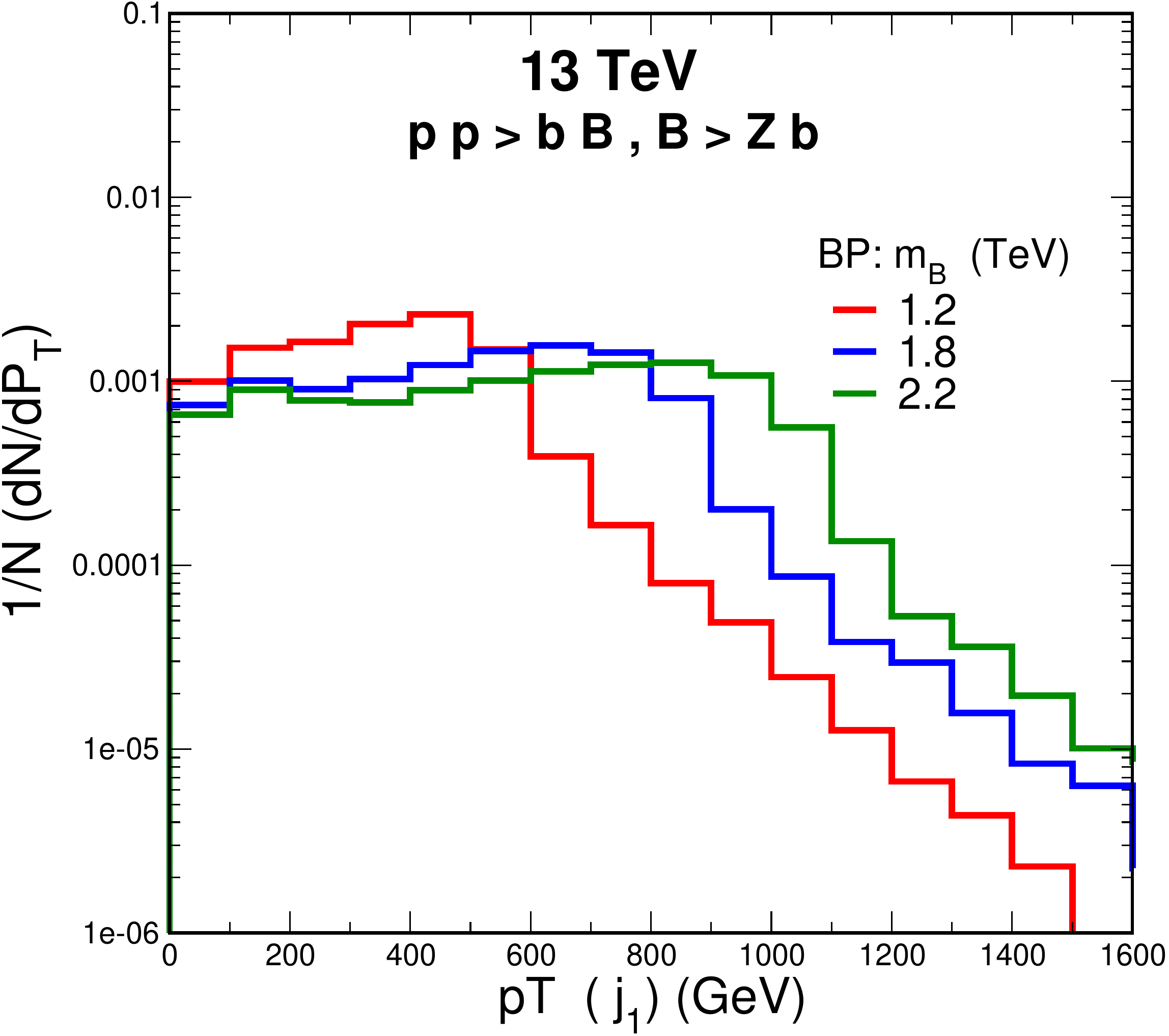}
\includegraphics[width=5.5cm,height=5.0cm]{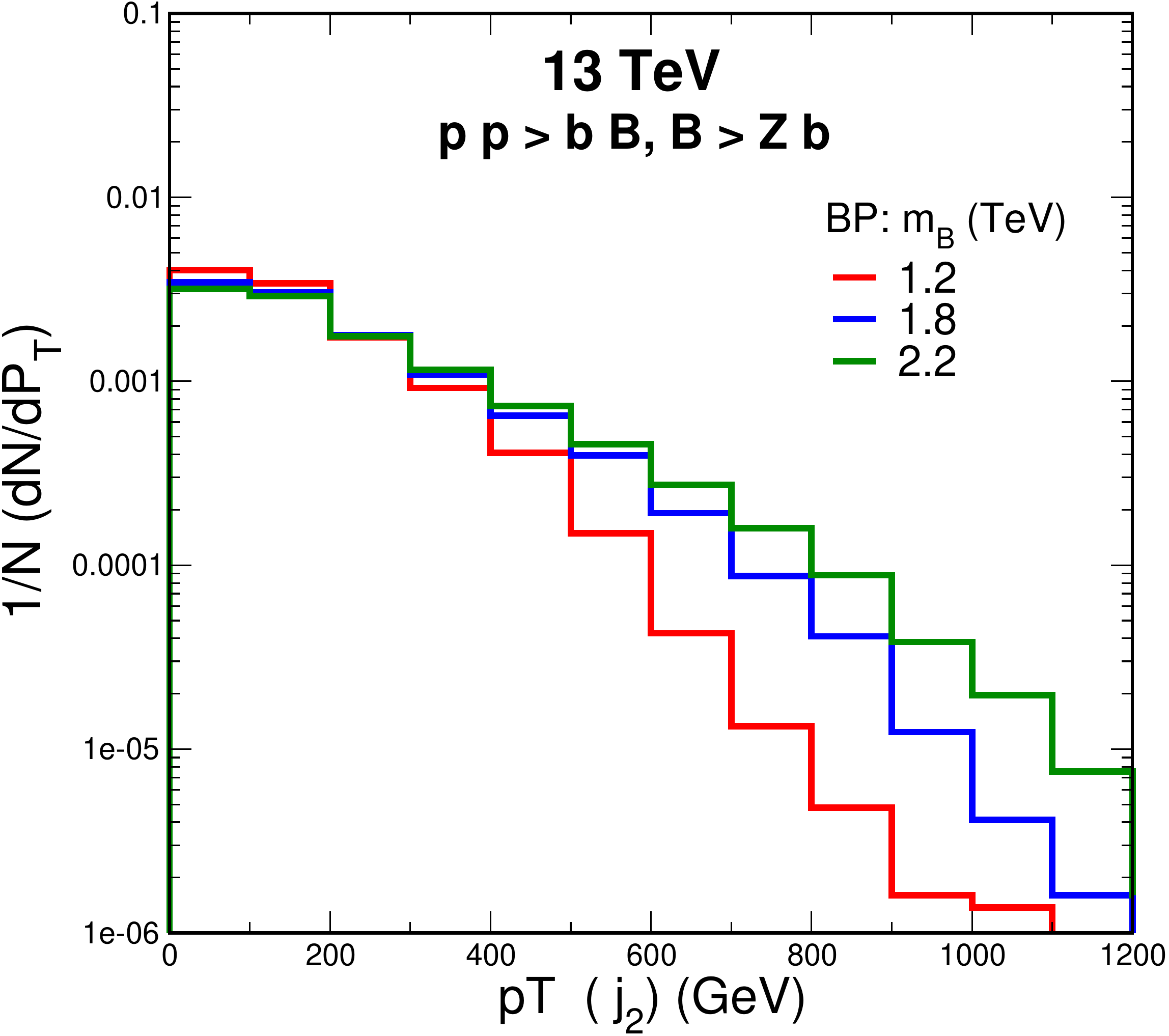}
\includegraphics[width=5.5cm,height=5.0cm]{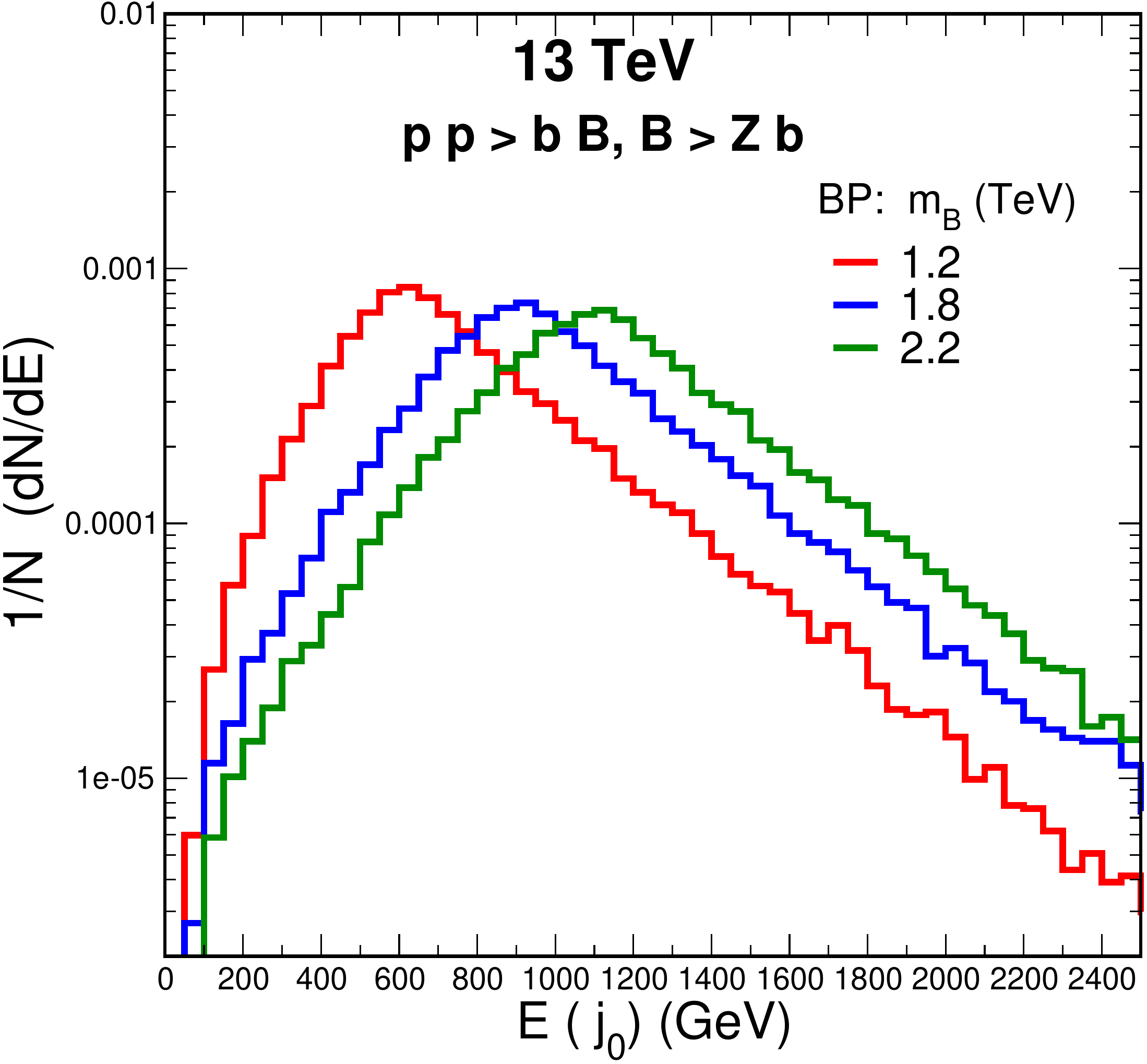}
\includegraphics[width=5.5cm,height=5.0cm]{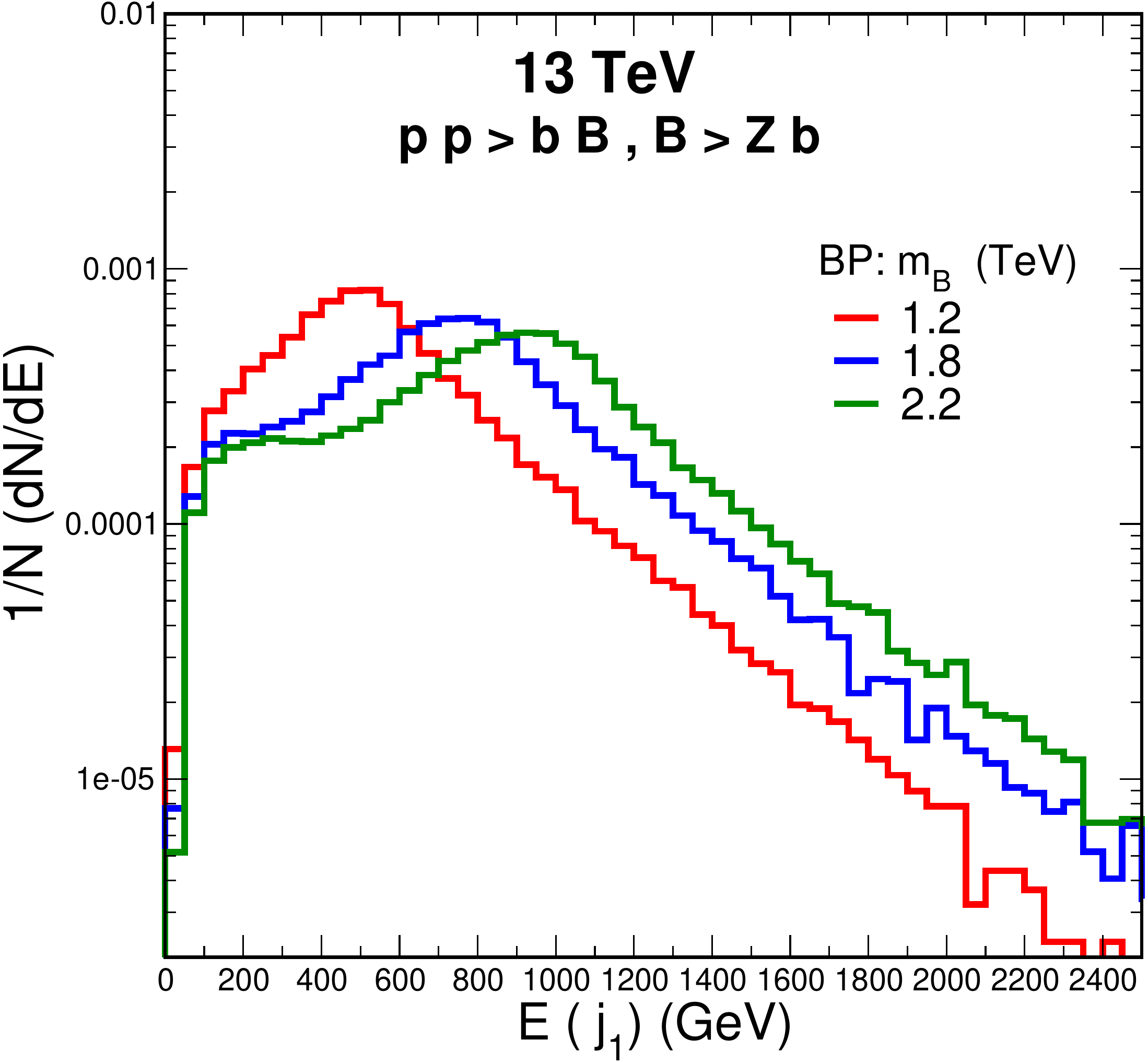}
\includegraphics[width=5.5cm,height=5.0cm]{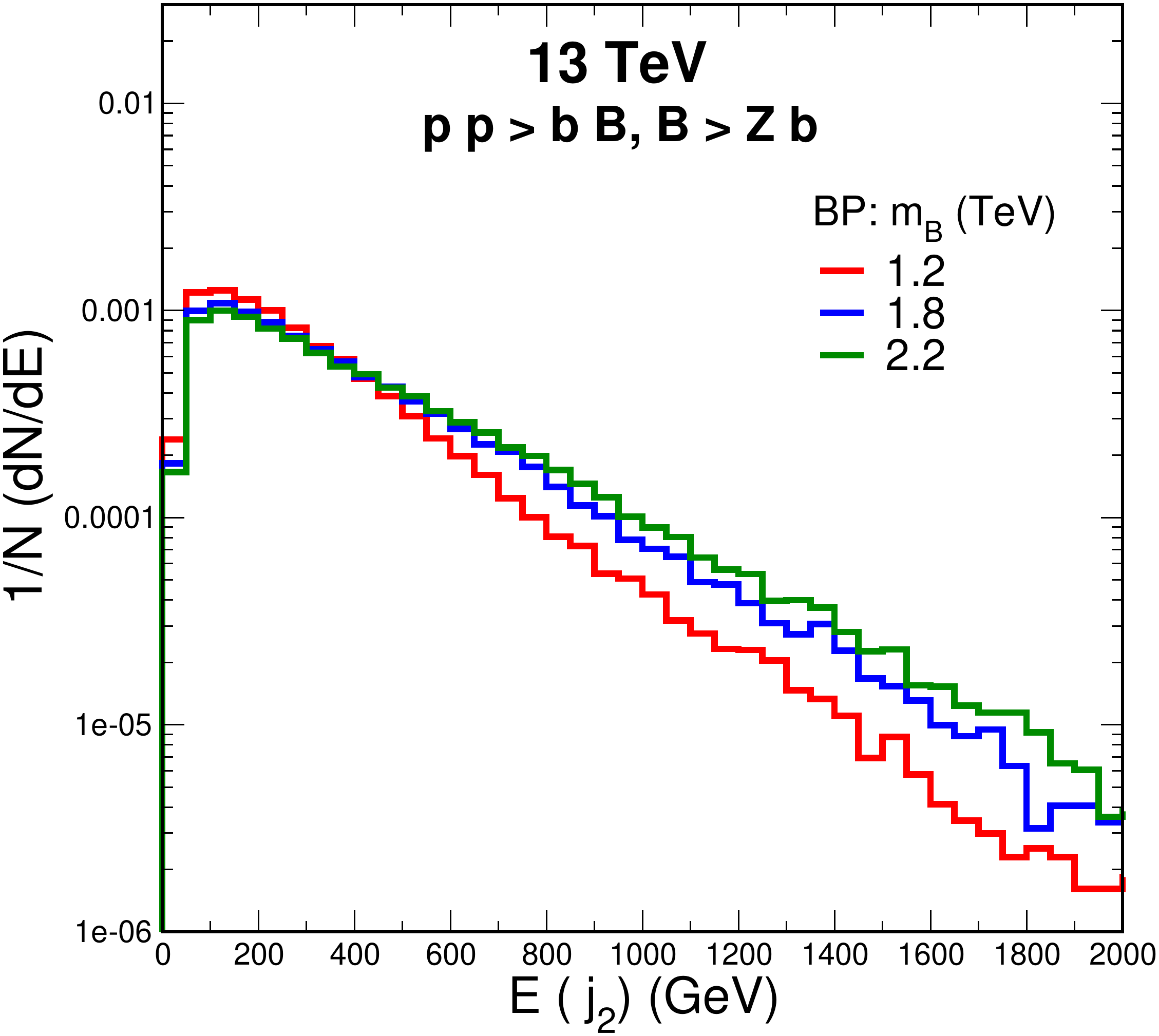}
\includegraphics[width=5.5cm,height=5.0cm]{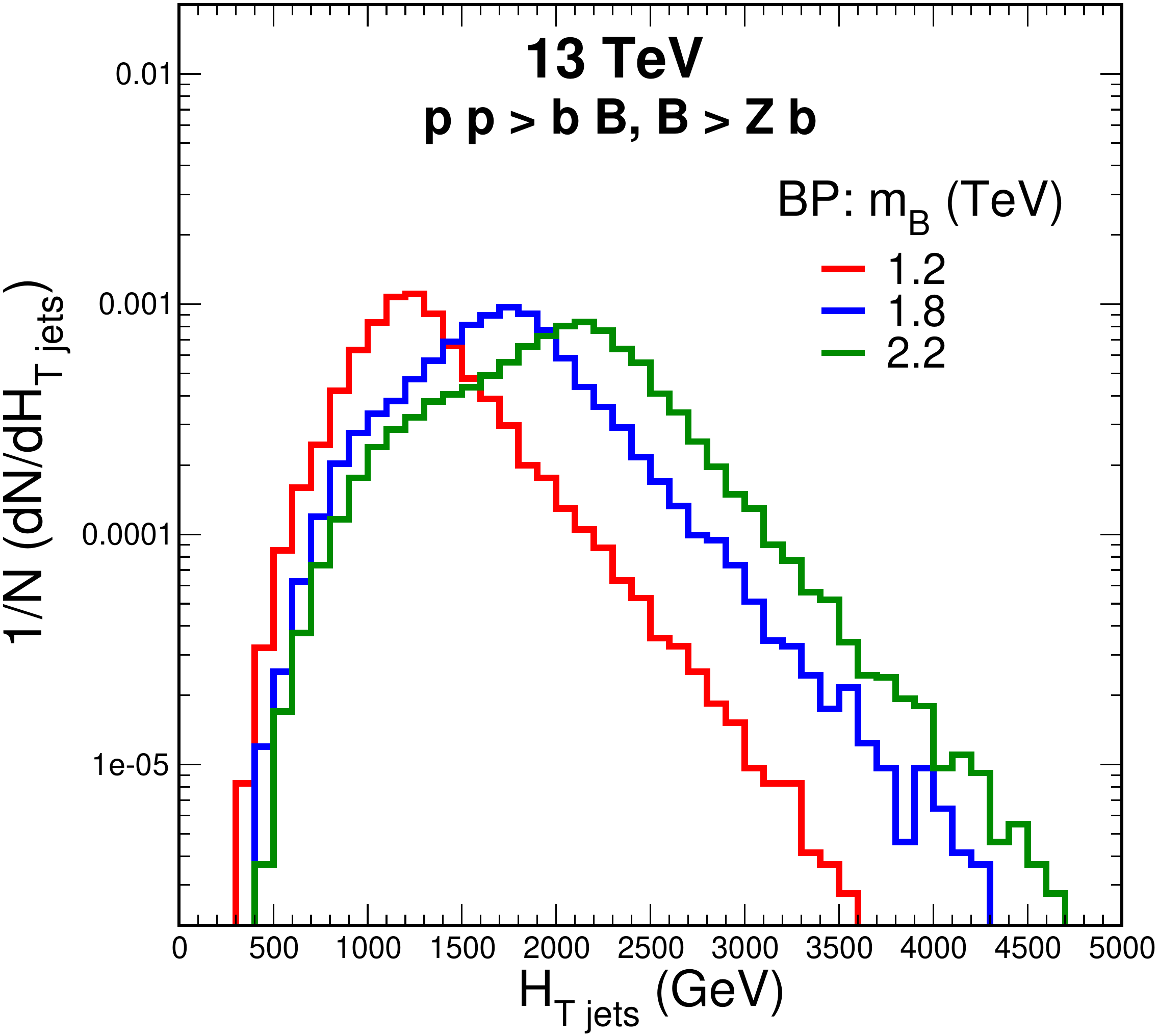}
\includegraphics[width=5.5cm,height=5.0cm]{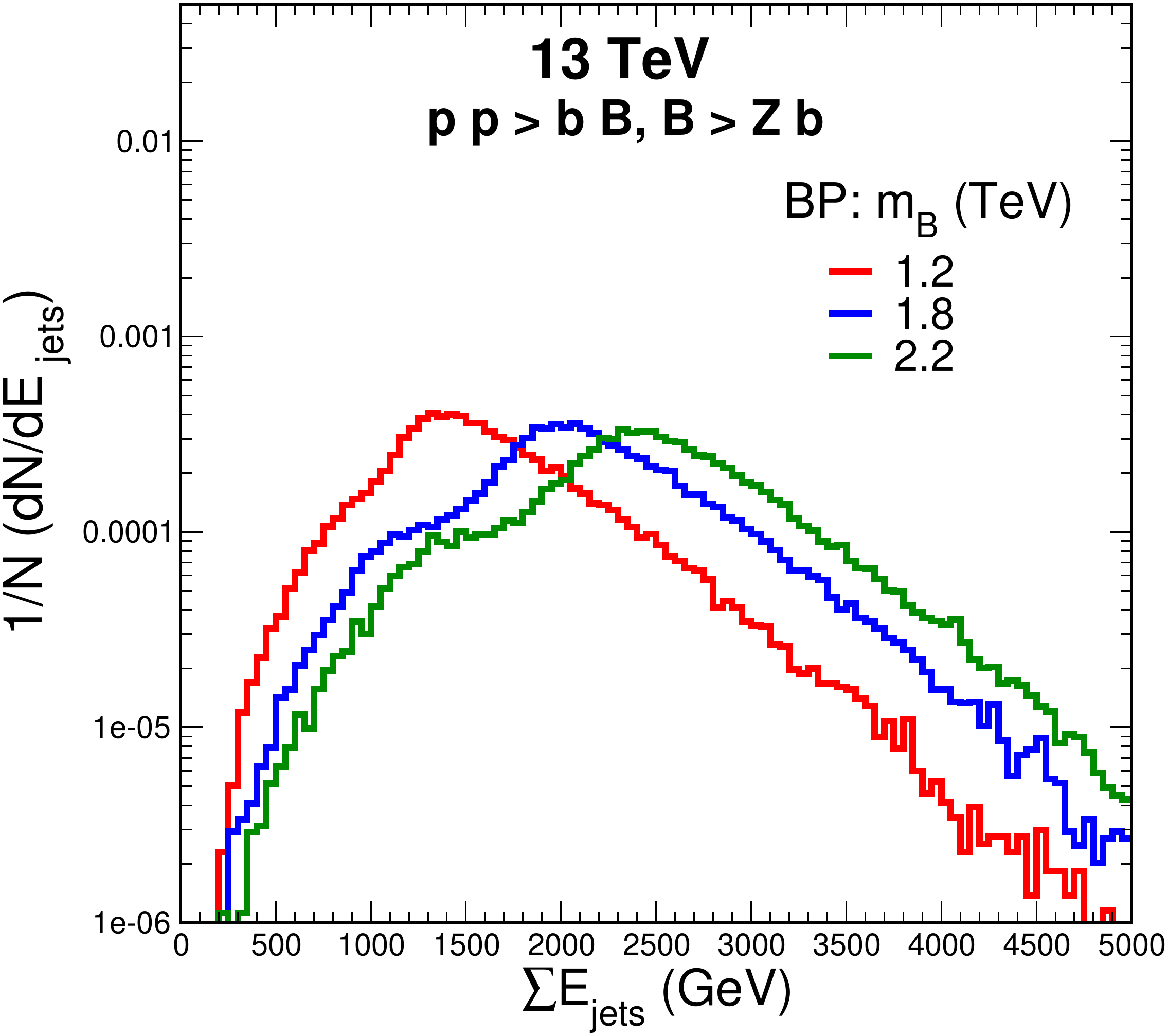}
\includegraphics[width=5.5cm,height=5.0cm]{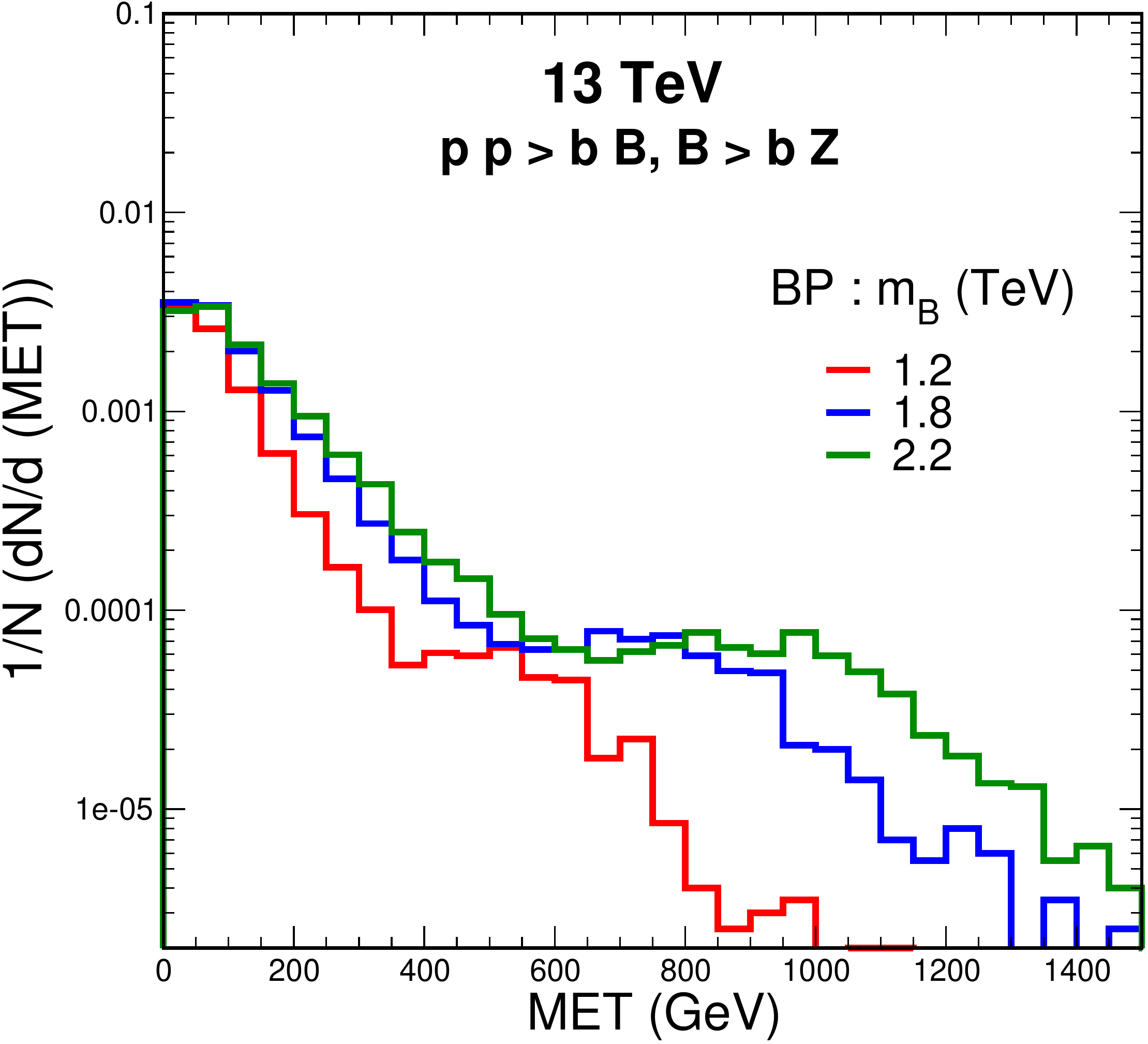}
\end{center}
\caption{\em Kinematic distributions for the signal with jets defined using a
radius $R = 0.5$. Each of the three benchmark points are depicted. Top
panel: $p_T$ distributions for the three leading jets $j_0$ (left),
$j_1$ (center) and $j_2$ (right). Middle panel: corresponding
distributions for jet energy. Bottom panel: the scalar sum of the
$p_T$s of all jets (left), the sum of all jet energies (center) and
the missing transverse momentum (right).}
\label{fig:4a}
\end{figure}

Since the hard process corresponding to our signal comprises of a fat
jet accompanied by another pair of  $b$-tagged jets,
we preselect events with a {\em minimum of three} relatively central
($|\eta| < 2.5$) jets, each with a minimum transverse momentum\footnote
{We impose a stronger $p_T$ cut later on.}  $p_T > 30$ GeV, demanding
that they be separated by at least $\Delta R(j,j) > 0.4$. Arranging
them in the descending order of $p_T$, as $j_0, j_1, j_2, j_3 \dots$,
we display, in the top panel of Fig. \ref{fig:4a}, the $p_T$
distributions of the leading $(j_0)$, sub-leading $(j_1)$ and
sub-sub-leading ($j_2)$ jets for each of the three benchmark points.
For a more massive $B$-quark, its decay products would be more
energetic in its rest frame, resulting in a harder $p_T$ spectrum. For
the two leading jets, this is reflected by the top left and top center
panels of Fig. \ref{fig:4a}.  jets. The distribution for the
sub-sub-leading jet (top right in Fig. \ref{fig:4a}) is a little more
intricate. For most signal events, this jet would originate from the
primary $b$ (produced in association with the $B$). This explains the
similarity in the three normalized spectra. It should also be realized
that, as $m_B$ grows large, it provides a scale for the kinematics,
and a somewhat large $p_T(B) = p_T(\bar b)$ would not cause
significant suppression. This explains the slightly harder $p_T(j_2)$
spectrum for larger $m_B$.

While the decay of a very massive $B$ would result in very energetic
daughters, not all of the energy would be manifested in the form of
transverse momenta. In view of this, we also investigate the energy of
the three jets(as defined in the laboratory frame). Although such a
variable is not very popular in the context of a hadronic collider
(owing to the lack of information of the longitudinal momentum of the
subprocess center-of-mass), we would find this to be useful. The
middle panel of Fig.\ref{fig:4a} exhibits the distribution in $E_j$
for each of the three leading jets. Qualitatively, the features of the
$p_T$ distributions are replicated except that the spectra are
hardened (understandably) and sharpened.

The relative hardness of the individual $p_T$ spectra is also
reflected (Fig.\ref{fig:4a}, bottom left) by that for $H_T \,
(\equiv \sum_{\rm jets} p_T)$, the scalar sum of the individual jet
$p_T$s. Note, though, that the difference between those for the three
benchmark plots is somewhat obscured by the inclusion of {\em all}
jets in $H_T$, not just the leading three. Much the same story is
repeated by the distribution in the sum of all jet energies
(Fig.\ref{fig:4a}, bottom center). Another potentially interesting
kinematic variable is the missing transverse energy (MET). In the
present context, it could arise from two sources, the first being the
neutrinos that result from leptonic decays, whether that of an entity
involved in a hard process or from the hadrons in the cascade. A
second, and given the
large jet energies, often more important source is the mismeasurement
of jet energies (as described in the preceding subsection).  A third
possible source is the event of a jet constituent not being registered
in the detector, including the possibility of entities passing through
detector gaps. Short of a full detector simulation, though, we cannot
include the last-mentioned contribution. However, it is not expected
to be a large effect. As we see in Fig.\ref{fig:4a}, the MET,
typically, tends to be not too large (a consequence of there not being
a hard invisible particle in the event) and its distribution has only
a relatively small dependence on $m_B$.

\begin{figure}[!h]
\begin{center}
\includegraphics[width=5.5cm,height=5.0cm]{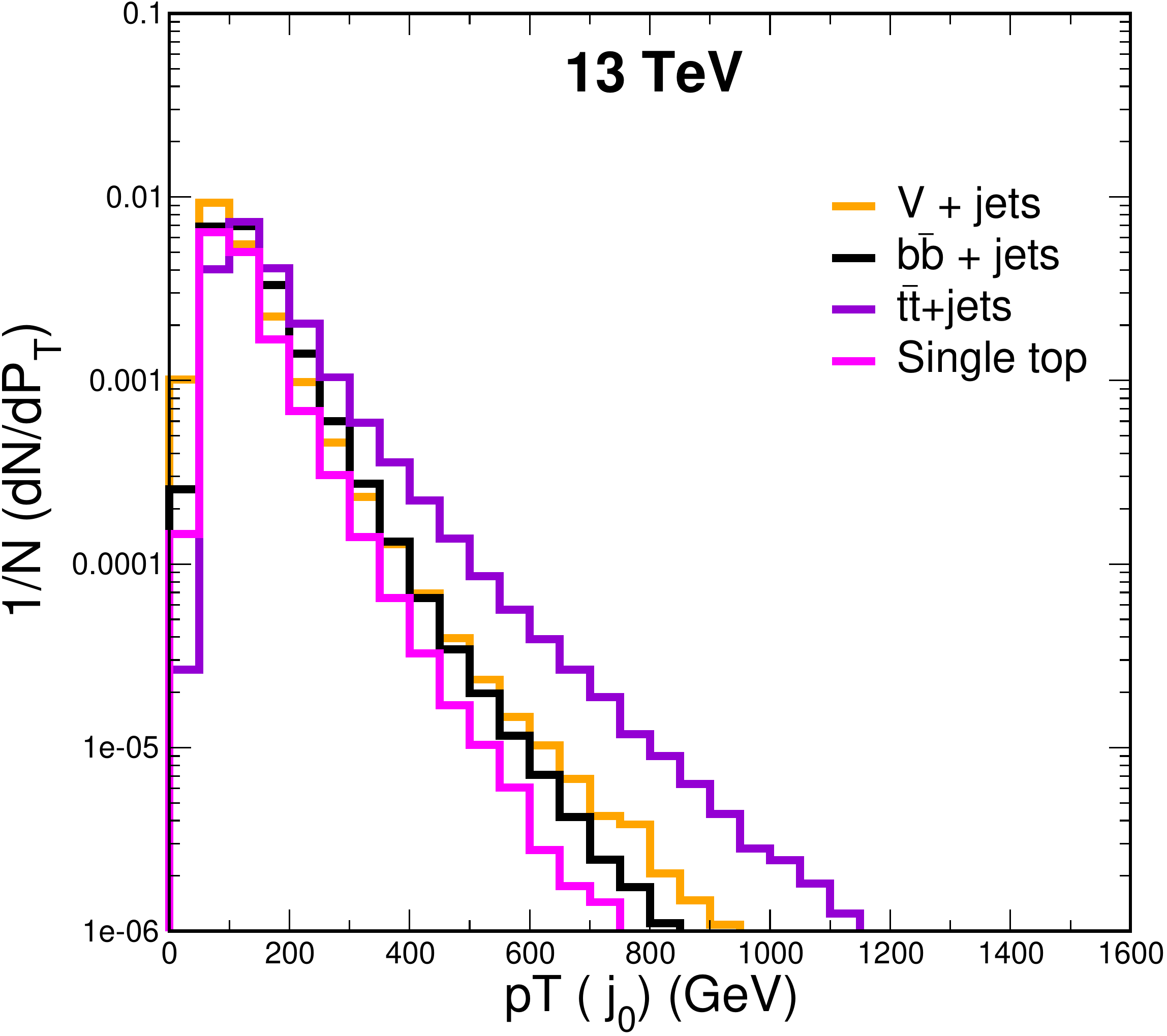}
\includegraphics[width=5.5cm,height=5.0cm]{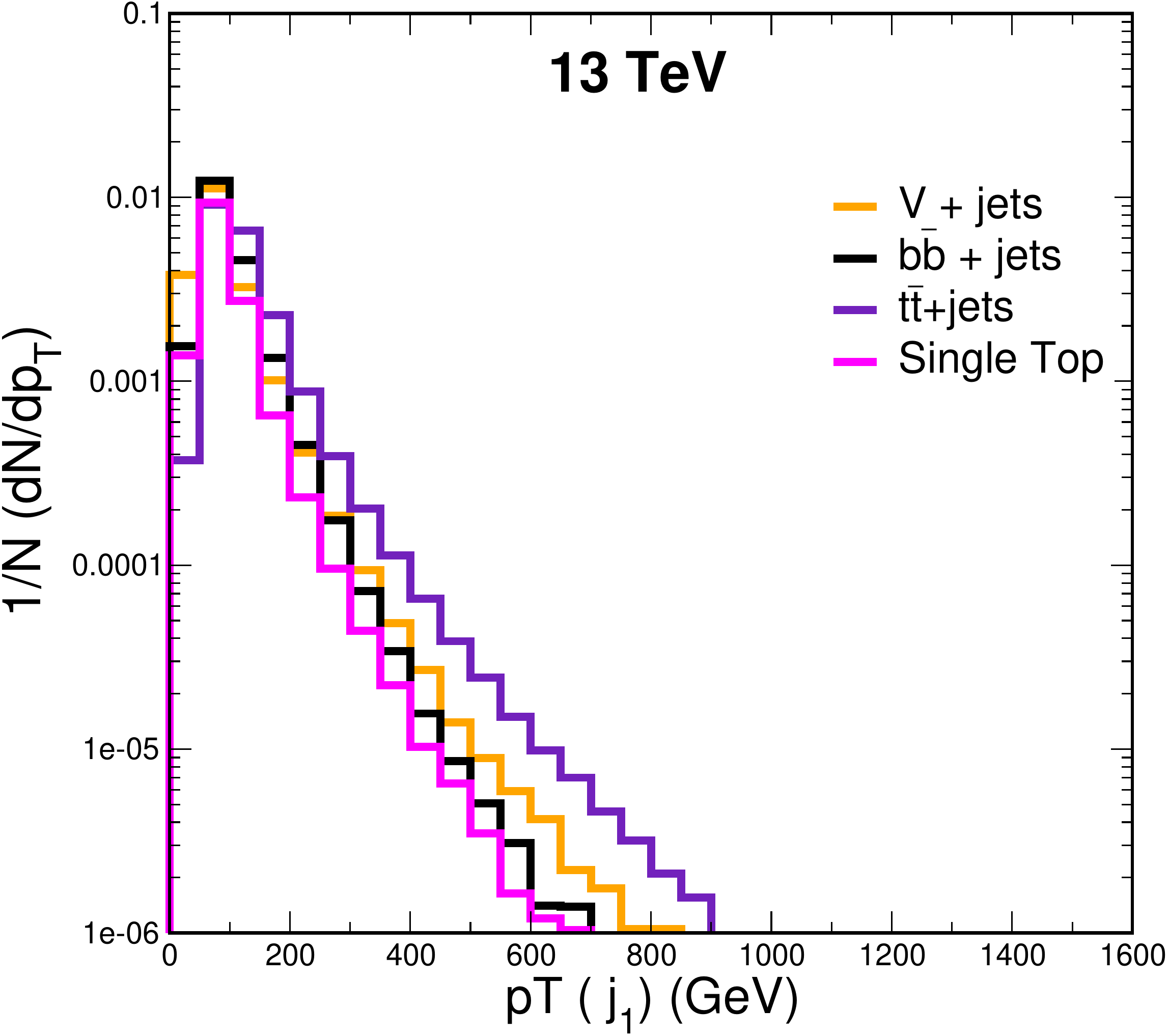}
\includegraphics[width=5.5cm,height=5.0cm]{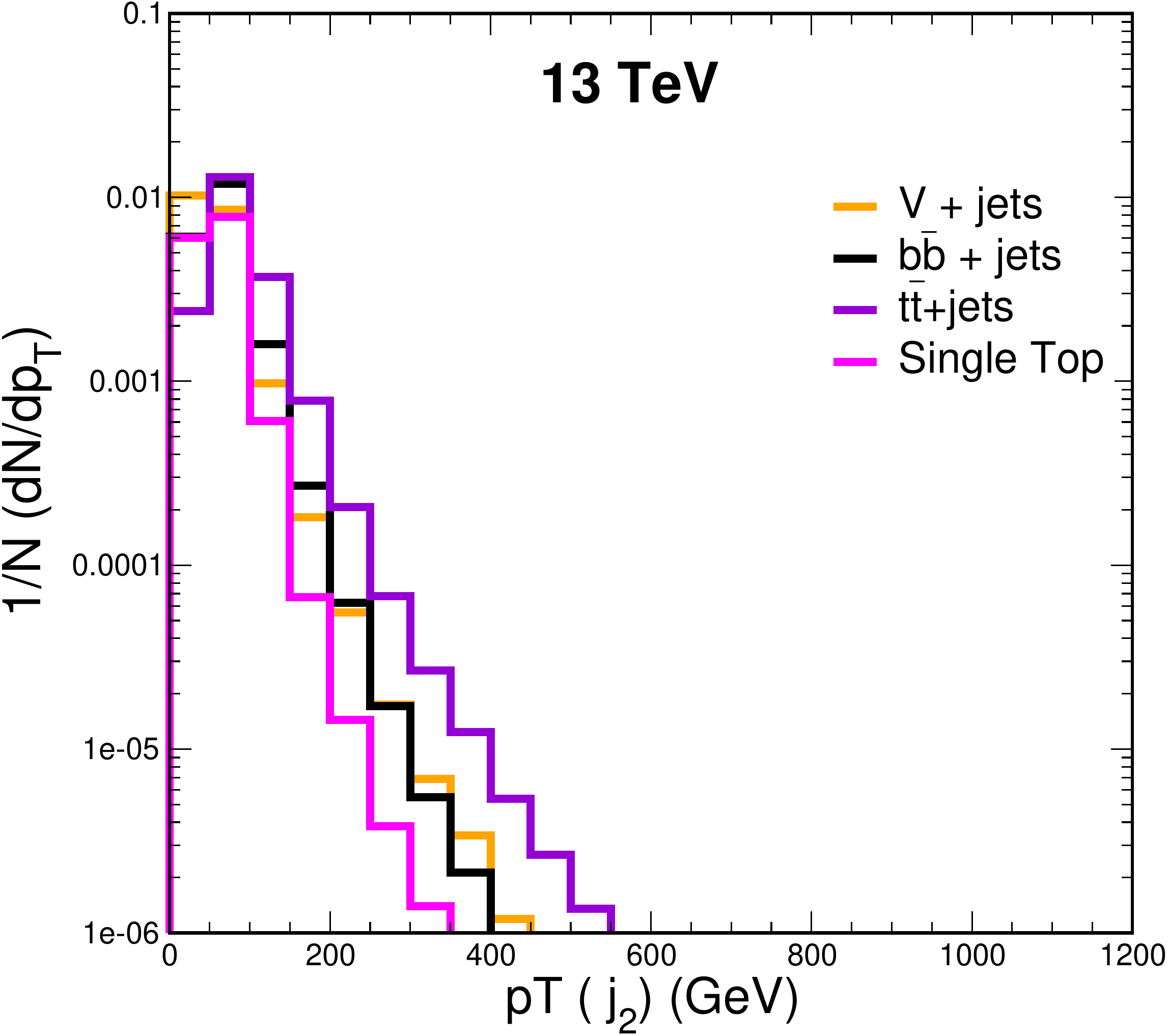}
\includegraphics[width=5.5cm,height=5.0cm]{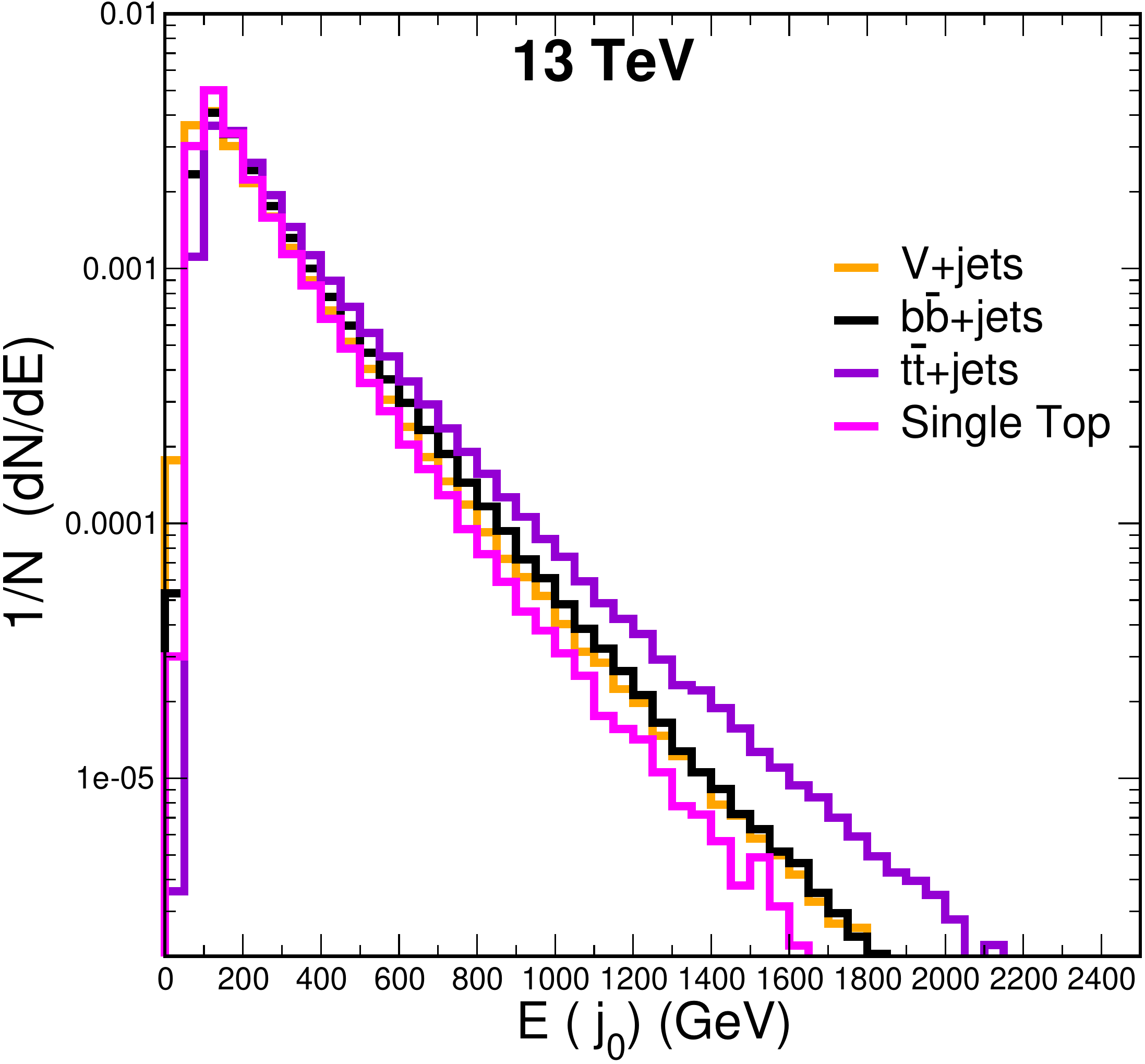}
\includegraphics[width=5.5cm,height=5.0cm]{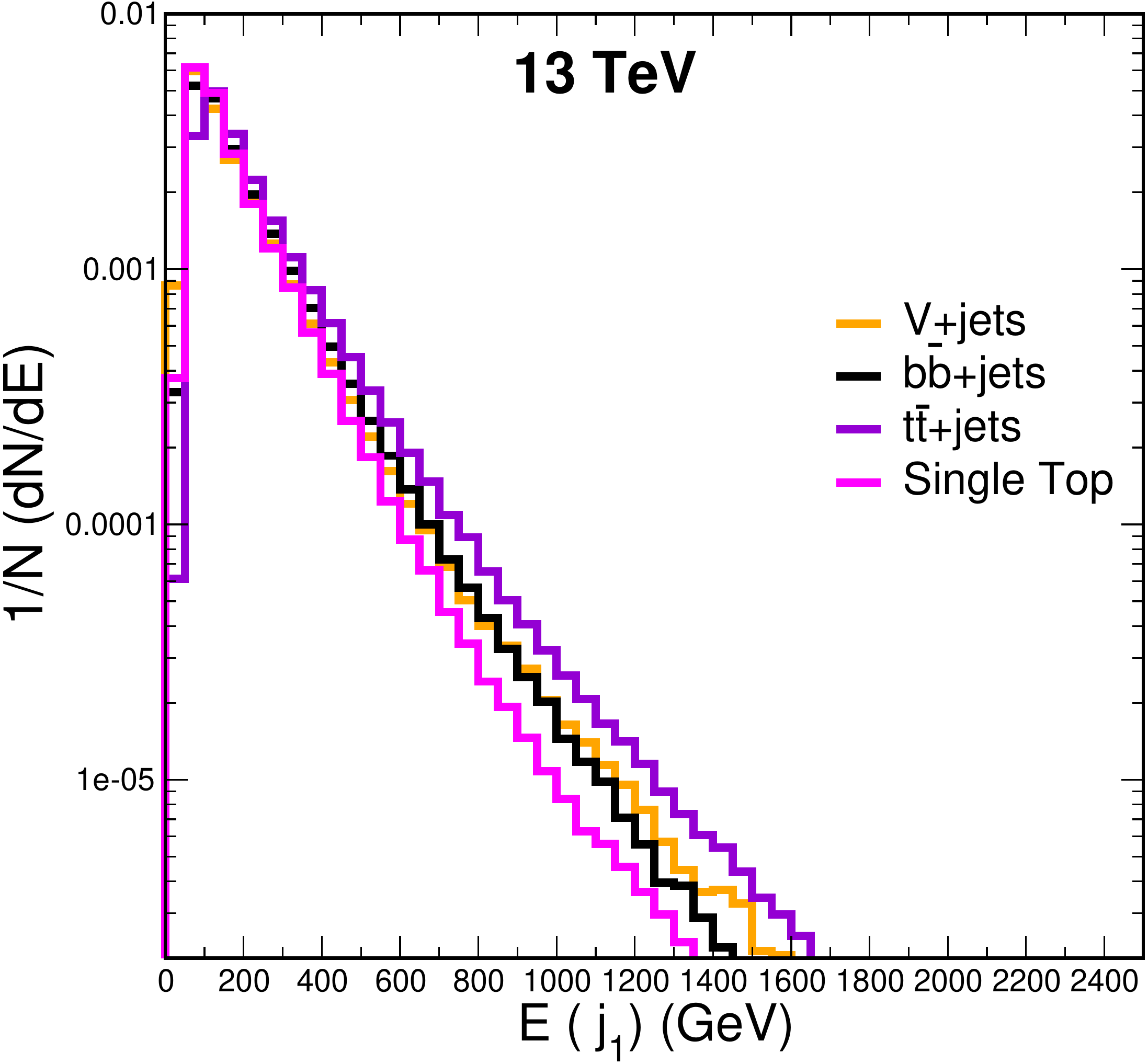}
\includegraphics[width=5.5cm,height=5.0cm]{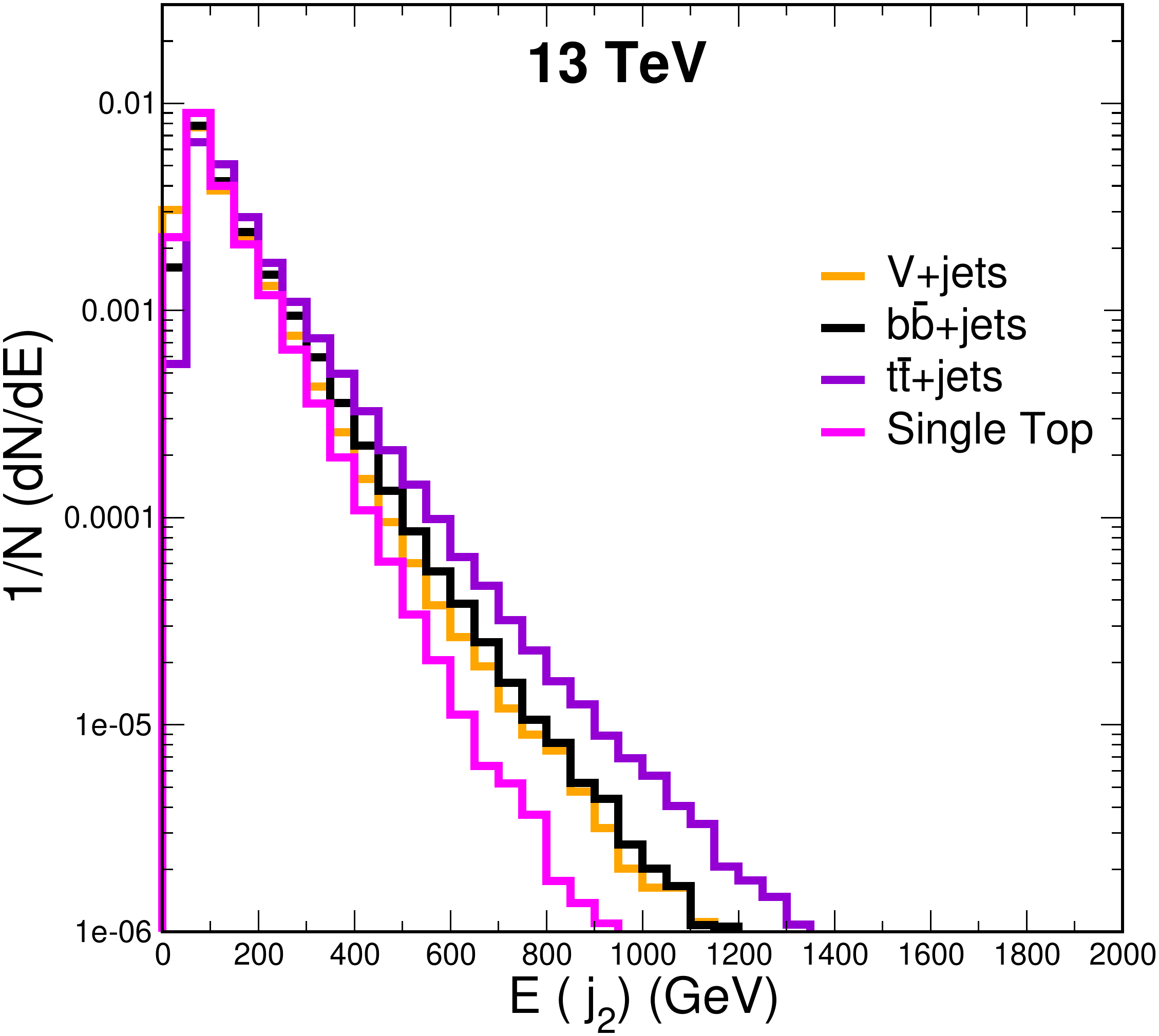}
\includegraphics[width=5.5cm,height=5.0cm]{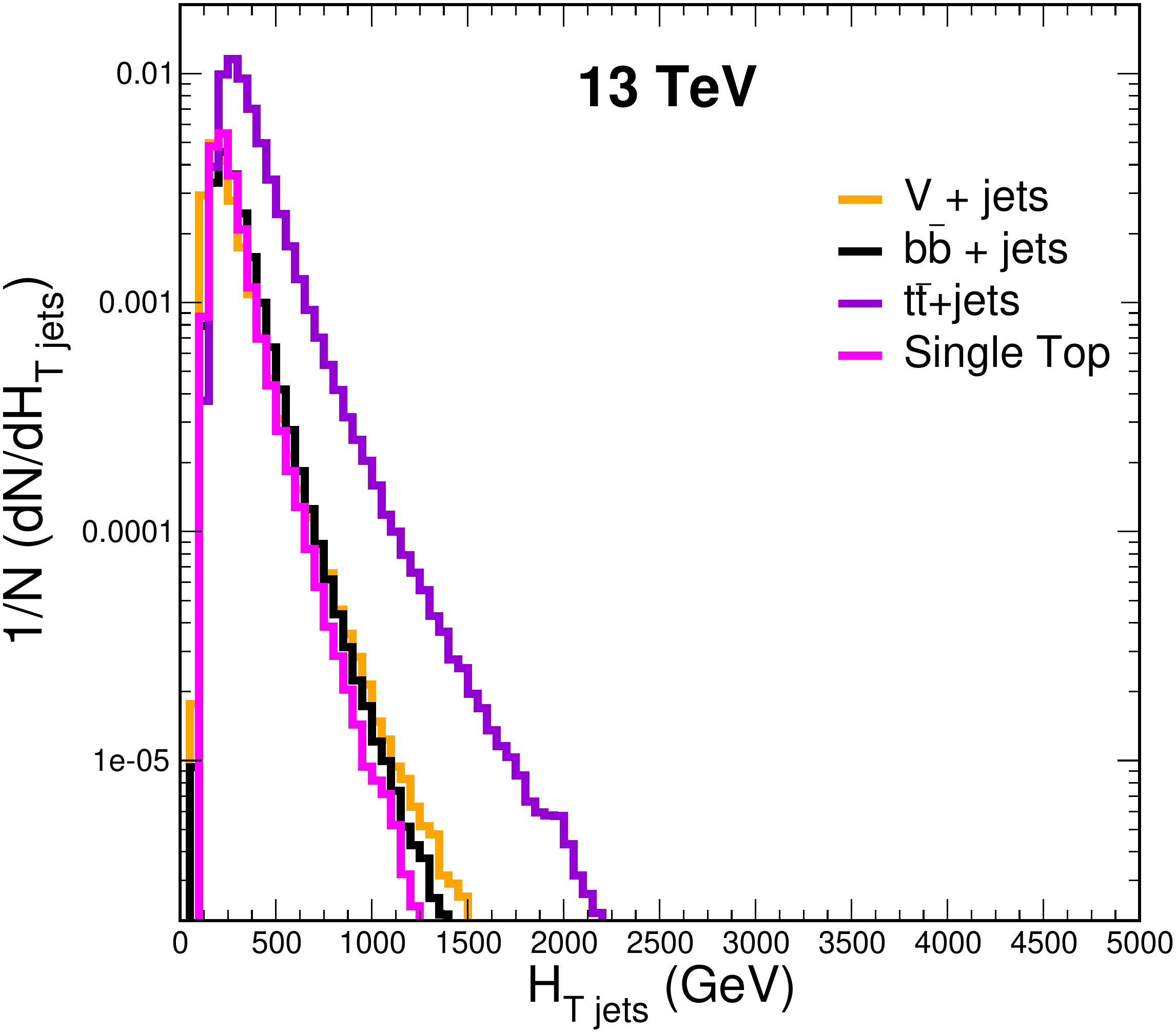}
\includegraphics[width=5.5cm,height=5.0cm]{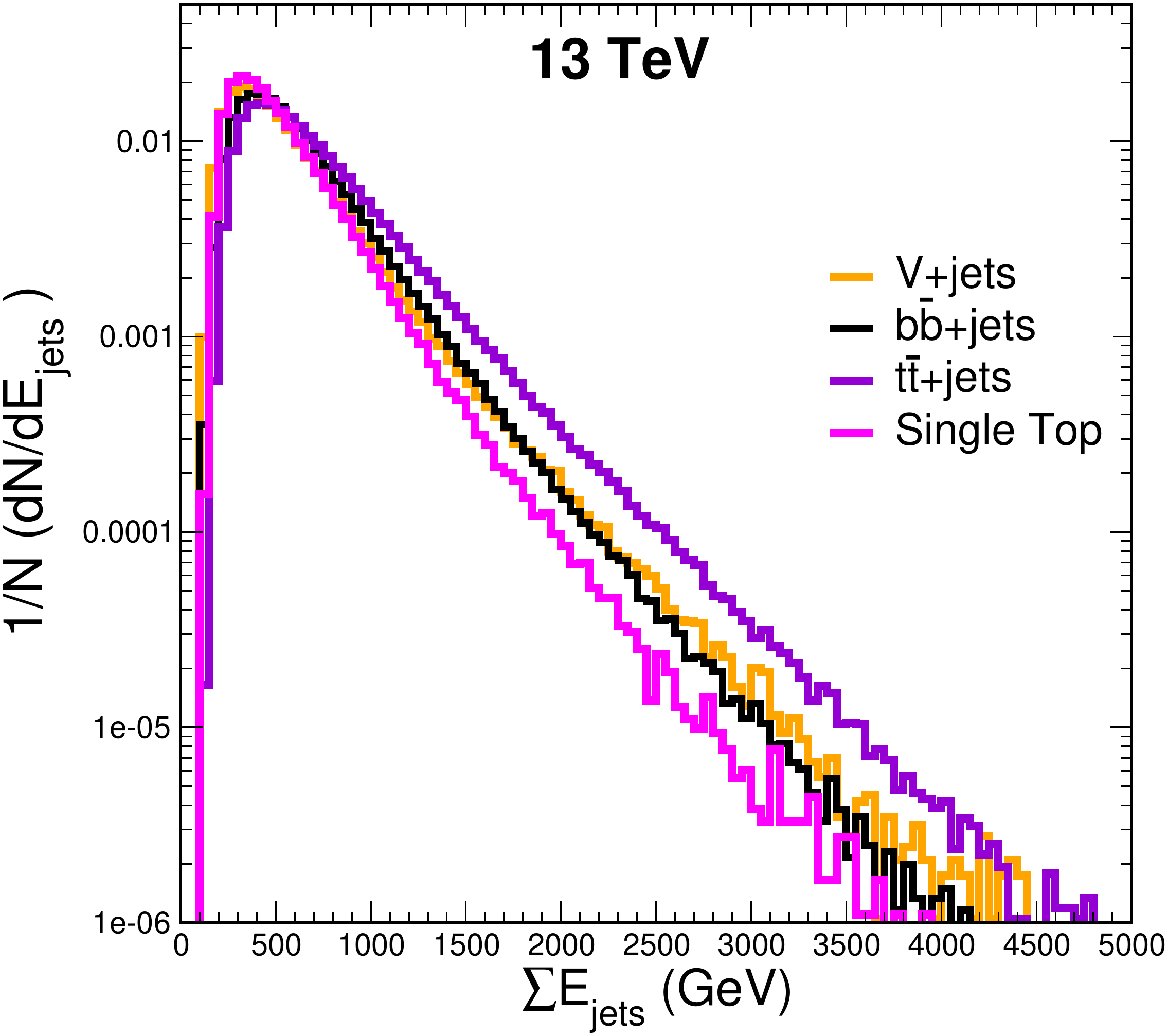}
\includegraphics[width=5.5cm,height=5.0cm]{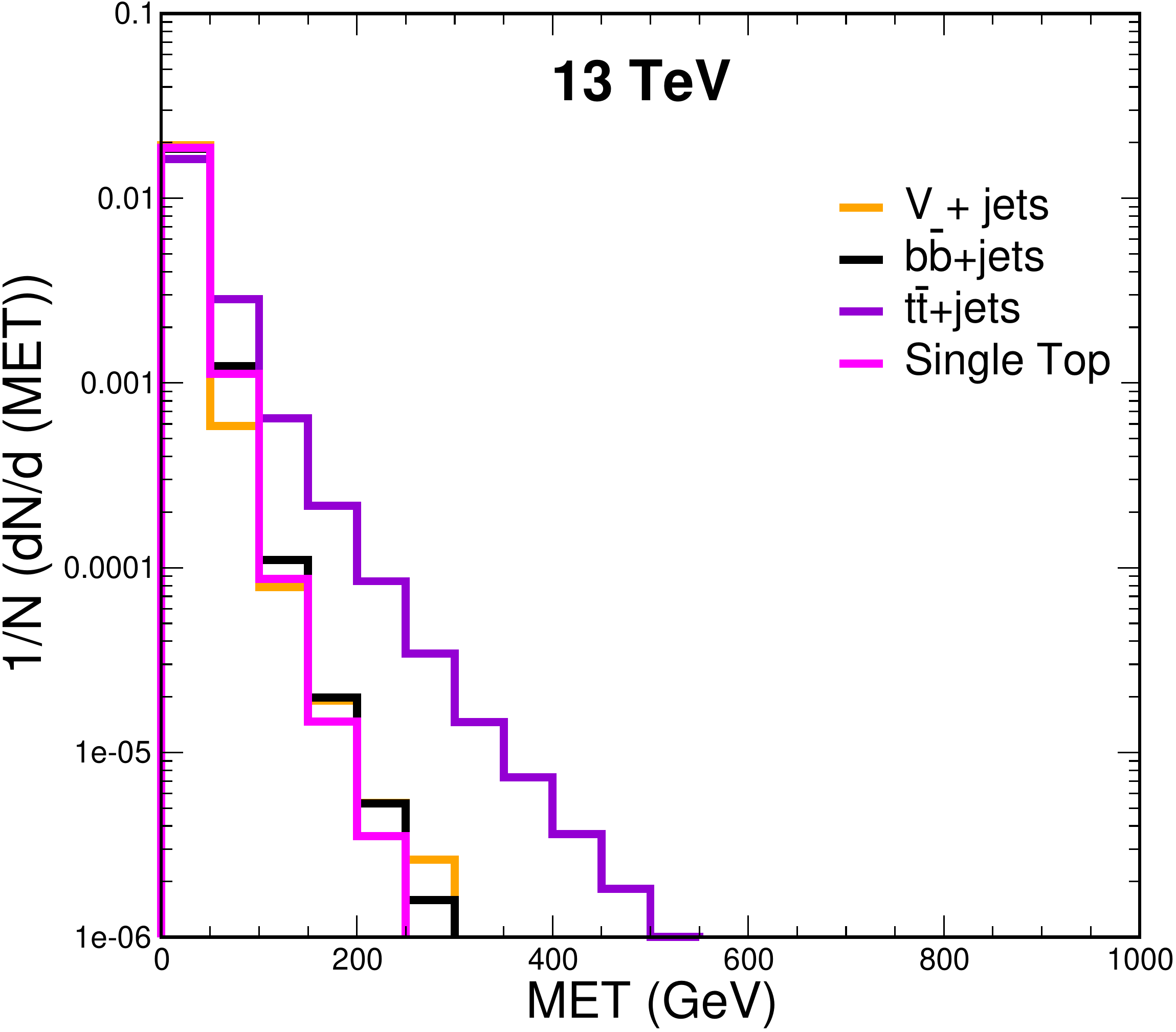}
\end{center}
\caption{Kinematic distributions for the major backgrounds with jets defined using a
radius $R = 0.5$. Top
panel: $p_T$ distributions for the three leading jets $j_0$ (left),
$j_1$ (center) and $j_2$ (right). Middle panel: corresponding
distributions for jet energy. Bottom panel: the scalar sum of the
$p_T$s of all jets (left), the sum of all jet energies (center) and
the missing transverse momentum (right).}
\label{fig:5}
\end{figure}

We turn, now, to a discussion of the various contributions to the
background.  In Fig.\ref{fig:5}, we display the corresponding
kinematic distributions.  The $p_T$ and energy distributions for the
backgrounds are much softer than those for the signal. This, of
course, is expected as far as the two leading jets $j_{0, 1}$ are
concerned, simply on account of the large $m_B$ that we are interested
in. As far as $j_2$ is concerned, the relative hardness of the signal
profile can be understood by realizing that it is $m_B$ that provides
the dominant scale in the production and a momenta of a few hundred
GeVs does not exact too large a price. Consequently, harder cuts on
$p_{Tj}$ or $E_j$ are expected to improve the signal-to-background
ratio. It should be realized, though, that the $p_T$ distributions for
inclusive $t \bar t$ production are relatively wider than those for
the other backgrounds. This difference is even more stark for the
$H_T$ distribution and, thus, a cut on $H_T$ would not preferentially
remove this background over the signal. As this feature has its origin
in the large number of jets resulting from inclusive $t \bar t$
production\footnote{Note that it is the hadronic decay of the top-pair
that is of concern here, for a lepton veto would remove the bulk of
the background from semileptonic decays.}, it might seem that a
restriction on the maximum number of hard jets might be useful in
improving the signal-to-background ratio. This, however, is not a good
option as, apart from the cut-definition sensitivity bringing into
question the infrared safety, an inordinately large fraction would be
lost from an already small signal size. Instead, a cut on $H_T$,
augmented by a minimum requirement for $\sum_j E_j$ (see
Fig.\ref{fig:5}, lower center) would serve the purpose better.  Also
displayed, in the lower right panel, is the distribution in the MET
for each of the individual contributions to the background.  It can be
easily ascertained that it is only the inclusive $t \bar t$ process
that has an MET spectrum significantly harder than that for the signal, the reason
being attributable to both the possible presence of neutrinos in the
former set of events as also the large number of jets, each with its
associated energy resolution.

The analysis presented so far corresponds to a particular choice of
the jet radius $R$ used for jet reconstruction, namely $R= 0.5$. Had
we chosen a somewhat different value, the aforementioned distributions
would suffer only relatively small quantitative changes and no major
qualitative ones. The rationale for our choice, as compared to the
more canonical $R = 0.8$ would become manifest soon.

At this stage, we make our first identification of a fat $Z$-jet,
selecting only those events that contain a jet, termed $J_Z$, that satisfies
the twin conditions of\footnote{The identification of a fat $H$-jet, appropriate for the other channel, proceeds analogously, with $m(J_H) \in [110, 140]\gev$.}
\begin{equation}
m(J_Z) \in [80, 105]\gev \ , \qquad p_T(J_Z) > 200 \gev
\label{eq:fatjet_defn}
\end{equation}
As the $Z$ would need to be highly boosted for its daughters to
coalesce into a single (fat) jet, the requirement on its $p_T$ is not
expected to lead to a significant loss of signal. Note that, even for
the signal, not all events would contain such a jet. For one, some of
the putative $Z$-jets could merge with other jets (or incorporate
hadrons with a different origin), thereby augmenting its mass to
beyond 105 GeV. Similarly, some of its constituents may go missing in
the jet reconstruction algorithm. Finally, there is the probability of
a jet energy mismeasurement. As Table~\ref{tab:1} shows, only in
35\%--50\% of the signal events, is a $Z$ actually reconstructed as a
fatjet.  Fortunately, though, the probability of reconstruction is
much smaller for the background events (see Table \ref{tab:1}). This
suppression is, understandably, extreme for the potentially largest
contributors to the backround, namely inclusive\footnote{Note that we
have not yet imposed $b$-tagging requirement.}  $b \bar b$ and QCD
$n$-jets, as there are no gauge-bosons in such events. The large
suppression for the $t\bar t$ and diboson backrounds owes itself to
two factors. For one, the top-production events have only $W$'s and
not a $Z$, whereas, of the diboson events, only a fraction have
$Z$'s. Secondly, a relatively small fraction of these gauge
bosons have a $p_T$ sufficiently large for the daughter to coalesce
into a fatjet defined using $R = 0.8$. For our favoured choice, namely
$R = 0.5$, this fraction is smaller still.

\begin{table}[!h]
\begin{center} 
\begin{tabular}{|c|c|c|c|c|}
\hline
\multicolumn{5}{|c|}{Signal Events}\\
\hline
 $m_B$ (TeV)& $\%$ of $j_0$ & $\%$ of $j_1$ & $\%$ of $j_2$ & Total(\%)\\
\hline
1.2 & 24.5(28.8) & 9.2(14.4) & 1.2(3.0) &34.9(46.2)\\
1.8 & 30.8(30.8) & 12.8(15.4) & 1.5(3.0) &45.2(49.3)\\
2.2 & 32.8(31.3) & 13.3(15.5) & 1.5(2.9) &47.7(49.8)\\
\hline
\multicolumn{5}{|c|}{Background Events}\\
\hline
Process  & $\%$ of $j_0$ & $\%$ of $j_1$ & $\%$ of $j_2$ & Total(\%)\\
\hline
$V+jets$ & 0.1(0.6) & 0.02(0.17) & 0(0) &0.13(0.78)\\
$b\bar{b}+jets$ & 0.1(0.82) & 0.02(0.24) & 0(0.01) &0.13(1.08)\\
$t\bar{t}+jets$ & 1.36(6.72) & 0.28(1.9) & 0.01(0.17) &1.66(8.81)\\
Single top & 0.16(1.33) & 0.05(0.52) & 0.0(0.02) & 0.22(1.87)\\
Di-boson & 0.93(3.72) & 0.26(1.11) & 0.02(0.04) &1.22(4.88)\\
\hline										

\end{tabular}
\caption{\label{tab:1} {\it The fraction of events
  where one of the three leading jets can be identified as fatjet with
  $p_T > 200\gev$.  Jets are reconstructed with a raduius $R=0.5$ with
  the parenthetical numbers denoting results for $R=0.8$.}}
\end{center}
\end{table}

It is also instructive to examine the distribution of the fatjet
within the three leading jets. As for the signal events, {\em a
priori}, one would imagine that, the $Z$ is as likely to lead to $j_1$
as compared to $j_0$. While this is indeed true, the differing rates
in Table~\ref{tab:1} can be understood by realizing that if $J_Z$ is
to be identified with $j_1$, it would require that $j_0$ must have a
$p_T$ substantially larger than 200 GeV. Naturally, only a smaller
draction of events would satisfy this. Similar arguments also explain
the much fewer incidences of $j_2$ (or, $j_{3,4}$ etc, if the exist)
being identified as the fatjet.

Particularly important, in this analysis, is our choice of the jet
radius $R$. With the gauge bosons from such background events not
being highly boosted, the choice $R = 0.5$ implies that the
daughters of the gauge boson are more likely to be reconstructed as
two independent jets rather than a single fatjet. This has to be
contrasted with the case of the signal where the $Z$ tends to be
highly boosted. Sure enough, an increase to $R = 0.8$ would
significantly enhance the probability for
background as opposed to a very small
increase for the signal events (see Table \ref{tab:1}). This
vindicates our use of $R = 0.5$.

\begin{figure}[!h]
\begin{center}
\includegraphics[width=5.5cm,height=5.0cm]{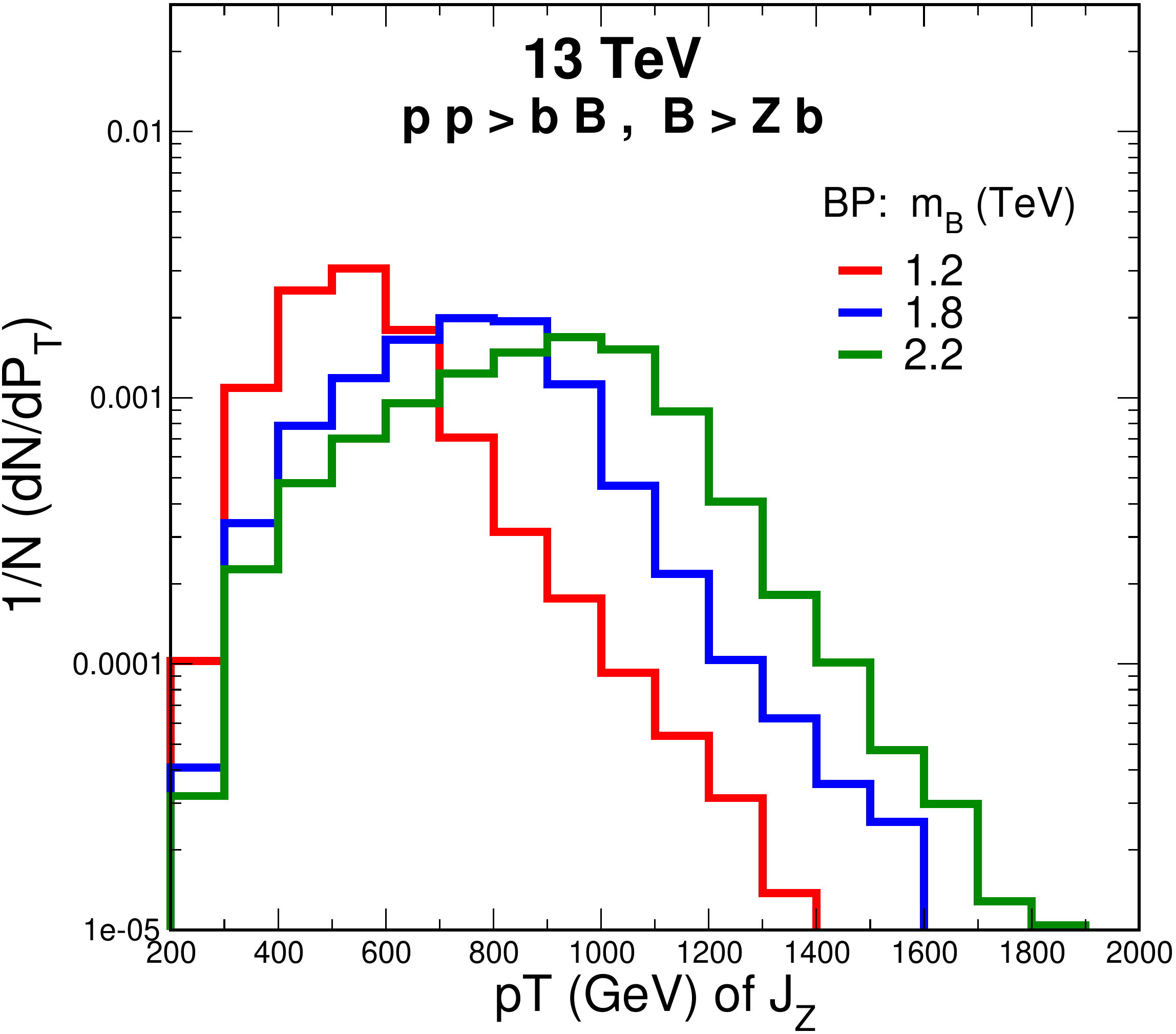}
\includegraphics[width=5.5cm,height=5.0cm]{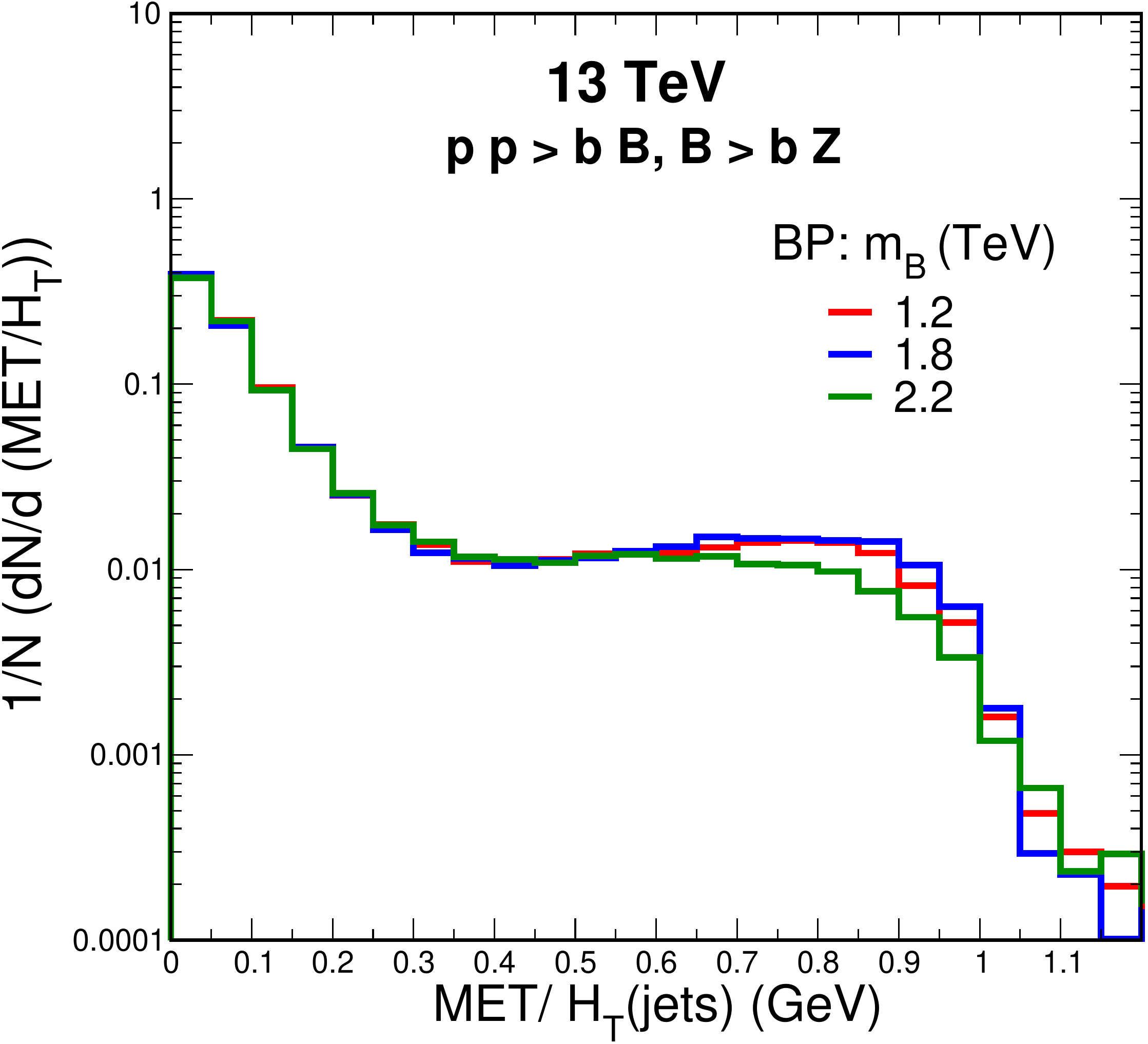}
\includegraphics[width=5.5cm,height=5.0cm]{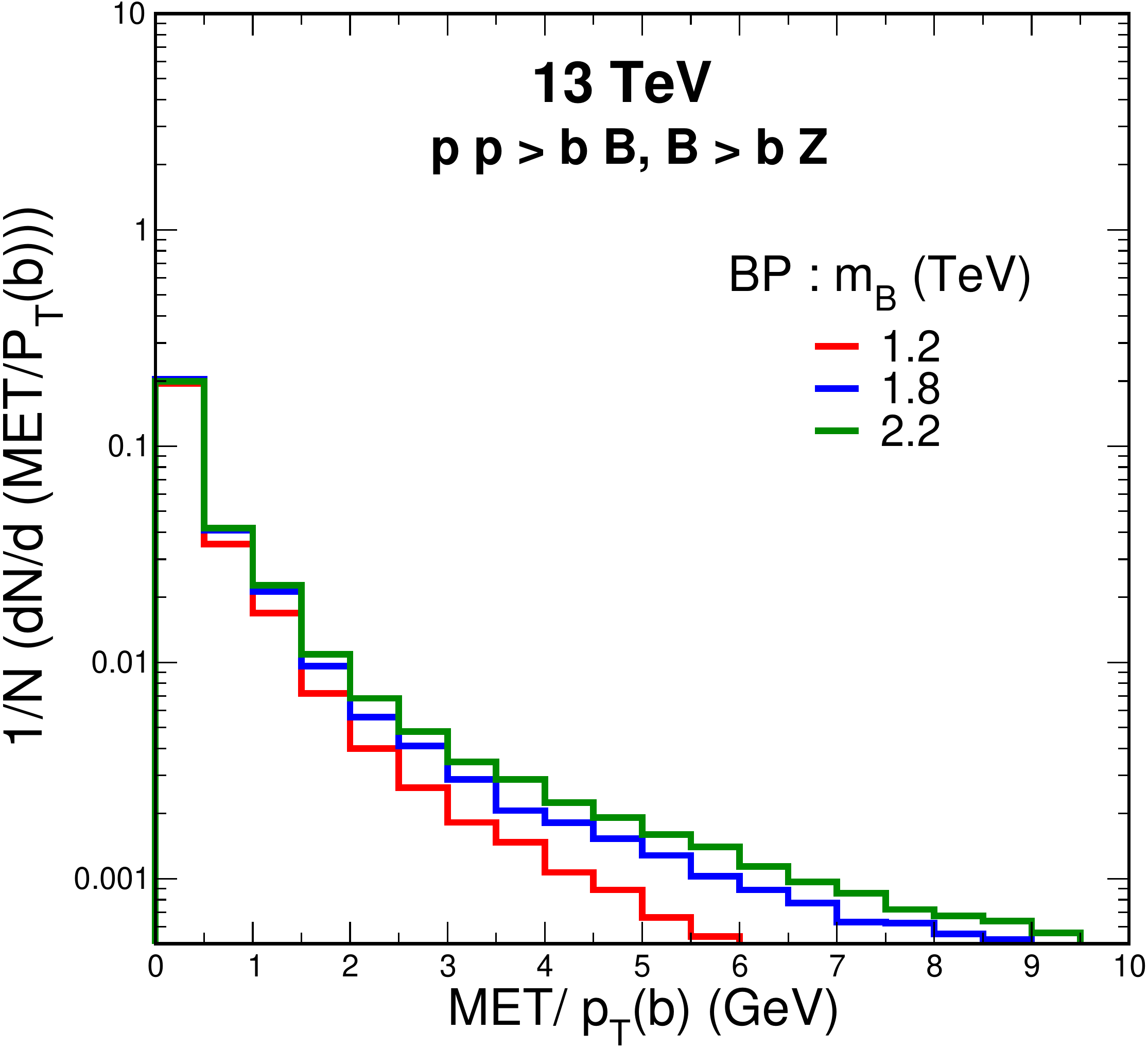}
\includegraphics[width=5.5cm,height=5.0cm]{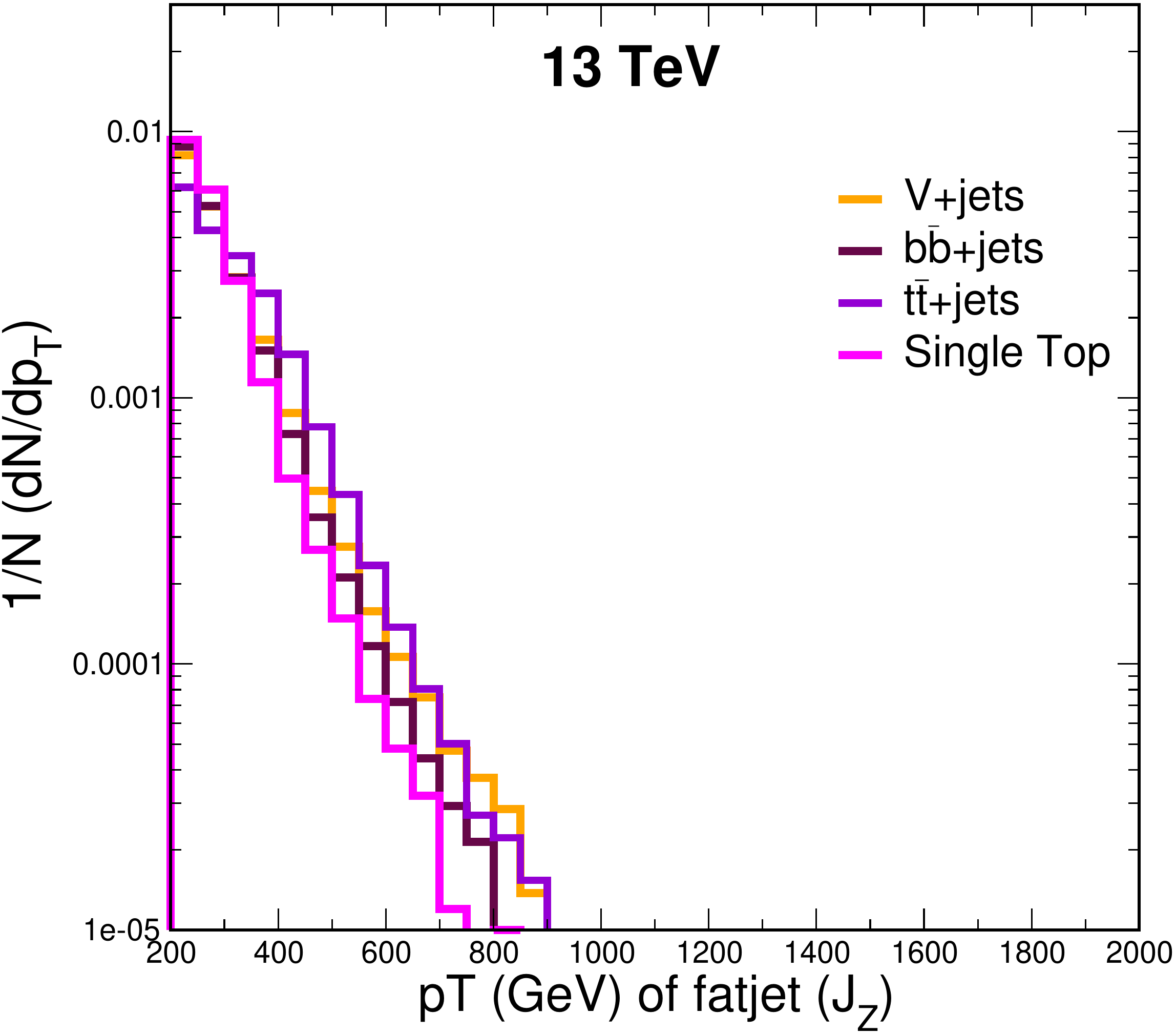}
\includegraphics[width=5.5cm,height=5.0cm]{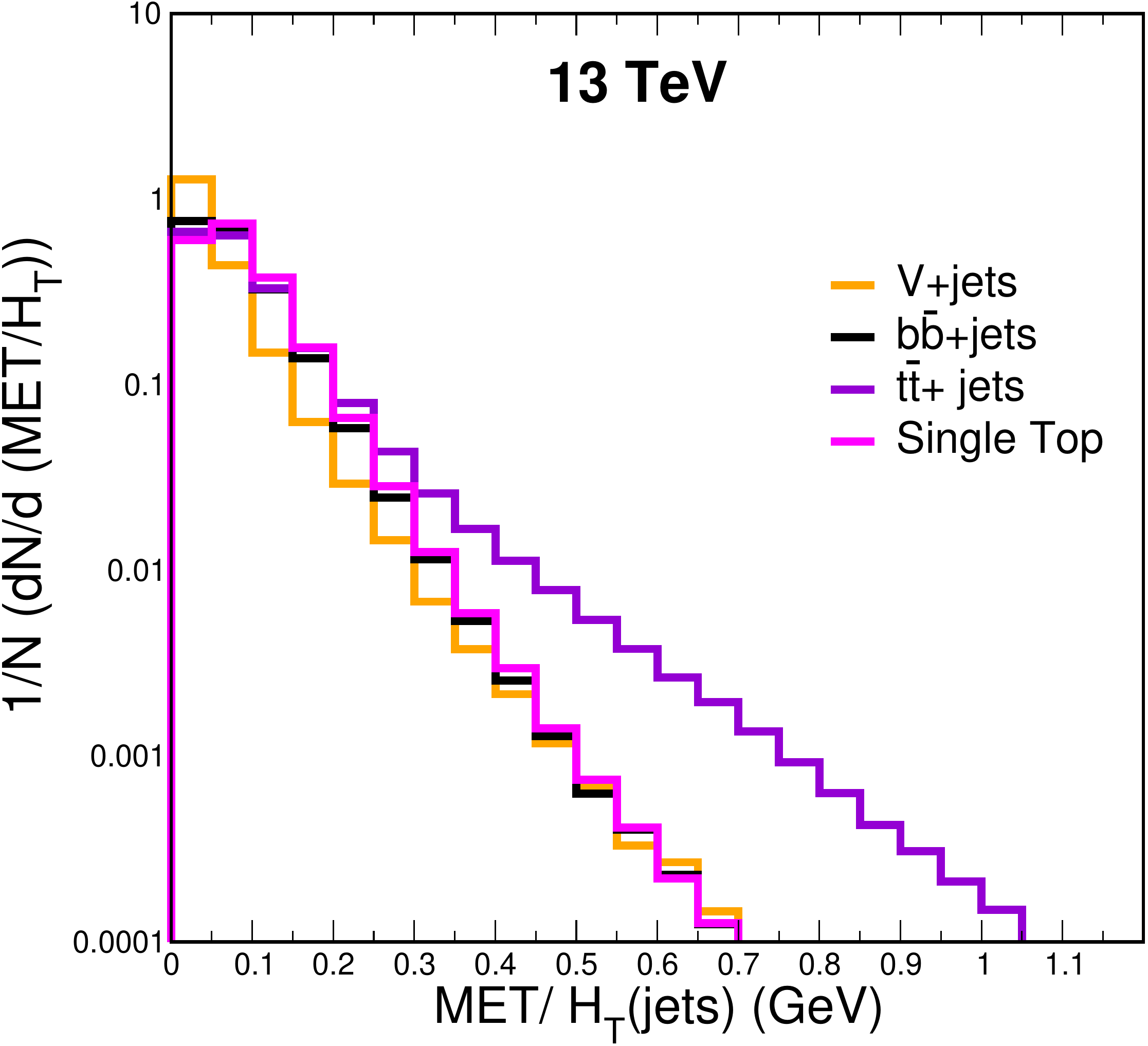}
\includegraphics[width=5.5cm,height=5.0cm]{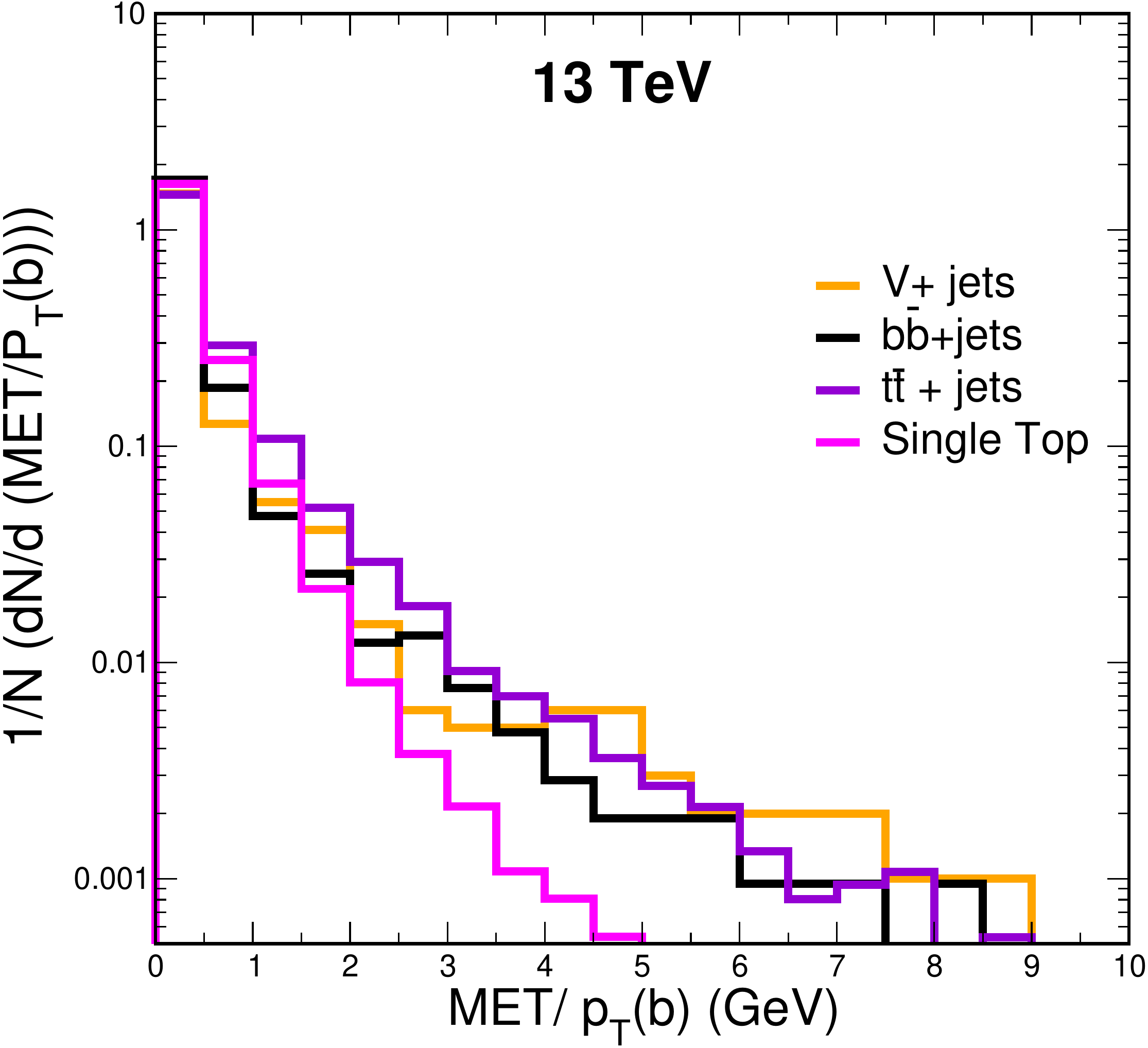}
\end{center}

\caption{\em Kinematic distributions after demanding the existence of a
$Z$-fatjet
(where jets are reconstructed with $R=0.5$) with $p_T(J_Z) > 200\gev$ and
$m(J_Z) \in [80,105]\gev$.}

\label{fig:6}
\end{figure}

In Fig.\ref{fig:6}, we display a few kinematic distributions of
importance constructed for events that do contain a fatjet as defined
by eq.(\ref{eq:fatjet_defn}). Even a cursory comparison of the two
leftmost panels establishes that a much stronger restriction on the
$J_Z$ transverse momentum, such as $p_T(J_Z) > 400 \, (500) \gev$
would significantly reduce the background without impinging much on
the signal strength.  Also shown in Fig.\ref{fig:6} are ratios of the
missing transverse momentum MET with $H_T$ (central panels) and the
$p_T$ of the leading $b$-tagged jet. Despite the former having
remarkably different structure for the signal and backgrounds, a restriction on this ratio so as to enhance the signal-to-background ratio would, simultaneously, reduce the total size of the signal drastically. On the other hand, the
ratio MET/$p_T$ would turn out to be an useful discriminant.

\begin{figure}[!h]
\begin{center}
\includegraphics[width=6.5cm,height=5.5cm]{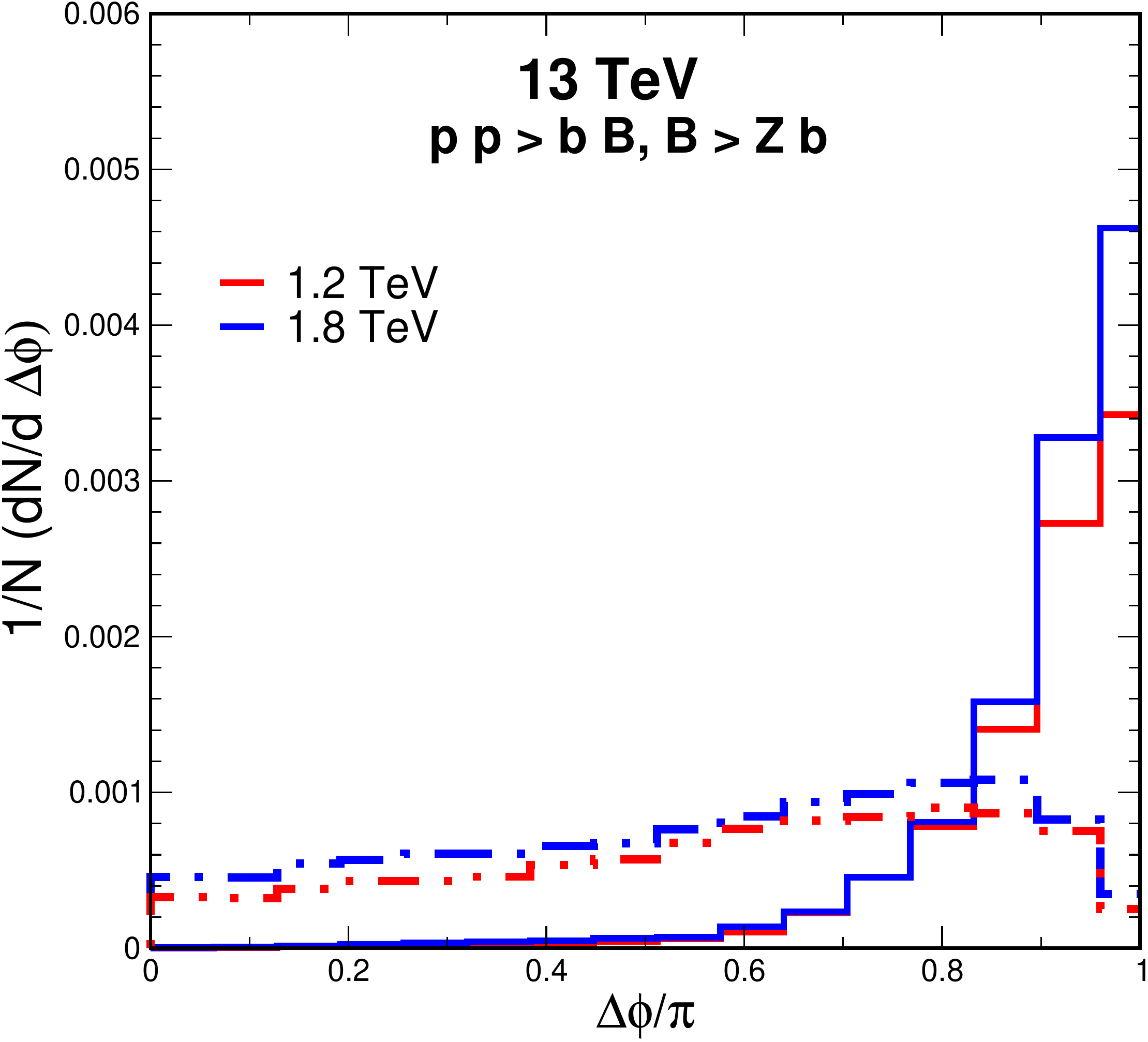}
\includegraphics[width=6.5cm,height=5.5cm]{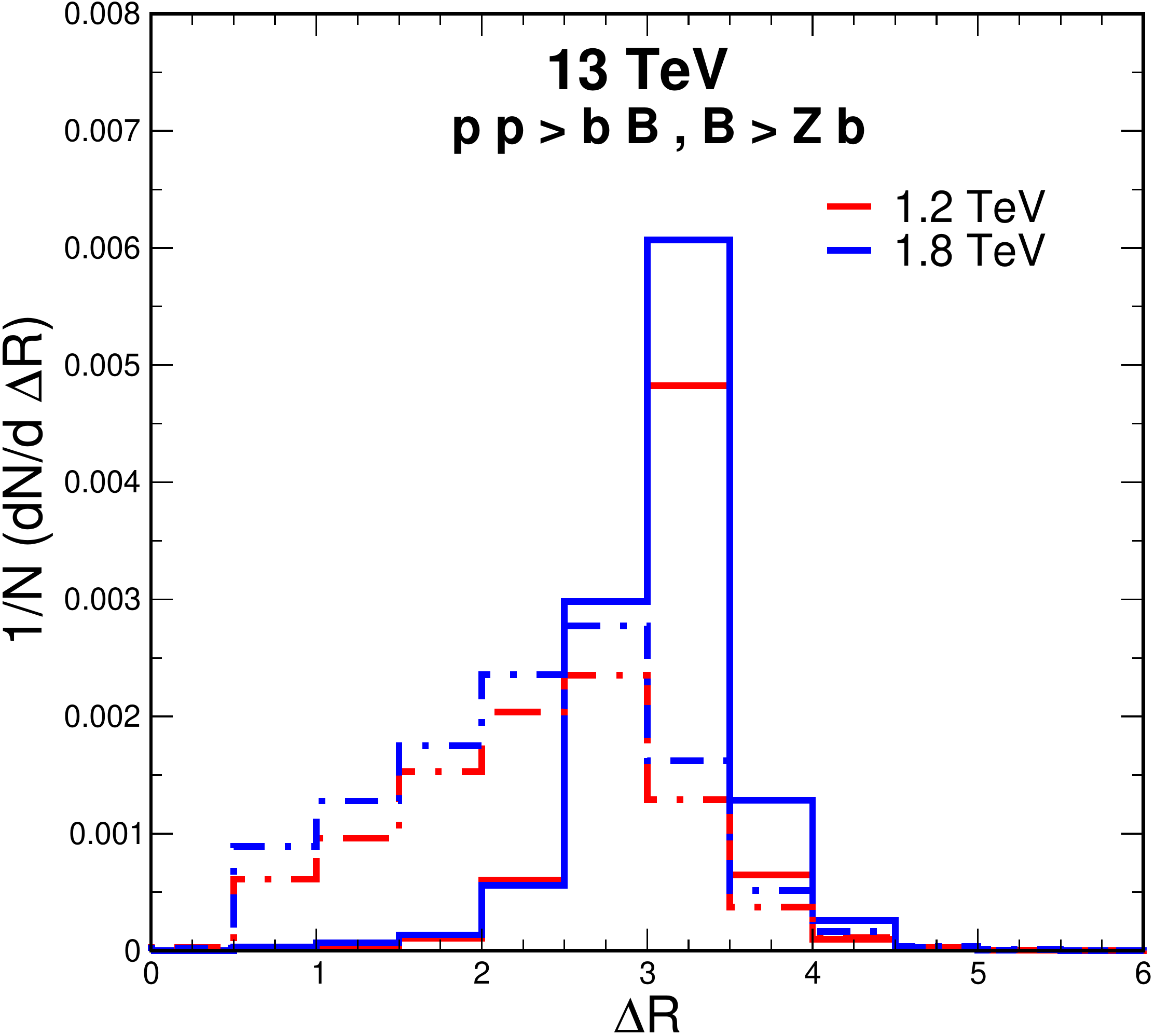}
\end{center}
\caption{\em Left: Solid (dashed) curves show the distribution in
azimuthal distance between $J_Z$ and $J_1 \, (J_2)$ for two signal
BPs.  Right: Analogous distributions for $\Delta R$.  The fatjet $J_Z$
are required to have $p_T > 200$ GeV and jetmass in the $[80,105]\gev$
range. The basic cuts as mentioned in Sec.4.3 are imposed too.}
\label{fig:del_phi_R}
\end{figure}

Once the fatjet $J_Z$ has been identified in the correct mass window,
we consider, next, the two hardest of the remaining jets. The one with
the maximal azimuthal separation with $J_Z$ is christened $J_1$ and
the other as $J_2$. This identification is particularly useful in
identifying the signal events. With the very heavy $B$, typically
being produced with a relatively small transverse momentum, {\em viz.}
$p_T(B) \ll m_B$, its daughters ($J_Z$ and the putative $J_1$) are
likely to be produced with $\Delta\phi(J_Z, J_1) \sim \pi$, as is seen
in Fig~\ref{fig:del_phi_R}(left). Consequently, the corresponding
$\Delta R$ too tends to be large as in
Fig~\ref{fig:del_phi_R}(right). Such a strong preference is not
expected of the background events. That $\Delta\phi(J_Z,
J_2)$ has a slight preference for large values as
well is understandable, for the $\bar b$ produced alongwith the
$B$ would have a small momentum too; consequently, momentum
conservation would stipulate that the transverse component would
prefer to be opposite that of the leading transverse momentum in the
event, namely that of $J_Z$. It might seem, though, that analogous
arguments would hold for the backgrounds too. This is only partially
true though, for some of the contributions tend to have a larger
sphericity. And, finally, in the event of multijet events, we find
that the probability of the true $b$--daughter of the decaying $B$ not
leading to $J_1$ is much less than 1\%, thereby establishing the
robustness of the identification.

\begin{figure}[!h]
\begin{center}
\includegraphics[width=6.5cm,height=5.5cm]{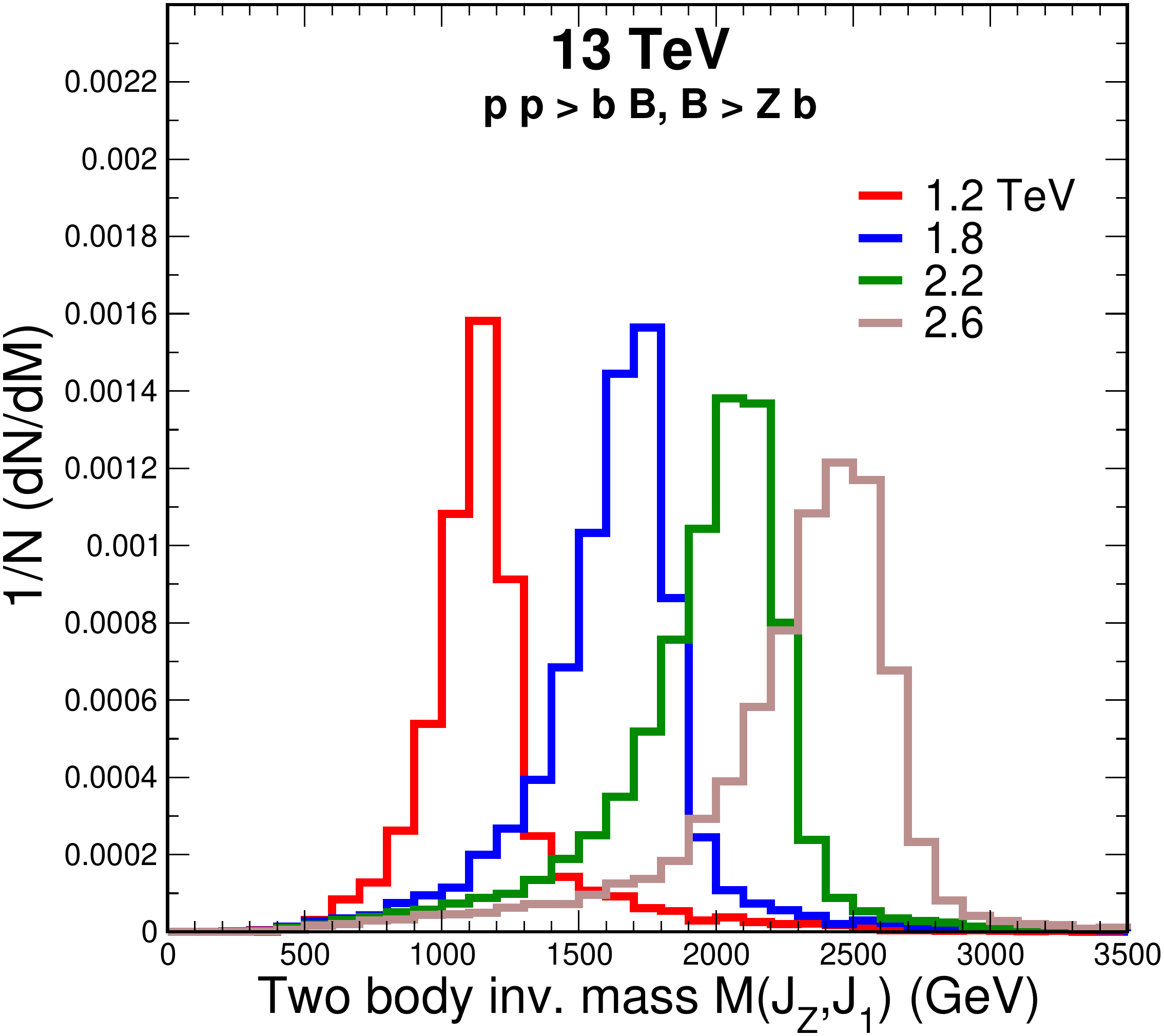}
\includegraphics[width=6.5cm,height=5.5cm]{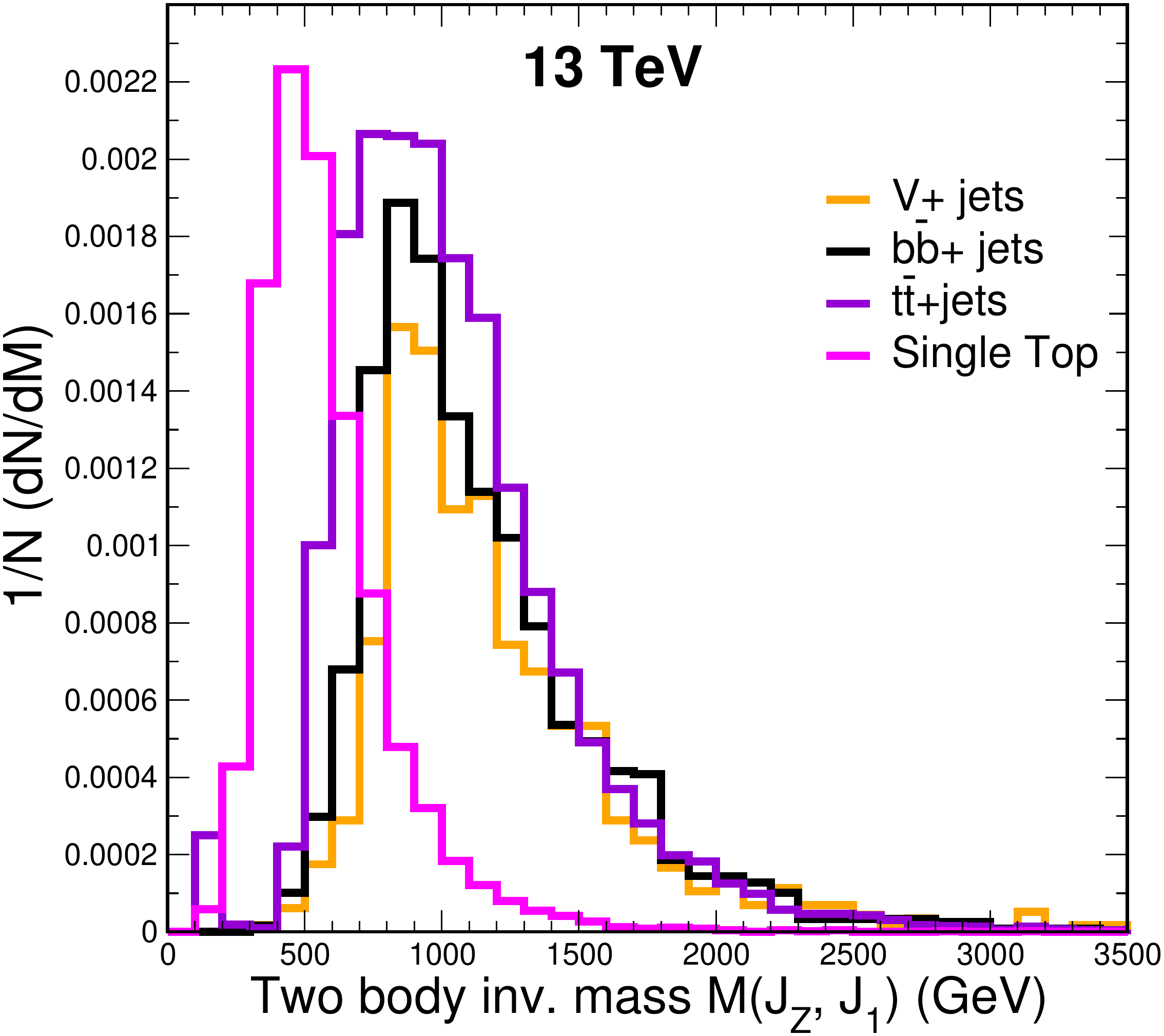}
\end{center}
\caption{\it The distribution in the two body invariant mass of the fatjet 
$J_Z$ and $J_1$ for the signal (left) at different BP's and the major
background (right).  Fatjets are selected with $p_T > 200$ GeV and
jetmass within $[80,105]\gev$, using $R=0.5$.  The basic cuts as
mentioned in Sec.4.3 are imposed.}
\label{fig:m_inv_jv_j1}
\end{figure}
This contention is further supported by Fig~\ref{fig:m_inv_jv_j1}
where we display the two body ($J_Z$ and $J_1$) invariant mass
distribution, for different signal BP's as well as individual
backgrounds. With the background falling away at large values of this
invariant mass, we expect that restricting ourselves to $|m_{\rm inv}
- m_B| \leq 3 \Gamma_B$  would further
accentuate the sensitivity for large $m_B$ (a region with small signal
cross sections).

\subsection{Fatjet Characteristics for Signal and Background} 
\label{subsec:fatjet_charac}
In the preceding subsection, we have demonstrated the importance of
demanding a fat jet($J_Z$) in the signal. The
criteria to define a jet to be a fat one, though, were {\em ad hoc} in
nature. We, now, reexamine such issues, aiming to best define it.  To this end, we take a few steps back, not only relaxing the
criteria on the fatjet mass (as described in the preceding
subsection), but also reconsidering the radius $R$ used to
define jets.

\begin{figure}[!h]
\begin{center}
\includegraphics[width=7.5cm,height=7cm]{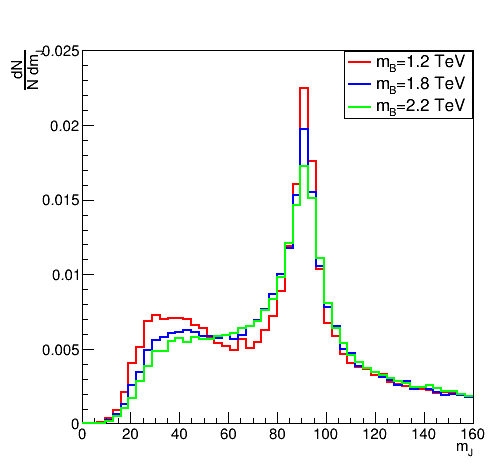}
\includegraphics[width=7.5cm,height=7cm]{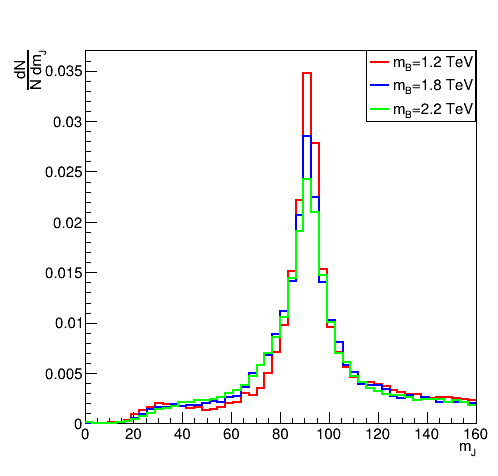}
\includegraphics[width=7.5cm,height=7cm]{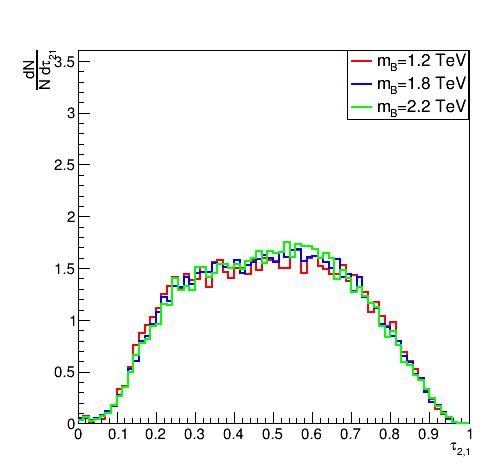}
\includegraphics[width=7.5cm,height=7cm]{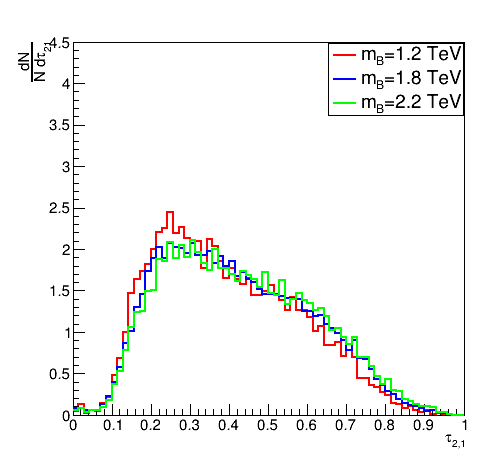}
\end{center}
\caption{\em The jet mass ($m_J$) and $\tau_{21}$ distributions
  for a fatjet (of $p_T > 500$) as constructed with a radius parameter
   $R=0.8$ and Top: Distributions of $m_J$ before any
   $\tau_{21}$ cut (left) and after $\tau_{21} < 0.5$ (right) for three
   signal BP's.  Bottom: Distributions in $\tau_{21}$ before any cut
   on $m_J$ (left), after a cut of $80<m_J<105$ GeV (right).}
\label{fig:fatjet_0.8}
\end{figure}

To begin with, we consider the popular choice of $R = 0.8$.  In
Fig.\ref{fig:fatjet_0.8}, we display two important characteristics of
the putative fatjet, requiring it only to satisfy $p_T > 500\gev$.
The left panels show the distributions without any further cuts, the
top one for the jet mass, and the bottom one for the ratio of two
subjettiness parameters (see eqn.\ref{N_subjettiness}), {\em viz.}
$\tau_{21} \equiv \tau_{2}/\tau_{1}$. While there exists a peak at
$m_Z$, there is a hint of a second peak a little below $40\gev$.
The latter can be understood in terms of relatively hard and asymmetric QCD
radiations. The expectation value of the mass of such jets
(calculated, for example, by considering the $q \rightarrow qg$
splitting function and then performing the $\theta$ and $z$
integrations) is given by\cite{Shelton:2013an}
\beq
\langle m^2 \rangle = \dfrac{3 \alpha_s}{8 \pi} C_F p_T^2 R^2
        \label{qcd_peak}
\eeq
and, hence, the mass scales linearly with jet $p_T$
and the radius parameter $R$. That the low-energy peaks in
Fig.\ref{fig:fatjet_0.8} are indeed QCD-driven is also attested to 
by the gradual drop around the central   value \cite{Shelton:2013an},
as distinct from sharp peak corresponding to an on-shell particle decay.
Such an origin also explains why the peak is 
more pronounced for smaller $m_B$.
 Notwithstanding our ability to explain the
secondary peak, its very presence and size seems to argue against the
identification of the fatjet as a two-pronged decay. Similarly, the
$\tau_{21}$ distribution does not show any inclination for
$\tau_{21}<0.5$, as a two-prong system should, ideally, favour.

\begin{figure}[!h]
\begin{center}
\includegraphics[width=7.5cm,height=7cm]{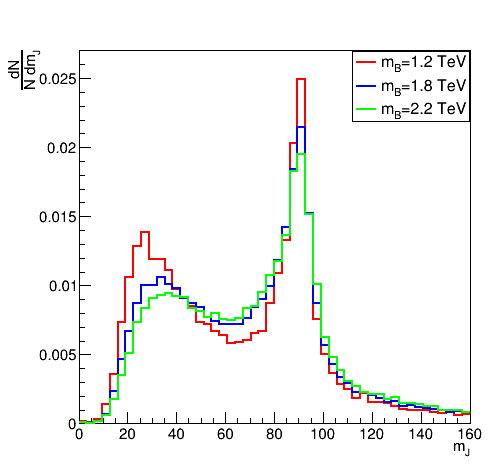}
\includegraphics[width=7.5cm,height=7cm]{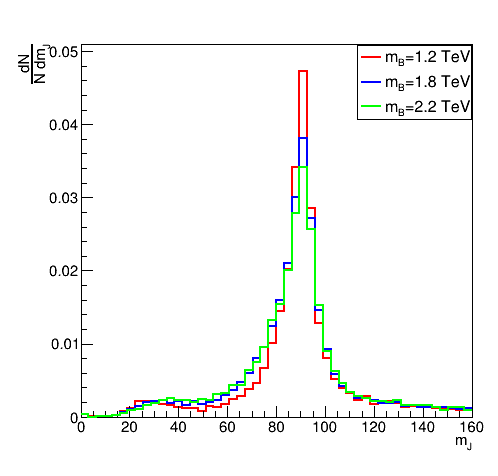}
\includegraphics[width=7.5cm,height=7cm]{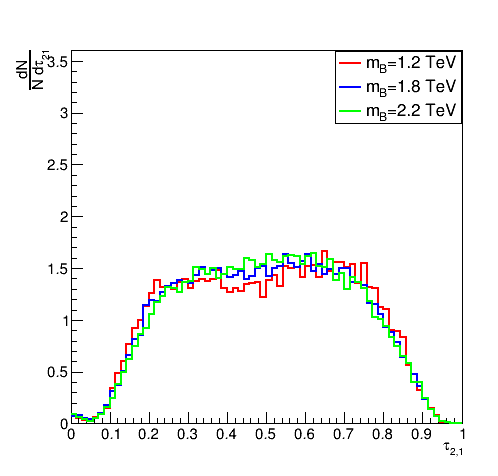}
\includegraphics[width=7.5cm,height=7cm]{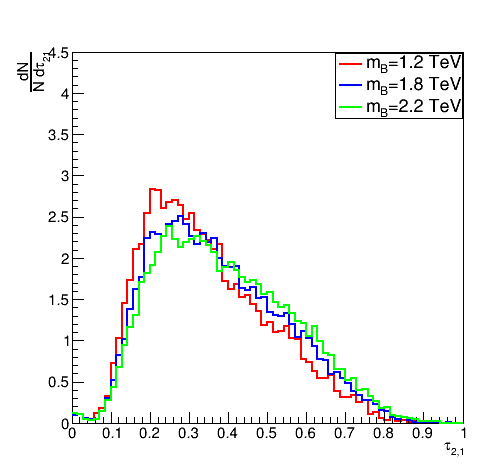}
\end{center}
\caption{\em The jet mass ($m_J$) and $\tau_{21}$ distributions
  for a fatjet (of $p_T > 500$) as constructed with a radius parameter
   $R=0.5$ and Top: Distributions of $m_J$ before any $\tau_{21}$ cut
   (left) and after $\tau_{21} < 0.4$ (right) for two signal BP's.
   Bottom: Distributions in $\tau_{21}$ before any cut on $m_J$
   (left), after a cut of $80<m_J<105$ GeV (right).}
\label{fig:fatjet_0.5}
\end{figure}

It is interesting to consider, at this stage, the situation for jet
reconstruction using $R=0.5$, as detailed in
Fig.\ref{fig:fatjet_0.5}. The aforementioned secondary peak (at $m_J
\lapp 40 \gev$) is even more pronounced now, thanks to the fact that
such an $R$ calls for narrower jets. Quite analogously, the preference
for $\tau_{21} > 0.5$ is strengthened. Both these arguments would seem
to call for $R=0.8$ being a better choice. This is, however, belied by
an examination of the right panels for both
Figs.\ref{fig:fatjet_0.8}\&\ref{fig:fatjet_0.5}. The top right panels
in both clearly demonstrate that a cut\footnote{The differing choices
for this cut (for the two values of $R$) is occasioned by consideration
of optimizing the signal-to-noise ratio.}  on $\tau_{21}$ strongly
accentuates the $Z$-peak, simultaneously obliterating the secondary
peak at low $m_J$. Analogously, requiring the jet mass to lie in the
$80 \gev \leq m_J \leq 105 \gev$ window results in the $\tau_{21}$
distribution showing a clear bias for $\tau_{21} < 0.5$ (bottom right
panels). Both these observations are as expected, for it is only such
events that should show up as two-prong decays of the $Z$. Indeed,
were we to consider a two-dimensional distribution (in the
$\tau_{21}$--$m_J$ plane), the event rate would show a clear
correlation between these two kinematic variables.

It is interesting to note that the $\tau_{21}$ distribution, after
effecting the fatjet mass window restriction, is slightly flatter for
a larger $m_B$. This holds true for both choices of $R$ (see
Figs.~\ref{fig:fatjet_0.8} \& \ref{fig:fatjet_0.5}) and is but a
consequence of the fact that as the fatjet is boosted more and more,
the distinction between one-prong and two-prong configurations is
progressively blurred.  This can be understood by  examining
the $Z$ decay into a $q \bar q$ pair which, through their subsequent
radiation, putatively lead to the two smaller cones within the
fatjet. In calculating $\tau_2$ (see eq.\ref{N_subjettiness}) two
subjet axes are assigned along the direction of these two cones.  For
an individual fatjet constituent, the minimum of the two $\Delta R$s
between the constituent and the assigned axes contrubutes to
$\tau_2$. Consequently, the value of $\tau_2$ tends to be small.  On
the other hand, in calculating $\tau_1$, only one subjet axis is
defined and assigned along the direction of the fatjet. Consequently
$\tau_1$ for $m_B = 1.8\tev$ is smaller than that for $m_B = 1.2\tev$
resuling in a flatter $\tau_{21}$ distribution for the former case

It should be further noticed that the $\tau_{21}$ distribution is
sharper in Fig.\ref{fig:fatjet_0.5} compared to
Fig.\ref{fig:fatjet_0.8}. Again, this was to be expected given that a
smaller value of $R$ will allow for smaller amount of secondary
radiations within the cone. As we shall shortly see,
the situation is markedly different for the top-decays contributing to
the background and, together, this prompts us to adopt $R=0.5$ as the
definitive requirement. As for the jet-mass requiremnt, clearly
the choice $[80, 105]\gev$ would be better than, say the $[65,
105] \gev$ one (as adopted in various analyses, on account of the
former discriminating against $W$-backrounds (whether from direct $W$
production or from top-decays). Finally, we find the $p_T > 500\gev$
requirement to be a nearly optimal one.

\begin{figure}[!h]
\begin{center}
\includegraphics[width=7.5cm,height=7cm]{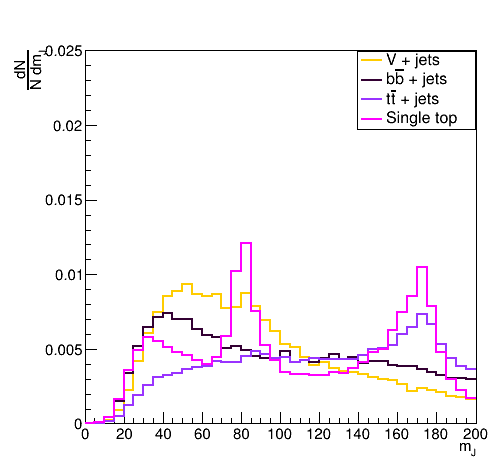}
\includegraphics[width=7.5cm,height=7cm]{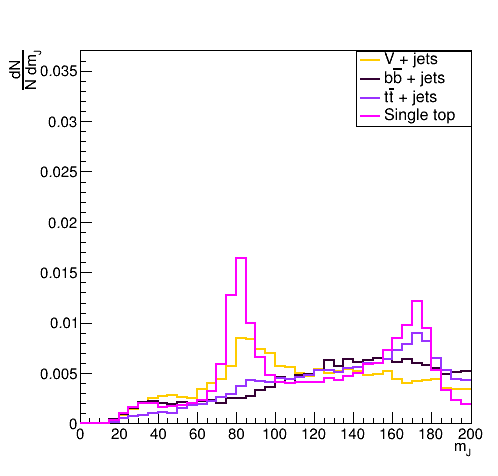}
\includegraphics[width=7.5cm,height=7cm]{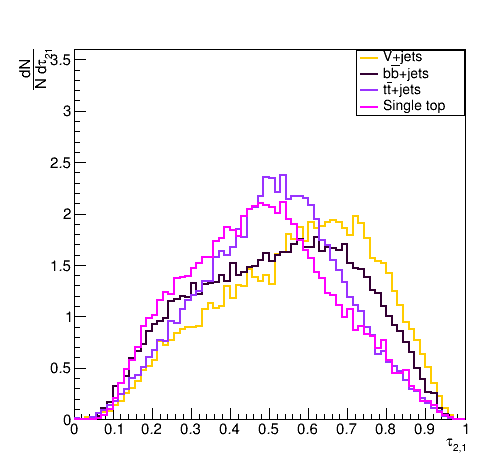}
\includegraphics[width=7.5cm,height=7cm]{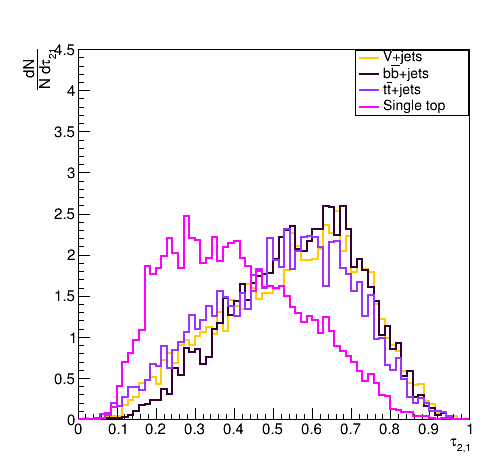}
\end{center}
\caption{\em As in Fig.\ref{fig:fatjet_0.8}, but for the leading contributions to the background with $R=0.8$.}
\label{fig:fatjet_bkgd_0.8}
\end{figure}
\begin{figure}[!h]
\begin{center}
\includegraphics[width=7.5cm,height=7cm]{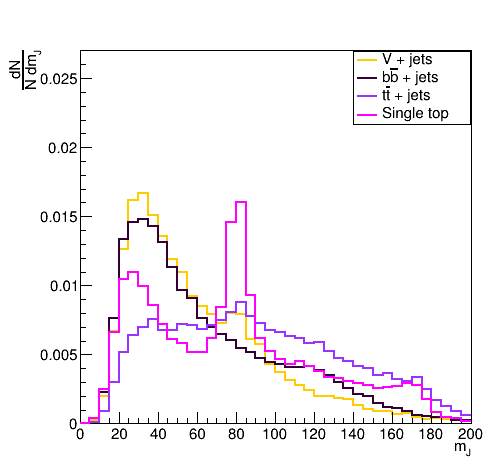}
\includegraphics[width=7.5cm,height=7cm]{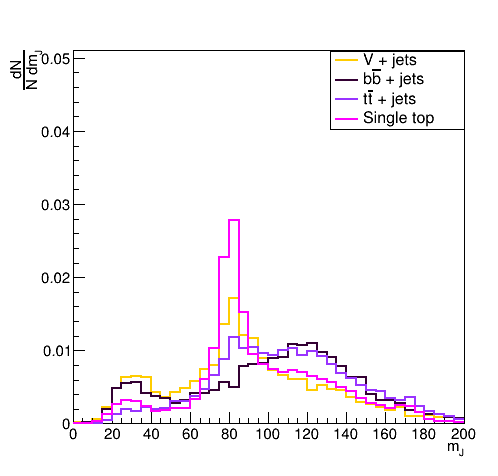}
\includegraphics[width=7.5cm,height=7cm]{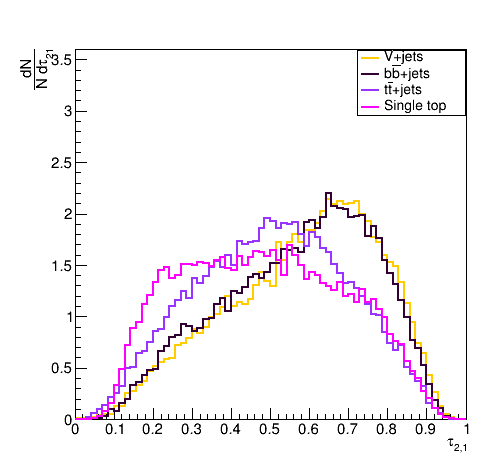}
\includegraphics[width=7.5cm,height=7cm]{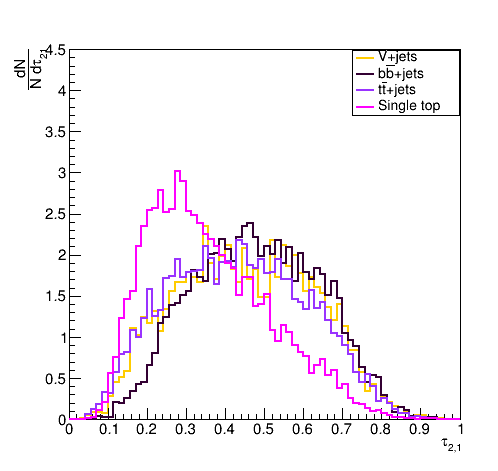}
\end{center}
\caption{\em As in Fig.\ref{fig:fatjet_0.5}, but for the leading contributions to the background with $R=0.5$.}
\label{fig:fatjet_bkgd_0.5}
\end{figure}

We, now, consider the corresponding fatjet characteristics for the
leading contributions to the background, once again for both $R = 0.8$
(Fig.\ref{fig:fatjet_bkgd_0.8}) and $R = 0.5$
(Fig.\ref{fig:fatjet_bkgd_0.5}). Concentrating on the top-left panel
of Fig.\ref{fig:fatjet_bkgd_0.8}, both the $t \bar t$ and the
single-top contribution clearly show a peak at
Fig.\ref{fig:fatjet_bkgd_0.8}. The latter, in addition, shows a
stronger peak close to the gauge boson masses. While, naively, one
would also have expected the $t \bar t$ events to also show the second
peak, courtesy the $W$ in the decay chain, note that the large $p_T$
demanded of the fatjet would mean that the $W$ and the $b$ would tend
to coalesce into a single jet. The use of $R = 0.5$, instead, altogether
removes the peak at $m_J \approx m_t$ (see
Fig.\ref{fig:fatjet_bkgd_0.5}). This can be understood by realizing
that the radius parameter is essentially set by $R \sim m_J/E_J$. For
$R = 0.8$, the large $p_T$ that has been demanded of the top (putative
fatjet) allows it to decay within $R$; and once the three prong nature
of the top is resolved, it is identified as a fatjet. For the
reconstruction of $t$-quark (as a fatjet) within a smaller radius of
$R=0.5$, a much higher energy is required of the top and only a small
fraction of events would satisfy this. Rather, the fraction of events
with $m_J$ close to $m_{W,Z}$ is enhanced.  Once again, the strong
peaks at small $m_J$ can be understood in terms of the secondary QCD
radiation (see discussion surrounding eqn.\ref{qcd_peak}). It is
interesting to note that neither of the two aforementioned plots shows
a peak at $m_J \approx m_t$ for the inclusive $t \bar t$ events. This
can be understood by realizing that only a small fraction of such
events would boast of a top with a $p_T$ sufficiently large enough for
the top to be manifested as a fat jet.

 In the bottom panels of both Fig.\ref{fig:fatjet_bkgd_0.8} and
Fig.\ref{fig:fatjet_bkgd_0.5}, we plot the distributions of
$\tau_{21}$ before any $m_J$ cut (left) and after requiring $m_J \in
[80,105]\gev$ (right).  Before the cut, none of the contributors to
the background show a preference for $\tau_{21} < 0.5$. Even on
imposition of the cut, it is only the single-top production that
exhibits this preference, primarily on account of the $W$. Similarly,
for the $V + {\rm jets}$ contribution, the $m_J$ cut better
accentuates the two-prong nature for $R=0.5$ (than for $R=0.8$). An
analogous enhancement does not occur for the $t \bar t$ background
(especially for $R = 0.8$) as a large fraction of the fatjet
reconstruction, before the cut, would be associated with the entire
top. The QCD jets being a diffuse spray of large angle radiation, the
$\tau_{21}$ distribution for the inclusive $b\bar b$ events is,
understandably, quite flat.

It might seem surprising that the top left panels of both
Fig.\ref{fig:fatjet_bkgd_0.8} and Fig.\ref{fig:fatjet_bkgd_0.5} do not
show a peak in the $m_J \in [80,105]\gev$  for the
$V$+jets background. This can be understood by realizing that the
gauge boson, very often, has a slightly smaller $p_T$ than the leading
QCD jet(s). Furthermore, the secondary radiation off the leading
parton often leads to these being identified as the fatjet,
especially when a strong cut is applied on the
fatjet $p_T$. Consequently, the peak is smeared to a great
extent. When the $\tau_{21}$ restriction is applied, the peak duly
stands out.

\begin{figure}[!h]
\begin{center}
\includegraphics[width=7.5cm,height=7cm]{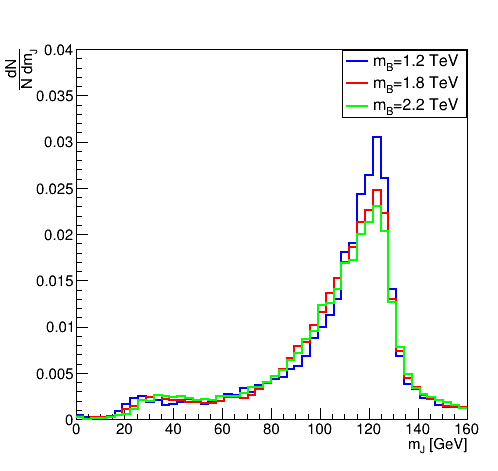}
\includegraphics[width=7.5cm,height=7cm]{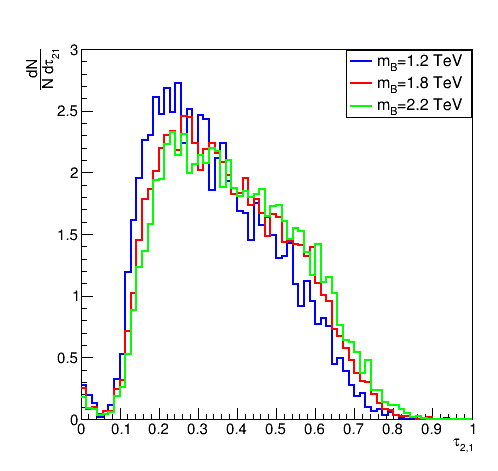}
\end{center}
\caption{\em Plots for Higgs-fatjet scenario, left one is jet-mass plot for $\tau_{21}<0.4$ and the right one 
is $\tau_{21}$ plot after a mass-cut of $110<m_j<140$ with $R=0.5$.}
\label{fig:Higgs_fatjet_0.5}
\end{figure}

With the event topology for $bH$ being similar to $bZ$, and given that
the difference $m_H - m_Z \ll m_B$, we would expect the Higgs too to
lead to a fatjet. So, in Fig~\ref{fig:Higgs_fatjet_0.5}, we
present the jetmass and the $\tau_{21}$ plots for the choice $R=0.5$ for the Higgs
scenario after putting a cut of $\tau_{21}<0.4$ on the former and a
cut of $110<m_J<140$ on the later. We see that the Higgs-mass and the
two-prong behavior are correctly identified. A comparison
of the results for the choices $R=0.5$ and
$R=0.8$ leads us to conclusions
 similar to those reached for the $Z$-fatjet case with the $\tau_{21}$ distribution for $R=0.8$ being flatter than the $R=0.5$ case in the Higgs fatjet mass window.
It is also evident from the background plots
(left figures of Fig.\ref{fig:fatjet_bkgd_0.8} ) that in the jet mass
region of [110-140] the backgrounds are expected to be much
smaller. The rest of the behaviour of the backgrounds remain same and,
hence, we do not show them separately. 
\subsection{Result}
\label{subsec:results}
We begin by discussing the $Z$-fatjet channel in detail. The analysis
for the $H$-fatjet channel (presented thereafter) follows analogously with some
subtle differences, which would be highlighted. Thereafter, we
the delineate the part of the parameter space that could be ruled
out by either channel or even lead to a discovery.
\subsubsection {The $Z$-fatjet Channel} 

As we have already discussed, the $R=0.5$ choice is not only more
commensurate with the two-prong nature of the signal, while
suppressing the contributions from the $t\bar t$ background, but also
allows for the imposition of a stronger cut on $\tau_{21}$, so as to
further reduce the backgrounds. Consequently, we would largely
concentrate on this choice, and comment only briefly on the results
for $R=0.8$.  Guided by the kinematic distributions for the signal and 
background events, as described earlier, we now discuss the selection
cuts, in the order these are imposed, alongwith the effects these
have.

\begin{itemize}
\item {\bf Selection 1 ($S_1$):} We require at least three jets, imposing a veto on isolated leptons satisfying
\beq
    p_T > 10\gev \ , \qquad |\eta| < 2.5 \ , \qquad {\rm and} \qquad
          \Delta R_{\ell j} > 0.4 \ .
\eeq
Ordering them according to the transverse momenta, the three leading jets
must not only satisfy
\beq
p_T (j_0) > 400\gev \ , \qquad p_T (j_1) > 250\gev \ , \qquad
p_T (j_2) > 50\gev
\eeq
but also marginally stronger requirements on their total energy (in the laboratory-fixed frame), namely
\beq
E(j_0) > 500\gev \ , \qquad E(j_1) > 300\gev \ , \qquad
E(j_2) > 100\gev \ .
\eeq
In addition, we require that
\beq
H_{T_{\rm jets}} > 1000\gev  \quad {\rm and} \quad \sum_{\rm jets}E > 1200\gev \ .
\eeq

\item {\bf Selection 2  ($S_2$):}
We require that at least one of the jets from the preceding stage gets
resolved into a fat jet $J_Z$ satisfying
\beq
    m_J \in [80, 105] \gev  \quad {\rm and} \quad p_T(J_Z) > 500\gev
\eeq
Of the remaining jets (from amongst $j_{0,1,2}$), we name the one with the
maximal azimuthal separation with $J_Z$  as $J_1$ and demand that
\beq
\Delta\phi (J_Z,J_1)>2.5.
\eeq

\item {\bf Selection 3  ($S_3$):} 
Of the jets that have {\em not} been identified as a fat jet, we
demand that at least two be $b$-tagged. In
other words, we consider the semi-inclusive final state  $N_b \geq
2$   (where we do not make any demand on the identity of the
others). In addition, we demand that the ratio of the MET and the
$p_T$ of the leading $b$-jet be less than unity.

\item {\bf Selection 4  ($S_4$):} At this stage, we optimize the fatjet
characteristics by effecting a single {\it fatjet selection} in terms of $\tau_{21}$.
We consider two choices, namely
\beq
(a) \, \tau_{21}< 0.5 \qquad {\rm or} \qquad  (b) \, \tau_{21}< 0.4\ .
\eeq
\end{itemize}

The devolution of the signal cross section, as
the selection criteria are applied
consecutively, are summarised in Table~\ref{tab:signal_selec1}.  The
corresponding numbers for the backgrounds are presented
too\footnote{A caveat needs to be enetered here. Keeping the
$S_3$ selection in mind, we deliberately omitted, from the table, the
multijet QCD events where the final state does not contain either of a
$b \bar b$ or a $c \bar c$ pair. As argued earlier, such backgrounds
are too small to be of any consequence here.} As even a cursory
comparison of these tables with the cross sections discussed in
Sec.\ref{subsec:bkgd} shows, $S_1$ itself results in a severe
preferential suppression of the background as compared to the signal.
The subsequent imposition of $S_2$ retains $\sim$45-70\% of the signal
events, while rejecting nearly $90\%$ of the total
background\footnote{It is obvious, though, that the suppression works
differently for the individual contributions to the background.}  that
survives $S_1$.
\begin{table}[!h]
  \begin{center}
\begin{tabular}{|c|c|c|c|c|c|}
\hline
\hline
$m_B$ (TeV)&$S_1$ & $S_2$ & $S_3$ & $S_4$ (a) & $S_4$ (b)\\
\hline
1.2 & 122.1  & 56.25 & 24.14 & 19.32 & 16.9\\
1.8 & 15.5 & 10.52 & 3.89  & 3.1   & 2.7\\
2.2 & 4.1  & 2.9  & 0.98 &  0.74 & 0.64\\
\hline										
\hline
\end{tabular}
\vskip 10pt
\begin{tabular}{|c|c|c|c|c|c|}
\hline
\hline
Backgrounds &$S_1$ & $S_2$ & $S_3$ & $S_4$ (a) &$S_4$ (b)\\
\hline
$b\bar b$+jets & $3.45\times 10^4$  & $4.68\times 10^3$  & 560  & 283 & 140\\
$V$+jets       & $7.5\times 10^3$  & 829 & 294  & 141 & 97\\
$t\bar t$+jets & $2.97\times 10^3$   & 973  & 433   & 325 & 238\\
$Others$       & 356    & 92  & 9.0  & 7.4  & 6.2\\
\hline
Total          & $4.53\times 10^4$  & $6.57\times 10^3$ & $1.3\times 10^3$ & 737 & 481\\
\hline
\hline
\end{tabular}
\end{center}
    \caption{\em {\em (Top)} The variation of the cross section for
    the $Z$-fatjet signal (B$\rightarrow$bZ) in (fb), for each of the
    BPs, as subsequent selection cuts are imposed for LHC operating at
    $\sqrt{s} = 13$~TeV. We use $\kappa =0.5$ and $R=0.5$. {\em
    (Bottom)} The corresponding variation for the various
    contributions to the background.} \label{tab:signal_selec1}
\end{table}

As is evident from the tables, the signal-to-background ratio improves
significantly once  $N_b \geq 2$ is demanded, with the improvement being driven
by that in the major channel, namely QCD-driven inclusive-$b\bar b$
production. This can be understood by realizing that the additional
jets in such processes would often appear close to one of the $b$-- or
$\bar b$-induced ones and, together, would constitute the fat jet
(thereby leaving only one $b$-tagged jet amongst the rest).  It
should be pointed out at this stage that the remaining component of
$S_3$, namely ${\rm MET}/p_T \leq 1$ for the leading $b$-tagged jet,
is particularly effective in reducing the background contributions.  A
relatively large value of this ratio, especially for the inclusive
$t\bar t$ production background, has two primary sources. For one,
such processes are associated with multiple
jets, with the associated MET accruing primarily
from jet-energy mismeasurements. As for the events associated with
semileptonic decays of the top, note that if the associated lepton be
the electron or the muon, then an event with a large $p_T$
carried by the neutrino would, most likely, be
eliminated by the isolated lepton veto. This, however, does not hold
for the tau-events, primarily because of the large semileptonic
branching fraction for the latter and also because the leptonic decays
tend to leave smaller momenta for the ensuing electron/muon. Thus, such
events account for a large fraction of the
high-MET background events.

As for selection $S_4$, since it pertains only to the fatjet, it is
naturally independent of the value of $N_b$. Furthermore, the
alternative $(a)$ is, understandably, less restrictive as compared to
$(b)$. As can be seen from a comparison of the two units of Tables
\ref{tab:signal_selec1}, 
imposing $S_4(a)$ would retain roughly $\sim 53\%$ of the total
background, but as much as 70--80\% of the signal (depending on
$m_B$). The corresponding numbers for $S_4(b)$ are $\sim 33\%$ and
64--70\% respectively. The somewhat larger improvement due to $S_4(b)$
can be traced to the fact that the stronger $\tau_{21}$ cut is better
able to selectively choose the two-prong fatjets. While a still
stronger cut would further improve the signal-to-background ratio, it
would be at the cost of signal strength and the consequent loss in the
significance. Indeed, $S_4(b)$ represents a
near-optimal choice. To understand the differing strength of
$S_4(a,b)$ when applied to the different contributions to the
background, we draw attention to the fact that the two prong nature is
dominant only for the single-top background (see the discussion in
Sec.\ref{subsec:fatjet_charac} and, in particular,
Fig.\ref{fig:fatjet_bkgd_0.5}). In comparison, the signal tends to
exhibit a more pronounced two prong nature (Fig.\ref{fig:fatjet_0.5}).
For $m_B \gapp 2\tev$, though, there is little advantage in applying
$S_4$, a consequence of the flattening of the $\tau_{21}$ distribution
as discussed in Sec.\ref{subsec:fatjet_charac}.

To further enhance the significance, we consider the invariant mass
$M_{\rm inv} \equiv m(J_Z, J_1)$, with $J_1$ defined as in selection
$S_2$.  Since, for the signal events, these two jets are expected to
have arisen from the decay of the $B$, we concentrate on intervals $
|M_{\rm inv} - m_B| \leq 3 \Gamma_B$, for a given $m_B$. For
$\kappa=0.5$, the total widths $\Gamma_B$ (as calculated in
Sec.\ref{subsec_decay_width}), for the benchmark points $m_B = 1.2,
1.8$ and 2.2 TeV are 32, 66.5 and 90 GeV respectively. As even a
cursory examination of Table~\ref{tab:signif_Z} shows, this
restriction on $M_{\rm inv}$ is extremely useful in
accentuating the signal-to-noise ratio.  Calculated for an integrated
luminosity of $300\,{\rm fb}^{-1}$, the consequent discovery
significances (defined as $\sigma\equiv S/\sqrt{B}$, where $S \, (B)$
represent the total number of signal(background) events) are also
presented in Table~\ref{tab:signif_Z}.

\begin{table}[!h]
\begin{center}
\begin{tabular}{|c|ccc||ccc|}
\hline
\hline
{\bf R=0.5, b-jet$\geq 2$} & After $S_3$&&& After $S_{4}(b)$&&\\
\hline
 $m_B \pm 3 \Gamma_B$ (TeV)& $1.2\pm 3\Gamma$ & $1.8\pm 3\Gamma$ & $2.2\pm 3\Gamma$ & $1.2\pm 3\Gamma$ &$1.8\pm 3\Gamma$ & $2.2\pm 3\Gamma$\\
\hline
Signal(fb)& 11.34 & 2.11 & 0.55 & 7.94  & 1.47 & 0.36\\
Background(fb)& 146.2 & 41.8 & 22.0 & 60.8  & 18.0  &  9.1 \\
Significance ($\sigma_{300}$)& 16.25 & 5.63 & 2.03 & 17.63 & 6.02 & 2.05\\
\hline
\hline
{\bf R=0.8, b-jet$\geq 2$} & After $S_3$&&& After $S_{4}(b)$ &&\\
\hline
\hline
Significance ($\sigma_{300}$) & 10.3 & 1.6 & 0.9 & 9.4 & 1.5 & 0.8\\
\hline
\hline
\end{tabular}
\end{center}
\caption{\em The signal and total background cross section after $S_3$ and $S_4$(b)
in the $|m(J_Z, J_1) - M_B| \leq 3 \Gamma_B$ bin for $\kappa=0.5$ is
presented for the $Z$-fatjet signal (B$\rightarrow$bZ). The comparison
among the significances($\sigma$), evaluated at 300 fb$^{-1}$ is also
shown for b-jet $\geq 2$. We also quote the
significance($\sigma$), evaluated at 300 fb$^{-1}$ for $R=0.8$ for
comparison.}
\label{tab:signif_Z} 
\end{table}

It is instructive to peruse Table~\ref{tab:signif_Z} carefully. For
one, the cut $S_4$(a) would result in smaller significance values as
compared to those for the alternative, namely, $S_4$(b). This is just
a vindication of our earlier argument that a stronger cut on the
subjettiness ratio $\tau_{21}$ preferentially removes the background
events.  An analogous (and expected) feature is afforded by a
comparison with the significance reach for $R=0.8$, presented in
Table~\ref{tab:signif_Z} for a cut of $\tau_{21} > 0.5$. With the
$Z$-fatjet getting progressively more collimated as $m_B$ increases,
an increase in $R$ (from 0.5 to 0.8) would not imply a significant
increase in the number of signal events, catching only a few extra
events with larger radiation. On the contrary, a much larger fraction
of the background events, especially those accruing from top decays,
would now be accepted by the algorithm, leading to a significant
decrease in the ensuing significance.

\subsubsection{The $H$-fatjet Channel}{\label{Higgs_result}}

We, now, consider the second possibility, namely when the vector-like
quark decays in the $B \to H b$ channel. As we have discussed earlier,
$Br(B \to H b) \approx Br(B \to Z b)$ and, with
the hadronic branching fraction of the Higgs not
being too different from that of the $Z$, nominally, the signal
strengths should be similar. On the other hand, many of the SM
backgrounds, such as those originating from $W/Z$ or top-production,
are expected to reduce drastically, on account of these peaking away
from $m_H$. In other words, the naive expectation would be that this
signal (comprising of one $H$-fatjet with two additional jets) would
be significantly more visible against the background, as compared to
that in the previous section. We would see, though, that this
is borne out to a great extent, though not
to the degree that a naive estimate would indicate.

With the mass difference $m_H- m_Z \ll m_B$, one would expect the
signal profile to be quite similar in the two cases. Consequently, we
retain all the earlier selection cuts, except for making the obvious
alteration in $S_2$, which, for the $H$-fatjet, now reads
\beq
    m_J \in [110, 140] \gev \quad {\rm and} \quad p_T(J_H) > 500\gev\
    .
\eeq
The devolution of the cross section, as the selection criteria are
applied consecutively, is summarised in Table~\ref{tab:signal_selec2}
for both the signal benchmark points as well as the backgrounds. Since
our methodology is virtually identical to that in the preceding
subsection, much of the details are very analogous, and we shall
just concentrate on identifying the major differences.

\begin{table}[!h]
  \begin{center}
\begin{tabular}{|c|c|c|c|c|c|}
\hline
\hline
$m_B$ (TeV)&$S_1$ & $S_2$ & $S_3$ & $S_4$ (a) & $S_4$ (b)\\
\hline
1.2 & 129.7 & 36.0 & 18.0 & 14.76 & 11.91\\ 1.8 & 17.23 & 7.27 &
3.15 & 2.42 & 1.88 \\ 2.2 & 4.5 & 2.1 & 0.82 & 0.61 & 0.46 \\
\hline
\end{tabular}
\vskip 10pt
\begin{tabular}{|c|c|c|c|c|c|}
\hline
\hline
Backgrounds &$S_1$ & $S_2$ & $S_3$ & $S_4$ (a) &$S_4$ (b)\\
\hline
$b\bar b$+jets& $3.45\times 10^4$ & $1.9\times 10^3$ & 304 & 167 &
131\\ $V$+jets & $7.5\times 10^3$ & 339 & 65 & 43 & 31\\ $t\bar
t$+jets & $2.79\times 10^3$ & 635 & 346 & 256 & 166 \\ $Others$ & 356
& 42 & 3.2 & 2.4 & 2.0 \\
\hline           
 Total & $4.53\times 10^4$ & $2.95\times 10^3$ & 719 & 468 & 330 \\
\hline
\end{tabular}
\end{center}
    \caption{\em {\em (Top)} The variation of the cross section for
    the $H$-fatjet signal (B$\rightarrow$bH) in (fb), for each of the
    BPs, as subsequent selection cuts are imposed for LHC operating at
    $\sqrt{s} = 13$~TeV. We use $\kappa =0.5$ and $R=0.5$. {\em
    (Bottom)} The corresponding variation for the various
    contributions to the background.}  \label{tab:signal_selec2}
\end{table}

That the backgrounds, on application of the cut $S_1$ alone, should
remain identical to those in Table~\ref{tab:signal_selec1} is
obvious. At this stage, the signal too remains very close to that in
the previous case, a testament to the near equality of the raw signal
strengths. This near equality is altered significantly on the
imposition of $S_2$, and is a consequence of the slightly higher mass
of the Higgs as compared to the $Z$. Note that the demand for a
two-prong jet of mass 110-140 GeV indirectly translates to a minimum
$p_T$ requirement (since $\Delta R \approx 2m_J/p_T$).  Consequently a
non-negligible fraction of events containing a putative fatjet with
$p_T > 500\gev$ would no longer be reconstructed as one. Of course,
this effect can be offset by allowing for a larger $R$. However, such
a choice would also entail a larger background count. Furthermore, the
somewhat flatter $\tau_{21}$ distribution that, say $R =
0.8$ entails (see Sec.~\ref{subsec:fatjet_charac}), renders the
$S_4(a/b)$ cuts less effective.

$S_3$, being only a requirement of the non-fatjet components of the
event being $b$-tagged, understandably has nearly equal efficiency for
the two channels. The small difference in efficiencies is attributable
to the slightly different phase space distributions of the $b$
emanating from the heavy $B$-decay.  And while the $S_4$ cut is
expected to have a different efficiency in the two cases, owing to a
difference in the corresponding $\tau_{21}$ distributions, the effect
is relatively small.
\begin{table}[!h]
  \begin{center} \begin{tabular}{|c|ccc||ccc|}
\hline
\hline
{\bf R=0.5, b-jet$\geq 2$} & After $S_3$&&& After $S_{4}(b)$ &&\\
\hline
 $m_B \pm 3 \Gamma_B$ (TeV)& $1.2\pm 3\Gamma$ & $1.8\pm 3\Gamma$ &
 $2.2\pm 3\Gamma$ & $1.2\pm 3\Gamma$ &$1.8\pm 3\Gamma$ & $2.2\pm
 3\Gamma$\\
\hline
Signal(fb)& 7.84 & 1.8 & 0.56 & 5.2 & 1.0 & 0.35\\ Background(fb)&
68.7 & 20.1 & 14.0 & 26.2 & 5.5 & 3.4 \\ Significance
($\sigma_{300}$)& 16.38 & 6.95 & 2.59 & 17.67 & 7.39 & 3.29\\
\hline
\hline
{\bf R=0.8, b-jet$\geq 2$} & After $S_3$&&& After $S_{4}(b)$ &&\\
\hline
\hline
Significance ($\sigma_{300}$) & 12.2 & 5.1 & 1.2 & 12.5 & 5.0 & 1.2\\
\hline
\hline
\end{tabular}
\end{center}
\caption{\em As in Table~\ref{tab:signif_Z} but for H-fatjet (B$\rightarrow$bH) channel}
\label{tab:signif_H} 
\end{table}

As in the preceding case, the significance can be further enhanced by
restricting the invariant mass $M_{\rm inv} \equiv m(J_H, J_1)$ to the
interval $ |M_{\rm inv} - m_B| \leq 3 \Gamma_B$.  The consequent
discovery significances, as calculated for an integrated luminosity of
$300\,{\rm fb}^{-1}$ are presented in Table~\ref{tab:signif_H}. Note
that we obtain marginally better significance in the
$J_H$ scenario for $m_B>1.8$ TeV, compared to
$J_Z$ channel, primarily due to 
extra suppression of the background. Once again, a stronger cut on the
subjettiness ratio $\tau_{21}$ preferentially removes the background
events and, for brevity's sake, we present the results only for
$S_4(b)$. And, as in the previous case, the use of a larger $R$ only
serves to dilute the significance.

\vskip 30pt
\subsubsection{Discovery Projection}
It is interesting to note that, notwithstanding the differences in the
effective efficiencies for the signal and background events, the two
channels have very similar sensitivities. With the search strategy
being very similar too, it is, thus, worthwhile to combine the two
sets of results to reach an enhanced sensitivity and this is what we
now embark on.
\begin{figure}[!h]
\begin{center}
\includegraphics[width=7.5cm,height=6.5cm]{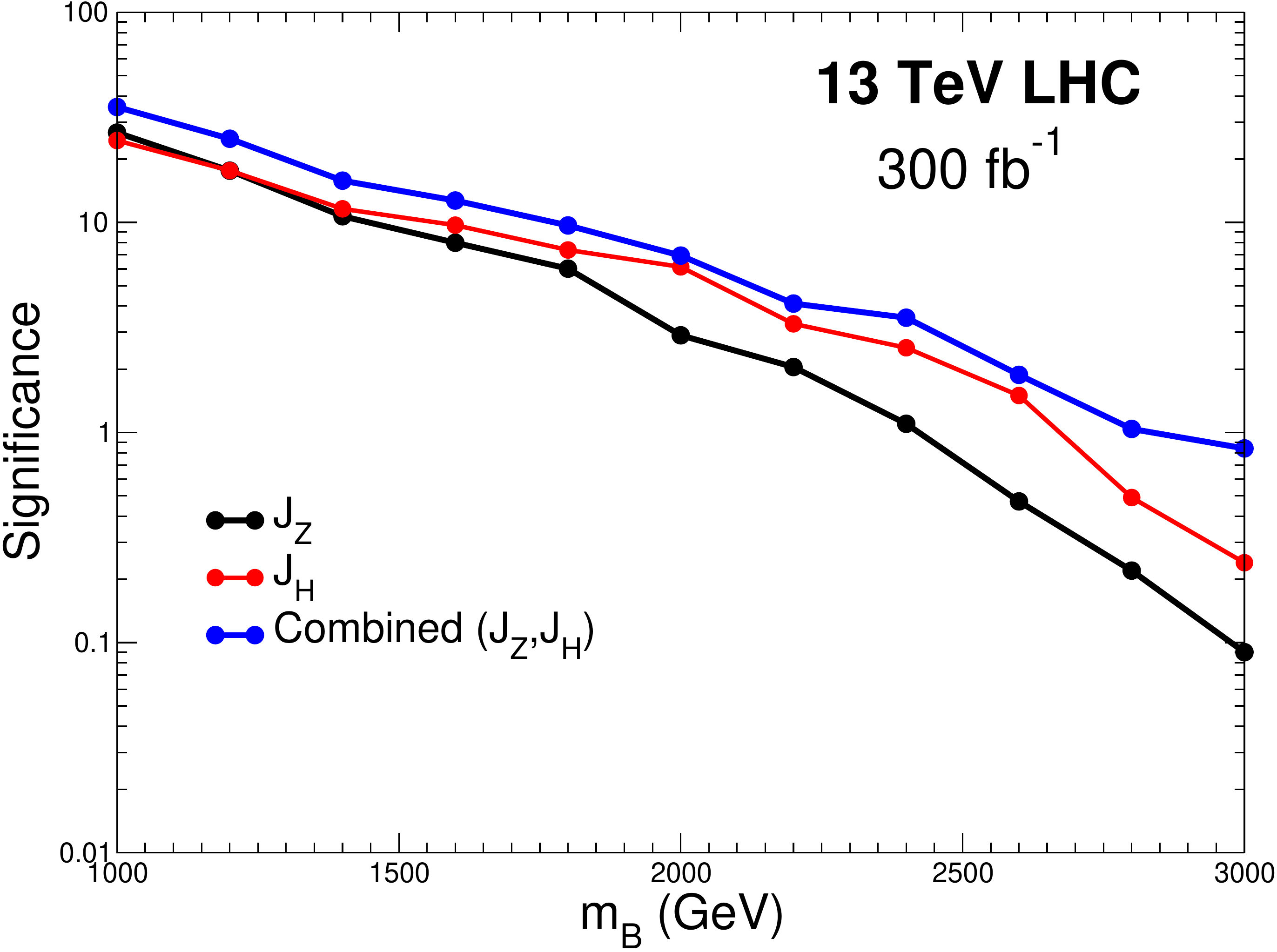}
\end{center}
\caption{\em Significance as a function of $B$ mass, for two different fatjet scenario, $Z$ and $H$,  
at 300 fb$^{-1}$ luminosities for b-jets $\geq 2$ and $S_4$(b). We
have assumed $R=0.5$, $\kappa=0.5$.}
\label{fig:12}
\end{figure}

\begin{figure}[!h]
\begin{center}
\includegraphics[width=7.0cm,height=7.0cm]{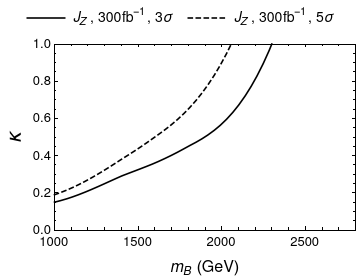}
\includegraphics[width=7.0cm,height=7.0cm]{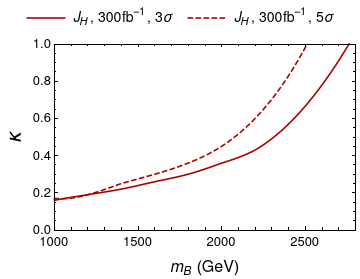}
\end{center}
\caption{\em Contours of $3\sigma$ and $5\sigma$ significances in the
  plane of $\kappa$ vs $m_B$, at luminosity of $300fb^{-1}$ for both
$J_Z$ (left) and $J_H$ (right) fatjet case.  Events are selected with
b-jets $\geq 2$ and $S_4$(b). We have assumed
$R=0.5$.}  \label{fig:13}
\end{figure}
We demonstrate the significance as a function of $m_B$ in
Fig~\ref{fig:12} for $\kappa=0.5$ for an integrated luminosity of $300\,{\rm fb}^{-1}$ 
It is $\geq 3\sigma$ for VLQ masses $< 2.2 (2.0)$ TeV
and $\geq 5\sigma$ for VLQ masses $< 2.05 (1.84)$ TeV in the
$J_H$($J_Z$) channel. Clearly, upto 1.6 TeV, both
the channels predict almost same significance, but 
thereafter the $H$-fatjet channel is clearly the more sensitive one. This small difference in significance at smaller $m_B$ values was expected because of the more restrictive nature of $S_2$ cut in the case of $H$, already discussed in Sec.\ref{Higgs_result}.
Combining the two channels, the joint significance reach is $\geq 5\sigma$ ($\geq 3\sigma$) for VLQ masses $<2.12 (2.42)$ TeV. 
In Fig~\ref{fig:13} we have plotted the contours of 3$\sigma$ and
5$\sigma$ significance in the $\kappa$ vs. $m_B$ plane for both $J_Z$
and $J_H$ channels. It is obvious from the plot that with the $J_H$ channel alone, it is possible to probe smaller values of $\kappa$ with 3/5$\sigma$ significance at a fixed $m_B$.
\section{Conclusions and Outlook}
\label{sec:outlook}
Non-chiral (or vector-like) fermions have been invoked in a multitude
of theories, to address a medley of issues. The more theoretical
concerns range from dynamical breaking of the electroweak symmetry or
alleviating the little hieracrchy problem in a class of theories to
models providing a portal for Dark Matter.  Concerns that are more
immediate include explaining muon or electron anomalous magnetic
moments or several puzzles in flavour physics.  In particular, topless
vector-like doublets ($B,Y$) have been shown to provide a redressal of the
tension between global fits to electroweak precision tests and the
forward backward asymmetry in $b\bar b$ production at LEP/SLC.

In our quest to examine the status of such resolutions, we begin by
delineating the constraints on the mixings of the $B$ with light
counterparts, as obtained from measurements of electroweak precision
variables and other flavour-sector processes. Such an exercise also
helps determine the branching fractions for the $B$-quark decays.  We,
then, investigate possible LHC signatures of such a $B$-quark.  The
very structure of such theories imply a sizable a transition
chromomagnetic moment $\kappa$, allowing for single-production such as
$g g \to \bar b B$. Despite the smallness of $\kappa$, such a
production channel could easily dominate QCD-driven pair production for
significantly large $m_B$. 

In this paper, we have concentrated on the the single production of
such a $B$ in association with a bottom-jet, with the $B$,
subsequently decaying to a $Z/H$ boson and another $b$-jet. While such
channels would be expected to be overwhelmed by large backgrounds
(unless the $Z$ decays hadronically), for a sufficiently large mass
$m_B$ (such as to evade the current experimental limits), the bosonic
daughter is boosted so as to often manifest itself as a
fatjet. Exploiting this feature, we have striven to identify both the
kinematic features as well as the optimal jet reconstruction
algorithms. In particular, the two-prong nature of the $Z/H$--fatjet
is best evinced by the choice $R= 0.5$ for the radius parameter,
rather than the more conventional $R=0.8$.  Complemented by the use of
the subjettiness ratio $\tau_{21}$ as well as the imposition of a
differential of cuts on the transverse momenta (a very stiff one for
the fatjet and more moderate ones for the two subleading jets), the
signal-to-background ratio can be improved to a great extent. As a
final discriminator, we use the fact that the fatjet $J_{Z/H}$ and the
jet with the maximal azimuthal separation from it should,
preferentially, reconstruct the mass of the $B$.

The consequent significance ratio is quite handsome in either of the
channels ($Z/H$) and, together, amount to more than $3\sigma$ for
$m_B \sim 2.5\tev$ with an integrated luminosity of only $300\, {\rm
fb}^{-1}$. The HL-LHC option would, understandably, push up the limit
significantly. It is worthwhile to notice that it is the Higgs channel is
the more sensitive one, a consequence of the twin facts that the
branching fractions $Br(B \to b + Z/H)$ are very similar and that the
SM backgrounds for the Higgs-fatjet channel are significantly smaller.

\begin{figure}[!h]
\begin{center}
\includegraphics[width=8.1cm,height=7.0cm]{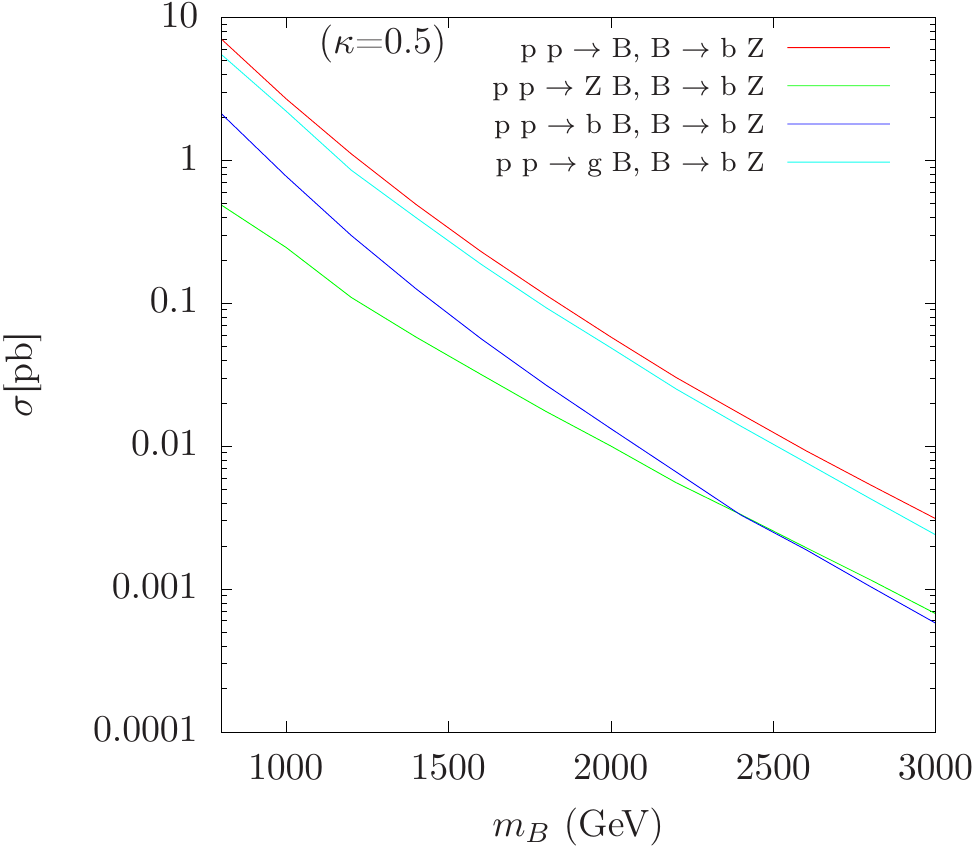}
\includegraphics[width=8.0cm,height=7.0cm]{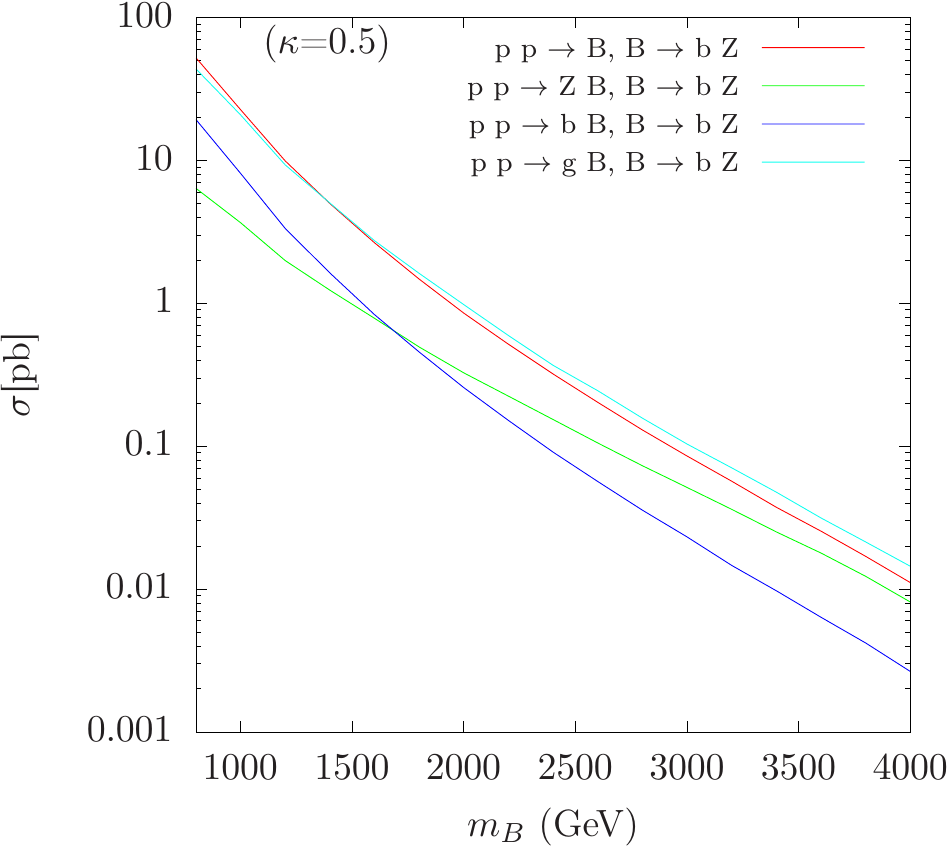}
\end{center}
\caption{\em (Left) LO cross sections in 5FS using (NNPDF23LO1) for
 various relevant processes at 13 TeV. (Right)
 Same for a center of mass energy of 27 TeV.
}
\label{fig:all_channels}
\end{figure}

A key component of our search strategy is the requirement that at
least two of the high-$p_T$ jets, other than the fatjet, should be
$b$-tagged ({\em i.e., $N_b \geq 2$}) While this was particularly
useful in suppressing the SM backgrounds, the adoption of this
criterion also meant that some of the potential signal processes too
had to be left out as discussed in Sec.\ref{sec:collider}. 
Indeed, as Fig.~\ref{fig:all_channels} shows, the resonant
production channel $p p \to B$ and even $p p \to g B$ have cross sections
significantly larger than that for the channel we worked with. However,
with the QCD-multijet background now increasing manifolds, the pursual of
such channels would require analyses much more sophisticated than what we
have used here. There is hope, though, that machine learning techniques could
unmask the corresponding signal and we hope to return to this question at a
later date.

Potentially much more interesting is the channel $pp \to B Z$. As can be seen from Fig.~\ref{fig:all_channels}, while this rather subdominant for small $m_B$, for $m_B \gapp 2.3 \tev$ it compares well with our channel and even overtakes it. While one $b$-jet is now lost, it is now compensated by a $Z$ ($H$), which has its own distinct signature, even for hadronic decays. This, potentially would substantially increase the signal-to-background ratio, primarily on account of a much reduced background.

Similarly, the production of a much heavier $B$ (than what we have
considered) is severely suppressed at the LHC. This, though, would not
be the case at a future circular collider. In paricular, note
that the $pp \to B Z$ mode now dominates over our chosen mode starting
with a much smaller $m_B$. However, such a machine would have its own
imprint on multiple issues such as the nature of QCD radiation, the
efficiencies etc., and a naive extension of our analysis would not be
tenable. In particular, new jet substructure observables might be
called for. These are only a subset of open questions that we hope to
address in future.

\vskip 20pt

\section*{Acknowledgments}
\vspace{0.5cm}
The authors thank Brajesh Choudhary for access to computers
brought under the aegis of the grant SR-MF/PS-02l2014-DUB (G) of the
DST (India). KD acknowledges Council for Scientific and Industrial Research (CSIR), India for JRF/SRF fellowship with the award letter no. 09/045(1654)
/2019-EMR-I. N.K. acknowledges the support from the Dr. D. S. Kothari Postdoctoral scheme (201819-PH/18-19/0013).
\bibliographystyle{JHEPCust.bst}
\bibliography{vecB.bib}
\end{document}